\documentclass[lettersize,journal]{IEEEtran}
\usepackage{amsmath,amsfonts}
\usepackage{amsthm}
\usepackage{subcaption} % 更现代的subfig替代
\usepackage[hyphens]{url}       % 正确处理 URL/DOI 换行
\usepackage[ruled,norelsize,vlined,linesnumbered]{algorithm2e}
\makeatletter
\newcommand{\removelatexerror}{\let\@latex@error\@gobble}
\makeatother
\usepackage{amssymb}
\usepackage{array}
\usepackage{caption}
\usepackage{textcomp}
\usepackage{stfloats}
\usepackage{url}
\usepackage{verbatim}
\usepackage{graphicx}
\usepackage{cite}
\usepackage{multirow}
\usepackage{makecell}
\usepackage[colorlinks,
            linkcolor=blue,
            anchorcolor=blue,
            citecolor=blue,
            urlcolor=black]{hyperref}
\makeatletter
\renewcommand{\eqref}[1]{%
    \textcolor{blue}{\hyperref[#1]{(\ref*{#1})}}%
}
\makeatother
\usepackage{booktabs} % 三线表支持
\usepackage{caption}  % 表格标题控制
\usepackage{tabularx}    % 自动调整宽度
\usepackage{graphicx}
\usepackage[mathscr]{eucal}

\usepackage[labelformat=simple]{subcaption}
\begin{document}

\title{Dependency- and Layer-Aware Microservice Workflow Offloading and  Service Image Caching for Edge Environments}

\author{Zhongxiao Wang, Yueshen Xu*,~\IEEEmembership{Member,~IEEE,} Qingshan Li, Xinkui Zhao, Wei Shao,~\IEEEmembership{Member,~IEEE,} Shuiguang Deng,~\IEEEmembership{Senior Member,~IEEE,} Rui Li,~\IEEEmembership{Member,~IEEE}
        % <-this % stops a space
\IEEEcompsocitemizethanks{
\IEEEcompsocthanksitem This paper is funded by National Key Research and Development Program of China (2023YFF0905100), National Natural Science Foundation of China (62472338 and 62172320), and Open Foundation of Yunnan Key Laboratory of Software Engineering (2023SE301) (\textit{Corresponding author: Yueshen Xu, and Yueshen Xu contributes equally with Zhongxiao Wang, so he is also the co-first author}). 
\IEEEcompsocthanksitem Zhongxiao Wang, Yueshen Xu, Qingshan Li, and Rui Li are with the School of Computer Science and Technology, Xidian University, Xi'an 710126, China. Yueshen Xu is also with Yunnan Provincial Key Laboratory of Software Engineering, Kunming 650504, China. E-mails: zhongxiaowang@stu.xidian.edu.cn, ysxu@xidian.edu.cn, qshli@mail.xidian.edu.cn, and rli@xidian.edu.cn.
\IEEEcompsocthanksitem Xinkui Zhao is with the School of Software Technology, Zhejiang University, Ningbo 315048, China. E-mail: zhaoxinkui@zju.edu.cn.
\IEEEcompsocthanksitem Shuiguang Deng is with the College of Computer Science and Technology, Zhejiang University, Hangzhou 310027, China. E-mail: dengsg@zju.edu.cn.
\IEEEcompsocthanksitem Wei Shao is with the School of Computer Science and Engineering, University of New South Wales, Sydney 2052, Australia. E-mail: phdweishao@gmail.com.
}
%\thanks{Manuscript received December 25, 2022; revised August 16, 2021.}
% \thanks{Manuscript received April 19, 2025; revised August 16, 2025}

}

\markboth{IEEE Transactions on Parallel and Distributed Systems,~Vol.~XX, No.~XX, XX~2025}%
{Shell \MakeLowercase{\textit{et al.}}: A Sample Article Using IEEEtran.cls for IEEE Journals}


\maketitle

\begin{abstract}
The microservice architecture has been applied broadly in many mainstream computing environments. As one of the most prevalent environments, edge computing also widely employs microservices to handle diverse requests and tasks. In practical scenarios, microservices usually constitute workflows that are built based on service dependencies to execute tasks. This offers an opportunity to explore the offloading technology to better harness resources and accelerate task execution. Unfortunately, this problem has not been paid attention to by existing research, and thus some valuable resources (e.g., microservice image cache) remain obscure. To fill this gap, we innovatively study the problem of joint optimization for microservice workflow offloading and service image caching in edge.
This problem is challenging due to several issues such as the complexity of its solution space, intricate relationships among shared image layers, long-range dependencies between workflow tasks, and the absence of a real-world collection of service images. To address these issues, this paper proposes an innovative dependency- and layer-aware workflow offloading and image caching framework for microservices. Our framework consists of 1) a novel mechanism, \textit{Layer-Aware Cross-Attention (LACA)}, for deeply exploring image layer sharing relationships, 2) a new mechanism, \textit{Dependency-Aware Multi-Head Cross-Attention (DAMH-CA)}, for comprehensively mining long-range dependencies among workflow tasks, and 3) leverages a Hierarchical Deep Reinforcement Learning (HDRL) to decouple the solution space for better convergence. We collected a real-world collection of microservice image layer data and published both this collection and experimental codes on GitHub. Extensive results demonstrate that our framework achieves superior performances, for example, a 22. 38\% reduction in average task completion time compared to baselines and significantly-increased image hit rates.
\end{abstract}


\section{Introduction}
\IEEEPARstart{I}{n} recent years, with the unprecedented expansion of intelligent devices around the world and the rapid development of advanced communication technologies, latency-sensitive and computation-intensive services and applications are exponentially proliferating (e.g., Large Language Models/LLMs \cite{LLM_TMC_2025}, Internet of Vehicles \cite{IoV_TITS_2024}, and intelligent robotics \cite{Robot_TVT_2023}). Faced with the conflict between the increasing demand for Quality-of-Service (QoS) and limited resources, edge computing \cite{UserMobile_TNSM_2025} as a new computing paradigm demonstrates significant potential to drive and support these emerging applications. In this paradigm, end users can offload their tasks to edge servers deployed around the network infrastructure (e.g., 5G base stations, smart access gateways, and wireless access points) to obtain low-latency computing services\cite{TransRate_IOTJ_2022}. 

\begin{figure}[htbp] % h:此处, t:顶部, b:底部, p:单独页面
    \centering
    \includegraphics[width=0.48\textwidth]{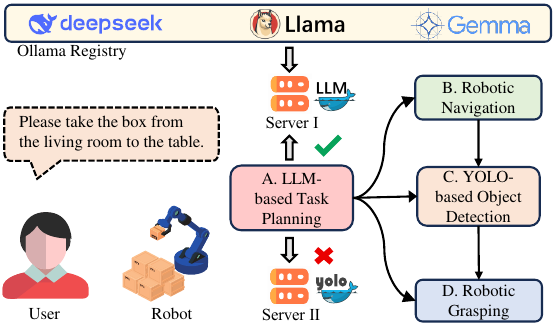}
    \caption{A motivation example to show that service deployment is essential for workflow task offloading, and we can use the microservice architecture to perform service deployment.} % 自动编号
    \label{fig:robot} % 用于交叉引用
\end{figure}

\textbf{In practical applications of edge computing, deploying task-dependent services becomes essential for edge servers to execute offloaded tasks successfully} \cite{NeedCacheService_INFOCOM_2020}. We offer an example of an LLM-based robotic object retrieval task\cite{LLMbasedRobot_CCC_2024}, as shown in Fig. \ref{fig:robot}. This task can be modeled as a workflow and decomposed into four interdependent tasks, including A: LLM-based task planning, B: robotic navigation, C: YOLO-based object detection, and D: robotic grasping. When employing edge computing to reduce task completion time, the task A should be offloaded to edge server I (deploying the LLM service) to execute successfully, while task C should be offloaded to edge server II (deploying the YOLO service) to complete its computation. However, traditional virtual machine-based service deployment approaches are difficult to achieve rapid service migration and deployment in edge server clusters due to several reasons (e.g., substantial resource consumption and cumbersome environment configuration processes). Some works introduce the microservice architecture into edge environments to perform dynamic service deployment in edge networks \cite{Microservice_Edge_1}\cite{Microservice_Edge_3}. The microservice architecture is a lightweight solution based on container technology that decomposes traditional monolithic applications into multiple loosely coupled microservices \cite{Containers_TSE_2024}. Each of these microservices operates within an isolated container environment and is often invoked as a component part of a workflow\cite{workflow_TPDS_2025}. In contrast to virtual machines, containers share the operating system kernel but are isolated in terms of process visibility (through namespaces) and resource usage (through control groups)\cite{namespace_TIFS_2024}. Therefore, containers enable faster service deployment with clearly lower resource consumption compared to virtual machines\cite{fasterdeploy_TPDS_2021}. 

Docker\footnote{https://www.docker.com/} is one of the most widely adopted container technologies. It encapsulates all runtime environments of a microservice into a portable microservice image, enabling users to deploy and remove microservices exclusively through Docker commands. \textbf{These runtime environments are stored as image layers and can be shared among different images, which give microservice images the specialty of layer sharing} \cite{Layer_INFOCOM_2021}. When we need to deploy a microservice to a server, only those image layers not locally cached need to be pulled from the registry center. This significantly reduces image pull delay during microservice deployment. For instance, Ollama\footnote{https://ollama.com/}, an open-source framework for operating LLMs, is implemented through Docker containers. To deploy the LLM service required by task A on server I in the example shown in Fig. \ref{fig:robot}, users can simply utilize the Ollama commands to pull the non-local image layers of the LLM service from a registry to server I, thereby facilitating the rapid local LLM deployment.

\begin{figure}[htbp]
\centering
\begin{subfigure}{0.24\textwidth}
    \includegraphics[width=\linewidth]{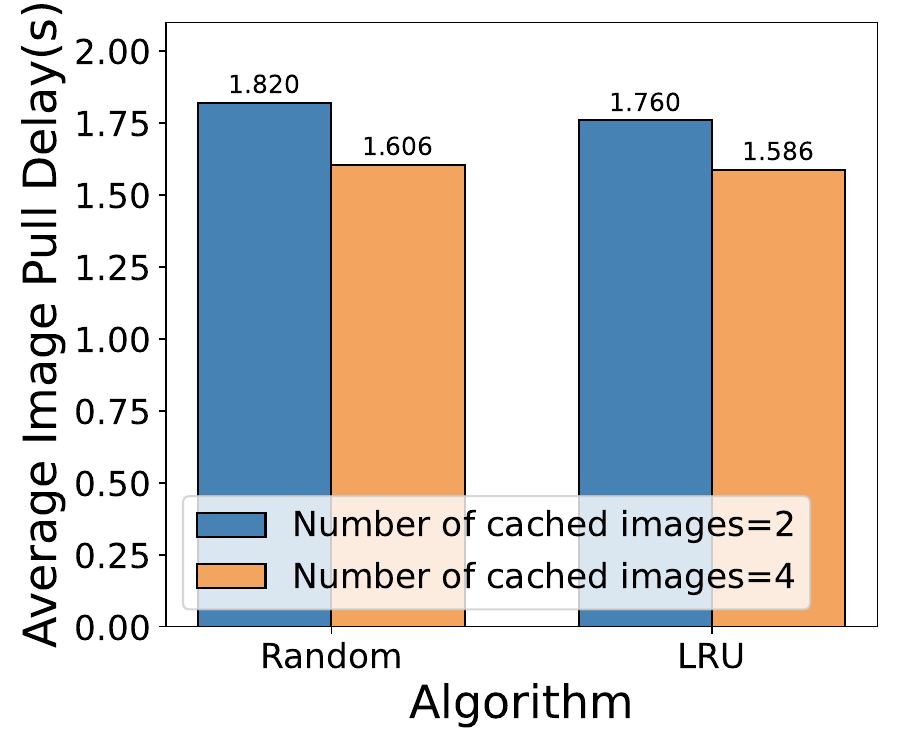}
    \caption{Average image pull delay}
    \label{Motivation:sub1}
\end{subfigure}
\hfill
\begin{subfigure}{0.24\textwidth}
    \includegraphics[width=\linewidth]{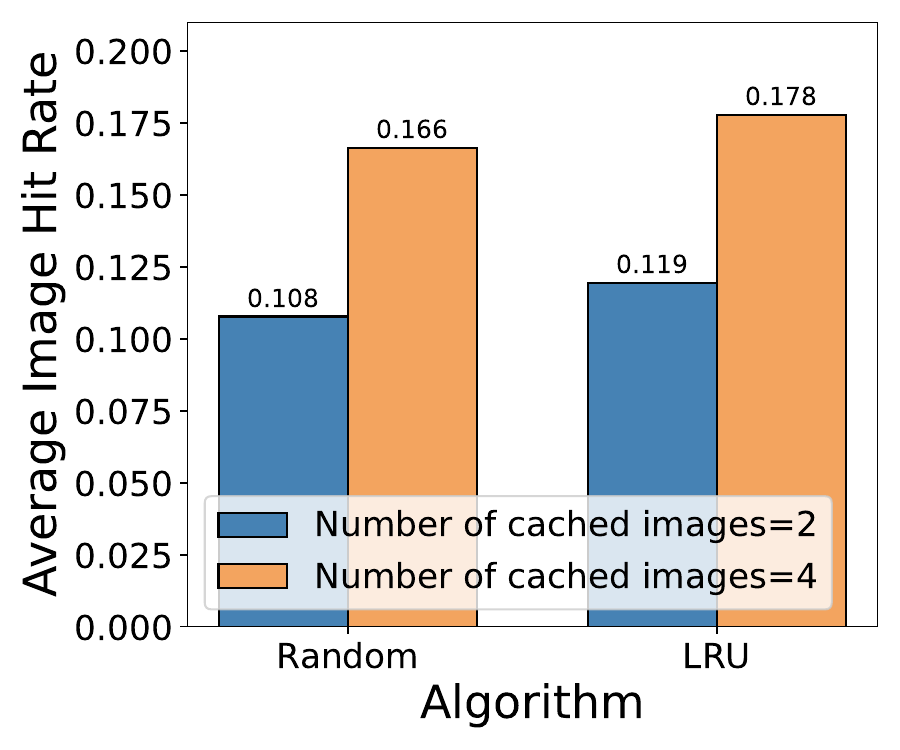}
    \caption{Average image hit rate}
    \label{Motivation:sub2}
\end{subfigure}

\caption{A preliminary experiment investigating the impact of the number of cached images on image pull delay.}
\label{Motivation}
\end{figure}

Some studies attempt to employee layer sharing among microservice images to help deploy microservices in edge networks \cite{Layer_TSC_2023}\cite{Layer_TCE_2025}. However, these studies frequently assume that only a single microservice image can be cached on an edge server simultaneously.  A running microservice usually occupies scarce memory resources, while its image is stored in low-cost storage devices. We analyzed the sizes of the 50 most popular (i.e., the 50 most frequently downloaded) microservice images in DockerHub\footnote{https://hub.docker.com}, and the sizes range from 7.8MB to 1306MB. Given that edge servers typically provide hundreds of gigabytes of storage capacity, it becomes a technically viable solution to dedicate a portion of this storage to microservice image caching. To reveal the impact of the number of cached images on image pull delay, we conducted preliminary experiments and results are illustrated in Fig. \ref{Motivation}, where we studied two strategies: Random caching and Least Recently Used (LRU) caching. We find that with an increasing number of cached images, both approaches exhibit the reduced average image pull delay while significantly improving the image hit rate. These experimental results decisively demonstrate that leveraging idle storage resources in edge servers for microservice image caching can effectively reduce image pull delay.

Based on the aforementioned discussion, we discover that optimizing microservice workflow offloading and service image caching in edge networks is a vital issue. However, this issue poses several challenges that should be addressed. 1) First, the issue constitutes a joint optimization problem involving both microservice workflow offloading and service image caching. Thus, the complexity of this problem causes traditional methods to fail to deliver satisfactory performance within their large-scale solution space. 2) Second, layer sharing relationships among different microservice images are complicated. This necessitates the development of a specific mechanism capable of comprehensively leveraging layer sharing for reducing image pull delay. 3) Third, workflow tasks in edge environments often exhibit topological heterogeneity, while the dependencies among microservices become increasingly complex along with an expanding number of users. 4) Fourth, our investigation finds the absence of a public dataset describing the real-world layer sharing relationships among microservice images, which presents an obstacle to reliably evaluating the practical value of research in this domain. This paper makes the following primary contributions to address the above challenges:
\begin{itemize}
    \item[$\bullet$] \textbf{To the best of our knowledge, we are the first to address the microservice workflow offloading and service image caching for edge environments.} We formulate this problem as an integer nonlinear programming problem with an optimization objective of minimizing task completion time and propose a Dependency and Layer-aware Hierarchical Deep Reinforcement Learning (DLA-HDRL) approach to efficiently solve this problem through decomposing the solution space.
\end{itemize}

\begin{itemize}
    \item[$\bullet$] \textbf{To fully leverage the layer sharing relationships among microservice images, we propose a Layer-Aware Cross-Attention (LACA) mechanism.} This mechanism computes attention vectors between task images and server-cached images, enabling the proposed DLA-HDRL approach to extract layer sharing features, thereby reducing task offloading completion time.
\end{itemize}

\begin{itemize}
    \item[$\bullet$] \textbf{To address the challenges arising from the complexity of dependencies among different microservices with increasing user numbers, we propose a Dependency-Aware Multi-Head Cross-Attention (DAMH-CA) mechanism.} Specifically, the cross-attention component computes attention vectors among different image sequences, while the multi-head attention component extracts long-range dependency features. This integrated mechanism enables DLA-HDRL to make efficient image caching decisions.
\end{itemize}

\begin{itemize}
    \item[$\bullet$] \textbf{Given the absence of a public dataset that contains layer sharing relationships among microservice images, we collected layer configuration information from 50 of the most popular microservice images on DockerHub.} This collection has been published on GitHub. To evaluate the performance of our DLA-HDRL approach, we constructed an experimental platform based on several real-world datasets. Both the trained models from this platform and the source code are equally available on GitHub.
\end{itemize}

\section{Related Work}
\label{Sec:RelatedWork}
In this section, we review three categories of related problems, which are microservice deployment in edge environments, microservice offloading, and microservice workflow in clouds.

\subsection{Microservice Deployment in Edge Environments}
Because the microservice architecture is widely employed in edge environments, some work studied the issue of microservice deployment. Lv et al. \cite{MicroserviceDeployEdge_TPDS_2022} proposed a reward sharing deep Q-learning method to solve the microservice deployment problem in edge computing with specifically considering the effect of different interaction frequencies among microservices. Taking into account both layer sharing and microservice dependencies, Zeng et al. \cite{Layer_INFOCOM_2023} proposed a randomized rounding-based microservice deployment algorithm to minimize the cost of microservice deployment in edge clouds. Wang et al.\cite{DependencyAwareMicroserviceDelpyment_TMC_2024} employed an attention mechanism to extract the computational capabilities of various devices within the network, and proposed an attention-modified soft actor-critic algorithm to address the microservice deployment problem in the context of microservice dependencies. Li et al. \cite{LayerDeployment_TSC_2025} proposed an online regularization and rounding algorithm to address the layer-aware container placement problem in edge computing. They achieved online joint optimization of request scheduling, layer-based microservice deployment, and resource provision. Xu et al. \cite{MultiObj_TCSS_2025} formulated the microservice deployment problem in edge computing as a multi-objective optimization problem and proposed a resource-aware multi-objective deep reinforcement learning algorithm based on the pareto front solutions set to efficiently solve it. Although these studies offer various solutions for the efficient deployment of microservices in edge computing, their approaches have not leveraged microservice architecture to rapidly deploy dependency services for computational tasks, thereby enhancing the QoS of task offloading.

\subsection{Microservice Workflows in Clouds}
Several studies investigated the scheduling and management of microservice workflows in clouds. Wang et al.\cite{WorkflowScheduling_TPDS_2021} proposed an elastic microservice scheduling framework to address the challenges of the two-layer resource structure consisting of virtual machines and containers. This framework aims to reduce the cost of container configuration in clouds while meeting deadline
constraints. Li et al. \cite{WorkflowScheduling_TPDS_2023} proposed a topology-aware scheduling framework that optimizes resource utilization, communication costs, and network quality in cloud environments by leveraging the topological structure of microservice workflows.  To execute microservice workflow tasks in the containerized hybrid cloud when computational resources are limited, Liu et al. \cite{WorkflowScheduling_TCC_2024} proposed a batch scheduling strategy for scheduling instances-intensive workflow tasks. Notably, these studies exclusively consider the microservice workflows in clouds while ignoring the mechanics of microservice workflows in edge environments.

\begin{table*}[t]
\caption{DISPARITIES BETWEEN OUR WORK AND EXISTING WORKs}
\label{DISPARITIES}
\centering
\begin{tabular}{@{}cccccccccc@{}}
\hline
% \textbf{Ref.} & \textbf{Dependency} & \textbf{Layer} & \textbf{Offloading} & \textbf{Caching} & \textbf{Edge}\\
Work &Microservice & \makecell{Dependency-\\workflow} & Layer & Deployment & Offloading & Caching & Edge\\
\hline
\cite{MicroserviceDeployEdge_TPDS_2022}\cite{MultiObj_TCSS_2025} & $\checkmark$ & $\checkmark$ & $\times$ & $\checkmark$ & $\times$ & $\times$ & $\checkmark$\\
\cite{Layer_INFOCOM_2023} & $\checkmark$ & $\checkmark$ & $\checkmark$ & $\checkmark$ & $\times$ & $\times$ & $\checkmark$\\
\cite{DependencyAwareMicroserviceDelpyment_TMC_2024} & $\checkmark$ & $\checkmark$ & $\times$ & $\checkmark$ & $\times$ & $\times$ & $\checkmark$\\
\cite{LayerDeployment_TSC_2025} & $\checkmark$ & $\times$ & $\checkmark$ & $\checkmark$ & $\times$ & $\times$ & $\checkmark$\\
\cite{WorkflowScheduling_TPDS_2021}\cite{WorkflowScheduling_TPDS_2023}\cite{WorkflowScheduling_TCC_2024} & $\checkmark$ & $\checkmark$ & $\times$ & $\times$ & $\times$ & $\times$ & $\times$\\
\cite{MicroserviceStore_GLOBECOM_2019} & $\checkmark$ & $\times$ & $\times$ & $\times$ & $\checkmark$ & $\times$ & $\checkmark$\\
\cite{MicroserviceOffloading_MASS_2023} & $\checkmark$ & $\checkmark$ & $\times$ & $\checkmark$ & $\checkmark$ & $\times$ & $\checkmark$\\
\cite{DeployMStoExecutionTasks_IOTJ_2024} & $\checkmark$ & $\times$ & $\checkmark$ & $\checkmark$ & $\checkmark$ & $\times$ & $\checkmark$\\
Ours & $\checkmark$ & $\checkmark$ & $\checkmark$ & $\checkmark$ & $\checkmark$ & $\checkmark$ & $\checkmark$\\
\hline
\end{tabular}

\end{table*}

\subsection{Microservice Offloading}
Offloading tasks to edge servers that host their corresponding services is critical to ensuring successful task execution, which has made microservice offloading a promising area of research. Gedeon et al. \cite{MicroserviceStore_GLOBECOM_2019} mitigated the excessive latency in pulling microservices from traditional cloud registries by designing an edge-located microservice store, which enabled efficient offloading in edge computing. Chen et al. \cite{MicroserviceOffloading_MASS_2023} proposed a digital twin-assisted deep reinforcement learning algorithm to address the microservice offloading problem with dependency constraints in collaborative edge computing. This approach employs digital twins to predict edge node workloads and network conditions. Tian et al. \cite{DeployMStoExecutionTasks_IOTJ_2024} fully leveraged the layer sharing among microservice images and proposed a deep reinforcement learning-based method to reduce the latency of long-term microservice deployment and task offloading in thing-edge-cloud computing. While these works leverage the rapid deployment characteristic of the microservice architecture to optimize the QoS for task offloading in edge computing, they have not fully exploited the idle storage resources within edge servers to cache microservice images.

In Table \ref{DISPARITIES}, we provide a summary of the disparities
between our work and existing research. It is evident
that existing research falls short in several aspects such as layer sharing among microservice images (\cite{MicroserviceDeployEdge_TPDS_2022}, \cite{DependencyAwareMicroserviceDelpyment_TMC_2024}, \cite{WorkflowScheduling_TPDS_2021} to \cite{MicroserviceOffloading_MASS_2023}), dependencies among workflow tasks (\cite{LayerDeployment_TSC_2025}, \cite{MicroserviceStore_GLOBECOM_2019}, \cite{DeployMStoExecutionTasks_IOTJ_2024}), offloading (\cite{MicroserviceDeployEdge_TPDS_2022} to \cite{WorkflowScheduling_TCC_2024}), and microservice caching (all). Thus, this paper fills these gaps.

\section{System Model and Problem Formulation}
\label{Sec:SystemModelandProblemFormulation}    
In this section, we first introduce a scenario on microservice workflow requests in edge environments. Next, we establish a mathematical model to characterize the layer sharing relationships among microservice images and the dependencies among different tasks.  Finally,  we formulate the problem of microservice workflow offloading and service image caching as a joint optimization problem. The major notations are summarized in TABLE \ref{table:Notation}.
\begin{table}[!htb]
  \centering
  \caption{MAIN NOTATIONS}
  \label{table:Notation}
  \resizebox{\linewidth}{!}{
    \begin{tabularx}{\linewidth}{l>{\raggedright\arraybackslash}X}
      \toprule
      Notation & Description \\
      \midrule
      $U,E,T$ & Set of users, servers and time slots \\
      $L, Z$ & Set of microservice images and image layers\\
      $G^{u,t}$ & The microservice workflow computation request generated by user $u$ at time $t$\\
      $MT^{u,t}$ &  Set of microservice-based computation tasks of $G^{u,t}$ \\
      $\psi_z$ &  Size of microservice image layer $z$ \\
      $l_{i,z}$ &  Whether the microservice image $l_i$ contains layer $z\in Z$ (i.e., $l_{i,z}=1$) or not (i.e., $l_{i,z}=0$). \\
      $R^\nu_{e_i}$ & Bandwidth to download the microservice image of $e_i$\\
      $\xi_{e_i},\sigma_{e_i}$ & Computing  capability and memory capability of server $e_i$\\
      $p_{e_i}$ & Geographical location (longitude and latitude) of $e_i$\\
      $\alpha_{e_i}$ & Radius of service area of server $e_i$\\
      $L^{e_i}$ &  Set of microservice images cached by server $e_i$\\
      $\beta$ & Number of images that a server can cache\\
      $d^{u,t}_n,\tilde{d}^{u,t}_n$ & Sizes of Uplink data and resulted data of task $mt^{u,t}_n$\\
      $\kappa^{u,t}_n,\sigma^{u,t}_n$ & Computational intensity and memory requirement of $mt^{u,t}_n$\\
      $l^{u,t}_n$ & Microservice image required by task $mt^{u,t}_n$\\
      $R^{u}_{u,\tilde{a}^{u,t}_n}$ & Data transmission rate between the terminals of $u$ and access server $\tilde{a}^{u,t}_n$\\
      $R^{u}_{\tilde{a}^{u,t}_n,x^{u,t}_n}$ & Data transmission rate between the network infrastructure around $\tilde{a}^{u,t}_n$ and the terminal of execution server $x^{u,t}_n$\\
      $T^c_{e_i}$ & Available time of the downlink channel of the server $e_i$\\
      $T^a_{x^{u,t}_n}$ & Arrival time of task $mt^{u,t}_n$ at execution server $x^{u,t}_n$\\
      $T^p_{x^{u,t}_n}$ & Image pull completion time of image $l^{u,t}_n$ on execution server $x^{u,t}_n$\\
      $T^s_{x^{u,t}_n}$ & Transmission completion time of the predecessor tasks’ results to $x^{u,t}_n$\\
      $T^e_{mt^{u,t}_n}$ & Execution completion time of task $mt^{u,t}_n$\\
      $T^r_{mt^{u,t}_n}$ & Transmission completion time of the result data to user $u$\\
      \bottomrule
    \end{tabularx}%
  }
\end{table}

\subsection{Overview}
\begin{figure}[htbp] % h:此处, t:顶部, b:底部, p:单独页面
    \centering
    \includegraphics[width=0.48\textwidth]{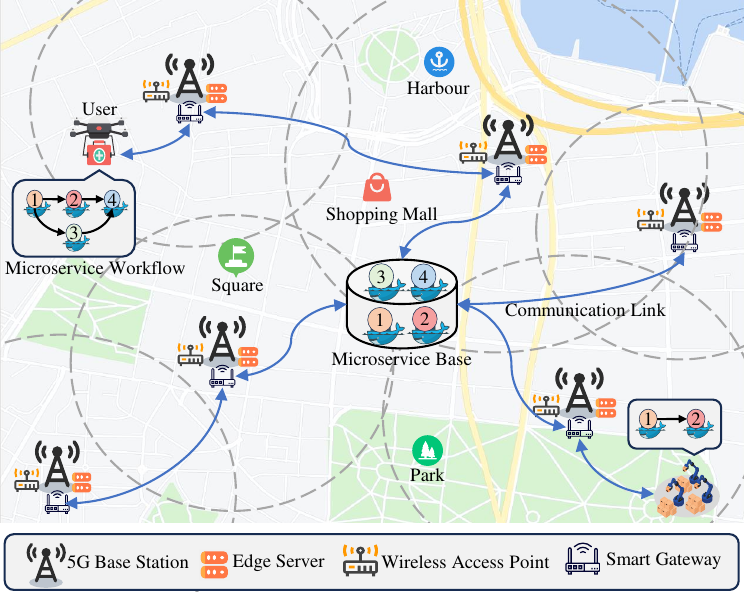}
    \caption{An illustrated scenario of microservice workflow offloading and service image caching.} % 自动编号
    \label{fig:architecture} % 用于交叉引用
\end{figure}

This paper focuses on the edge computing scenario in which users at different locations generate topologically heterogeneous workflow computation requests, where each task in the workflow depends on a specific microservice. We offer such a scenario shown in Fig. \ref{fig:architecture} comprising three core components:

\textbf{1) Microservice base and images:} 
We use $C$ to represent the microservice base (sometimes also named as \textit{microservice store} \cite{MicroserviceStore_GLOBECOM_2019}) which caches all microservice images $L=\{l_1,l_2,\dots,l_M\}$. Let the column vector $l_{i}=[l_{i,1},l_{i,2},\dots,l_{i,|Z|}]^\top \in \{0,1\}^{|Z|\times1}$ denote the layer structure of the image $l_i$, where $l_{i,z}$ indicates whether the microservice image $l_i$ contains layer $z\in Z$ (i.e., $l_{i,z}=1$) or not (i.e., $l_{i,z}=0$). The column vector $\psi=[\psi_{1},\psi_{2},\dots,\psi_{|Z|}]^\top \in \mathbb{R}^{|Z|\times1}$ represents the size of every microservice image layer $z$.

\textbf{2) A distributed edge server cluster:} 
We use $E=\{e_1, e_2,\dots,e_N\}$ to denote the set of $|E|$ edge servers with heterogeneous resources deployed in our scenario. Specifically, we use a 5-tuple $e_i=\left(R^\nu_{e_i},\xi_{e_i}, \sigma_{e_i},p_{e_i},\alpha_{e_i}, L^{e_i}\right)$ to represent the edge server $e_i$, where $R^\nu_{e_i}$ is the bandwidth to download the microservice image of $e_i$. $\xi_{e_i}$ and $\sigma_{e_i}$ are the computing and memory capability of $e_i$, respectively. $p_{e_i}=\left(lot_{e_i}, lat_{e_i}\right)$ is the geographical location (longitude and latitude) of $e_i$. $\alpha_{e_i}$ is the radius of service area of server $e_i$. Matrix $L^{e_i}=[l^{e_i}_1,l^{e_i}_2,\dots,l^{e_i}_{\beta}]\in \{0,1\}^{|Z|\times\beta}$ represents the set of microservice images cached by server $e_i$, where $l^{e_i}_j \in L$ is image vector, and $\beta$ denotes the maximum number of images that an edge server can cache.

\textbf{3) Microservice workflows with diverse topologies:} We use $U=\{1,2,\dots,\mathcal{U}\}$ and $T=\{1,2,\dots, \mathcal{T}\}$ to represent the set of users and time slots in our system. We define the microservice workflow computation request generated by user $u$ at time $t$ as a Directed Acyclic Graph (DAG) $G^{u,t} = \{MT^{u,t}, \mathcal{E}^{u,t}\}$, where $MT^{u,t}=\{mt^{u,t}_n|n\in\{1,2,\dots,\mathcal{M}\}\}$ is the set of microservice-based computation tasks of $G^{u,t}$ and the edge set $\mathcal{E}^{u,t}=\{\mathcal{E}^{u,t}_{i,j}\in\{0,1\}|i,j\in\{1,2,\dots,|MT^{u,t}|\},i<j\}$ specifies the dependencies among different tasks in $G^{u,t}$. Each task can be represented as a 6-tuple $mt^{u,t}_n=\left(d^{u,t}_n,\tilde{d}^{u,t}_n,p^{u,t}_n,\kappa^{u,t}_n,\sigma^{u,t}_n,l^{u,t}_n\right)$, where $d^{u,t}_n$ and $\tilde{d}^{u,t}_n$ are the sizes of the uplink data and resulted data of task $mt^{u,t}_n$, respectively. $p^{u,t}_n=\left(lot^{u,t},lat^{u,t}\right)$ is the geographical location (longitude and latitude) of the end user $u$ at time $t$. $\kappa^{u,t}_n$ and $\sigma^{u,t}_n$ denote the computational intensity and memory requirement of task $mt^{u,t}_n$, respectively. $l^{u,t}_n\in L$ represents the microservice image required by the current task.
\subsection{Communication Model}
Considering that communication resources are generally restricted on edge servers, we assume that each edge server in our architecture has only one downlink channel. The channel can receive exclusively one type of data (e.g., data from users or microservice image files) per time slot. Let $T^{c}_{e_i}$ represent the available time of server $e_i$'s downlink channel, which depends on the completion time of data transmission in the channel. According to \cite{NeglectSendChannel_TPDS_2025}, the result data transmitted through the uplink channel of the server is small enough so that its transmission cost can be considered negligible. Building on this foundation and following the approach in \cite{FDMA_TCCN_2025}, we implement Frequency Division Multiple Access (FDMA) in our architecture to divide the uplink channel into multiple orthogonal sub-channels, which enables parallel transmission of result data across arbitrary time slots. We next analyze the entire microservice workflow offloading process in detail.

The task $mt^{u,t}_n$ generated by user $u$ at time $t$ should first be transmitted to the network infrastructure (e.g., 5G base stations, smart access gateways, and wireless access points) deployed around the access edge server $\tilde{a}^{u,t}_{n} \in E$ that covers the user equipment. Then it is offloaded to the execution server $x^{u,t}_{n} \in E$, whose downlink channel is available (i.e., $\max\{t,T^c_{x^{u,t}_n}\}$). Therefore, we define the arrival time of $mt^{u,t}_n$ at $x^{u,t}_{n}$ as follows:
\begin{flalign}
\label{1}
T^a_{x^{u,t}_{n}} &= \max\{t,T^c_{x^{u,t}_{n}}\}+\tilde{T}^a_{x^{u,t}_{n}}\\
\label{2}
\tilde{T}^a_{x^{u,t}_{n}} &= \frac{d^{u,t}_n}{R^u_{u,\tilde{a}^{u,t}_{n}}}+\frac{d^{u,t}_n}{R^e_{\tilde{a}^{u,t}_{n},x^{u,t}_{n}}}
\end{flalign}
where $T^c_{x^{u,t}_{n}}$ is the available time of downlink channel on $x^{u,t}_{n}$. $\tilde{T}^a_{x^{u,t}_{n}}$ is the data transmission delay. $R^u_{u,\tilde{a}^{u,t}_{n}}$ and $R^e_{\tilde{a}^{u,t}_{n},x^{u,t}_{n}}$ are the data transmission rate between user $u$ and server $\tilde{a}^{u,t}_{n}$, server $\tilde{a}^{u,t}_{n}$ and server $x^{u,t}_{n}$, respectively.
Accounting for both path-loss interference and background noise between the equipment terminals, the data transmission rate $R^u_{u,a^{u,t}_{n}}$ between the user equipment and the network infrastructure deployed around $a^{u,t}_{n}$ can be derived as follows\cite{TransRate_IOTJ_2022}:
\begin{equation}
\label{deqn_ex1a}
R^u_{u,a^{u,t}_{n}} = B \cdot \log_2\left(1+\rho\frac{\left[H\left(p^{u,t}_n,p_{\tilde{a}^{u,t}_n}\right)\right]^\eta}{\lambda^2}\right)
\end{equation}
where $B$ is the channel bandwidth. $\rho$ is the transmission power of user equipments. $\lambda^2$ is the background noise power. $\eta$ is the path-loss exponent, and $H\left(p^{u,t}_n,p_{\tilde{a}^{u,t}_n}\right)$ is the Haversine distance \cite{Haversine_CVPR_2024} between $u$ and $\tilde{a}^{u,t}_n$. 

When the task $mt^{u,t}_n$ arrives at server $x^{u,t}_{n}$, the server must deploy the corresponding microservice to enable the task for execution. The deployment of microservice $l^{u,t}_n$ is achieved exclusively by pulling its non-cached image layers from the microservice store to server $x^{u,t}_{n}$. Given that the microservice startup time is predominantly determined by the image pull delay \cite{MSstartupTime_TSC_2025}, we formulate the image pull completion time accounting for the layer sharing as the following theorem.

% \begin{theorem}\label{thm:layer_pull}
\textit{Theorem 1 :}
The image pull completion time $T^p_{x^{u,t}_n}$, considering the advantages of layer sharing among microservice images, can be represented as:
\begin{flalign}
        \label{image_layer}
\tilde{T}^p_{x^{u,t}_n}=&\frac{\sum_{i=0}^{|Z|}\psi_i\left[l^{u,t}_{n,i}\left(1-\max\{L^{x^{u,t}_n}_{i,:}\}\right)\right]}{R^\nu_{x^{u,t}_n}}\\
    \label{5}
    T^p_{x^{u,t}_n}=&\max\{t,T^c_{x^{u,t}_n}\}+\tilde{T}^p_{x^{u,t}_n}
\end{flalign}
% \begin{equation}
%         \label{image_layer}
% \tilde{T}^p_{x^{u,t}_n}=\frac{\sum_{i=0}^{|Z|}\psi_i\left[l^{u,t}_{n,i}\left(1-\max\{L^{x^{u,t}_n}_{i,:}\}\right)\right]}{R^\nu_{x^{u,t}_n}}
% \end{equation}
where $\tilde{T}^p_{x^{u,t}_n}$ is the image pull delay. $|Z|$ is the total count of distinct image layers in our system. $\psi_i$ is the size of the image layer $i$. $l^{u,t}_{n, i}\in\{0,1\}$ indicates whether the microservice image $l^{u,t}_n$ contains layer $i\in Z$ (i.e., $l^{u,t}_{n, i}=1$) or not (i.e., $l^{u,t}_{n, i}=0$). $L^{x^{u,t}_n}_{i,:}$ is the $i$-th row vector of matrix $L^{x^{u,t}_n}$ and $R^\nu_{x^{u,t}_n}$ is the bandwidth of $x^{u,t}_n$ to download microservice image.

% When the current task $mt^{u,t}_n$ has been generated and the downlink channel of the execution server $x^{u,t}_n$ is available (i.e., $max\{t, T^c_{x^{u,t}_n}\}$), the microservice image of task $mt^{u,t}_n$ can be pulled to the execution server. Therefore, the image pull completion time can be represented as:
% \begin{equation}
%     \label{5}
%     T^p_{x^{u,t}_n}=\max\{t,T^c_{x^{u,t}_n}\}+\tilde{T}^p_{x^{u,t}_n}
% \end{equation}
% \end{theorem}

\textit{Proof :} 
The matrix $L^{x^{u,t}_n}=[l^{x^{u,t}_n}_1,l^{x^{u,t}_n}_2,\dots,l^{x^{u,t}_n}_{\beta}]\in \{0,1\}^{|Z|\times\beta}$ represents the set of microservice images cached by server $x^{u,t}_n$. The maximum value in the $z$-th row of $L^{x^{u,t}_n}$ indicates whether server $x^{u,t}_n$ caches image layer $z$ (i.e., $\max\{L^{x^{u,t}_n}_{1,:}\}=1$) or not (i.e., $\max\{L^{x^{u,t}_n}_{1,:}\}=0$). Therefore, the image layers cached by $x^{u,t}_n$ can be expressed as follows:
\begin{equation}
\label{6}
    \mu_{x^{u,t}_n}=
    \begin{bmatrix}
        \max\{L^{x^{u,t}_n}_{1,:}\}\\
        \max\{L^{x^{u,t}_n}_{2,:}\}\\
        \vdots\\
        \max\{L^{x^{u,t}_n}_{|Z|,:}\}
    \end{bmatrix}\in\{0,1\}^{|Z|\times1}
\end{equation}
Then, the server $x^{u,t}_n$ can startup the microservice $l^{u,t}_n$ immediately after pulling only those image layers absent from its local cache. The data volume of image layers yet to be pulled can be represented as follows:
\begin{flalign}
\label{7}
\nonumber
&D_{x^{u,t}_n}=\psi^\top \cdot\left(l^{u,t}_n-l^{u,t}_n\odot\mu_{x^{u,t}_n}\right)&&\\
\nonumber
&    =\begin{bmatrix}
        \psi_1\ldots\psi_{|Z|}
    \end{bmatrix} \cdot\Big(
    \begin{bmatrix}
        l^{u,t}_{n,1}\\
        l^{u,t}_{n,2}\\
        \vdots\\
        l^{u,t}_{n,|Z|}
    \end{bmatrix} - 
    \begin{bmatrix}
    l^{u,t}_{n,1}\\
    l^{u,t}_{n,2}\\
    \vdots\\
    l^{u,t}_{n,|Z|}
    \end{bmatrix}
    \odot
    \begin{bmatrix}
        \max\{L^{x^{u,t}_n}_{1,:}\}\\
        \max\{L^{x^{u,t}_n}_{2,:}\}\\
        \vdots\\
        \max\{L^{x^{u,t}_n}_{|Z|,:}\}
    \end{bmatrix}\Big)&&\\
\nonumber
&   = \begin{bmatrix}
        \psi_1\ldots\psi_{|Z|}
    \end{bmatrix} \cdot
    \begin{bmatrix}
        l^{u,t}_{n,1}\cdot \left(1-\max\{L^{x^{u,t}_n}_{1,:}\}\right) \\
        l^{u,t}_{n,2}\cdot \left(1-\max\{L^{x^{u,t}_n}_{2,:}\}\right)\\
        \vdots\\
        l^{u,t}_{n,|Z|}\cdot \left(1-\max\{L^{x^{u,t}_n}_{|Z|,:}\}\right)
    \end{bmatrix}&&\\
&   = \sum_{i=0}^{|Z|}\psi_i\left[l^{u,t}_{n,i}\left(1-\max\{L^{x^{u,t}_n}_{i,:}\}\right)\right]&&
\end{flalign}
Thus, the image pull delay can be calculated as Eq. \eqref{image_layer}. When the downlink channel of $x^{u,t}_n$ is available (i.e., $\max\{t,T^c_{x^{u,t}_n}\}$), the microservice image can be pulled to the server $x^{u,t}_n$. Therefore, the image pull completion time can be calculated as Eq. \eqref{5}, which ends the proof. $\hfill\blacksquare$ 
    
The server $x^{u,t}_n$ then follows the image caching strategy $y^{u,t}_n\in\{1,2,\dots, \beta \}$ to replace the cached image $l^{x^{u,t}_n}_{y^{u,t}_n}$ by the task image $l^{u,t}_n$. Due to task dependencies in the microservice workflow, $mt^{u,t}_n$ can only be executed after all results of its predecessor tasks have been successfully transmitted to the server $x^{u,t}_n$. The resulted data $\tilde{d}^{u,t}_j$ of a predecessor task can only be transmitted to $x^{u,t}_n$ when this predecessor task has been completed and the downlink channel of $x^{u,t}_n$ is available (i.e., $\max\{T^e_{mt^{u,t}_j}, T^c_{x^{u,t}_n}\}$). Therefore, we compute the transmission completion time of these predecessor tasks' results as follows:
\begin{equation}
    T^s_{x^{u,t}_n} = \max_{mt^{u,t}_j\in pre\left( mt^{u,t}_n \right)}\{\max\{T^e_{mt^{u,t}_j}, T^c_{x^{u,t}_n}\}+\frac{\tilde{d}^{u,t}_j}{R_{x^{u,t}_j,x^{u,t}_n}}\}
\end{equation}
where $pre\left( mt^{u,t}_n\right)$ is the set of the predecessor tasks of $mt^{u,t}_n$. $R_{x^{u,t}_j,x^{u,t}_n}$ is the data transmission rate between server $x^{u,t}_j$ and server $x^{u,t}_n$. Task $mt^{u,t}_j$ will start once the following four conditions all occur: its uplink data has arrived, its required microservice image has been downloaded, the resulted data from its predecessor tasks have been transferred, and computational resources on its execution server become available (i.e., $\max\{T^a_{x^{u,t}_j},T^p_{x^{u,t}_j},T^s_{x^{u,t}_j},\check{T}_{x^{u,t}_j}\}$). Therefore, we compute the execution completion time $T^e_{mt^{u,t}_j}$ as follows:
\begin{flalign}
    \label{4}
    T^e_{mt^{u,t}_j} &= \max\{T^a_{x^{u,t}_j},T^p_{x^{u,t}_j},T^s_{x^{u,t}_j},\check{T}_{x^{u,t}_j}\}+\tilde{T}^e_{mt^{u,t}_j}\\
    \label{10}
    \tilde{T}^e_{mt^{u,t}_j}&=\frac{d^{u,t}_j\kappa^{u,t}_j}{\xi_{x^{u,t}_j}}
\end{flalign}
where $\tilde{T}^e_{mt^{u,t}_j}$ is the computation delay of task $mt^{u,t}_j$, and $\check{T}_{x^{u,t}_j}$ is the available time of the computing resources of server $x^{u,t}_j$, which is equivalent to the execution completion time of the preceding task in $x^{u,t}_j$.

After the execution completion time $T^e_{mt^{u,t}_j}$, the resulted data $\tilde{d}^{u,t}_n$ will first be transmitted from the execution server $x^{u,t}_n$ to the server $f^{u,t}_n$ (i.e., $\tilde{d}^{u,t}_n/R_{x^{u,t}_n,f^{u,t}_n}$). The service area of $f^{u,t}_n$ should cover the user equipment. Finally, the resulted data will be transmitted to the user $u$ through server $f^{u,t}_n$ (i.e., $\tilde{d}^{u,t}_n/R_{f^{u,t}_n,u}$). Therefore, the completion time of the resulted data transmission $T^r_{mt^{u,t}_n}$ can be represented as follows:
\begin{equation}
\label{11}
    T^r_{mt^{u,t}_n} = T^e_{mt^{u,t}_n} + \frac{\tilde{d}^{u,t}_n}{R_{x^{u,t}_n,f^{u,t}_n}} + \frac{\tilde{d}^{u,t}_n}{R_{f^{u,t}_n,u}}
\end{equation}
\subsection{Problem Formulation}
Based on the aforementioned system model, we formulate a joint microservice workflow offloading and service image caching problem $\mathbf{P}$, aiming to minimize the long-term average system offloading completion time. The problem $\mathbf{P}$ is as follows:
% \begin{gather}
%     \mathbf{P}: \min_{X,Y}\lim_{\mathcal{T} \to \infty}\frac{1}{\mathcal{T}}\sum_{t\in T}T^r_{G^{U,t}}
%     \\\text{s.t.}\ T^r_{G^{U,t}}=\max_{mt^{u,t}_n\in G^{U,t}}\{T^r_{mt^{u,t}_n}\}\label{13}
%     \\\sigma_{ x^{u,t}_n}\geqslant \sigma^{u,t}_n,\forall u \in U, \forall t\in T
%     \label{14}
%     \\x^{u,t}_n\in\{1,2,\dots|E|\},\forall u \in U, \forall t\in T
%     \label{15}
%     \\y^{u,t}_n\in\{1,2,\dots|L^{x^{u,t}_n}|\},\forall u \in U, \forall t\in T\label{16}
% \end{gather}

\begin{alignat}{2}
\mathbf{P}: \min_{X,Y} \quad & \lim_{\mathcal{T} \to \infty}\frac{1}{\mathcal{T}}\sum_{t\in T}T^r_{G^{U,t}} & \label{12}\\
\mbox{s.t.}\quad
&T^r_{G^{U,t}}=\max_{mt^{u,t}_n\in G^{U,t}}\{T^r_{mt^{u,t}_n}\} & \label{13}\\
&\sigma_{ x^{u,t}_n}\geqslant \sigma^{u,t}_n,\forall u \in U, \forall t\in T &\label{14}\\
&x^{u,t}_n\in\{1,2,\dots|E|\},\forall u \in U, \forall t\in T &\label{15}\\
&y^{u,t}_n\in\{1,2,\dots|L^{x^{u,t}_n}|\},\forall u \in U, \forall t\in T & \label{16}
\end{alignat}
where constraint \eqref{13} is the nonlinear equality constraint, which indicates that the system offloading completion time in time slot $t$ is equivalent to the maximum of all task offloading completion time over all users. Constraint \eqref{14} ensures that tasks can be offloaded to servers with sufficient memory resources to meet their requirements. Constraint \eqref{15} is an integer constraint that guarantees each task can only be offloaded to one execution server, and constraint \eqref{16} is also an integer constraint that ensures the completion of microservice image caching for task execution. Obviously, the problem $\mathbf{P}$ is an Integer NonLinear Programming (INLP) problem. The INLP problem has been extensively proven to be NP-hard \cite{INLP1_TWC_2024}\cite{INLP2_TNSE_2024}. Thus, it is necessary to develop an applicable optimization algorithm. In this paper, we chose the Deep Reinforcement Learning (DRL) as the base algorithm.
\begin{figure*}[!t]
  \centering
  \includegraphics[width=0.99\textwidth]{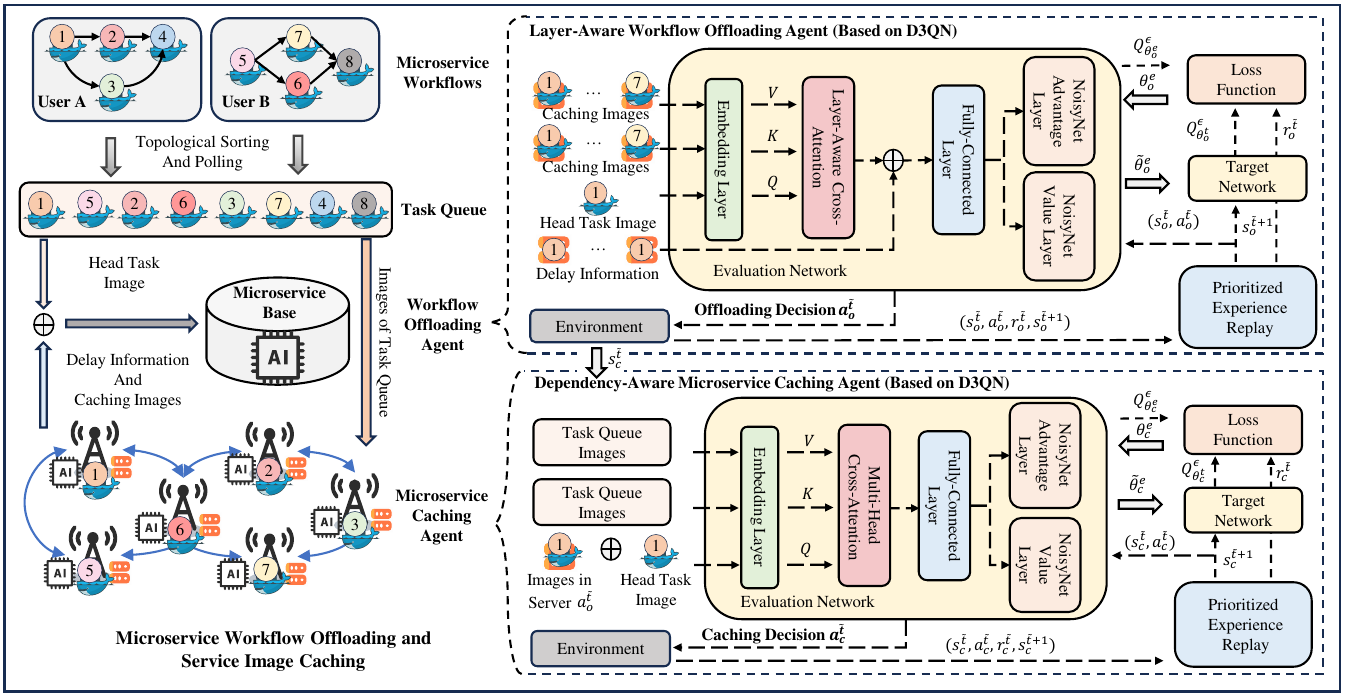}
  \caption{Framework of our proposed DLA-HDRL approach}
  \label{fig:Approach}
\end{figure*}

\section{Dependency- and Layer-aware Hierarchical DRL Approach for Microservice Workflow Offloading and Service Image Caching}
\label{DLA-HDRL}

Given the NP-hard nature of the problem $\mathbf{P}$, DRL-based approaches have emerged as an effective solution \cite{DependencyAwareMicroserviceDelpyment_TMC_2024}. However, traditional DRL-based approaches often suffer from disadvantages in adequately exploring the solution space due to the large-scale action space with diverse actions \cite{TraditionalDRLProblem_TCCN_2025}. Against this background,  researchers attempted to develop some alternatives, such as Hierarchical Deep Reinforcement Learning (HDRL) \cite{HDRL_TCCN_2024}\cite{HDRL_TOWC_2025}, which tries to decompose the solution space to solve joint optimization problems. Meanwhile, we found the vanilla HDRL has some disadvantages for solving our problem. Therefore, we turned to develop a new prompted approach. Finally, we introduced the attention mechanism and proposed a novel Dependency- and Layer-Aware HDRL approach (DLA-HDRL).

\subsection{Overview}
As shown in Fig. \ref{fig:Approach}, the proposed DLA-HDRL approach is composed of: 1) a workflow offloading agent deployed on a microservice base and 2) several microservice caching agents deployed on edge servers. The offloading procedure of microservice workflows in the DLA-HDRL is elaborated as follows:

\textbf{1) Microservice task queue construction:} 
When $|U|$ users generate the set of microservice workflows $G^{U,t}$ in time slot $t$, we propose and construct a task queue using a combination tactic of Topological Sorting and Polling (TSP), successfully
generating the microservice task queue $MQ^{U,t}$. The overall procedure of the proposed TSP is shown in Algorithm \ref{TSP}. Initially, the TSP algorithm applies topological sorting to each microservice workflow, effectively converting the original graph-based representation $G^{u,t}$ into linearly ordered data sequence $G^{u,t}_{top}$ for subsequent polling (lines 2-4). In the polling  phase, the TSP algorithm sequentially dequeues the head task $G^{u,t}_{top}\left[0\right]$ from each microservice workflow sequence $G^{u,t}_{top}$ and enqueues it into the microservice task queue $MQ^{U,t}$ until the set $G^{U,t}_{top}$ becomes an empty set, at which point the algorithm terminates (ines 5-12).
\begin{figure}[h]
    \removelatexerror
    \begin{algorithm}[H]
        \SetAlgoLined %显示end
        \caption{TSP Method}\label{TSP}
        \LinesNumbered
        \KwIn{Set of microservice workflows $G^{U,t}$}
        \KwOut{Microservice task queue $MQ^{U,t}$}
        $MQ^{U,t} \gets \varnothing$; $G^{U,t}_{top} \gets \varnothing$\;
        \tcp{\textit{Topological Sorting}}
        \ForEach{
            $G^{u,t} \in G^{U,t}$
        }{
        $G^{U,t}_{top} \gets G^{U,t}_{top} \cup toplogical(G^{u,t})$\;
        }
        \tcp{\textit{Polling}}
        \While{$G^{U,t}_{top} \neq \varnothing$}{
                \ForEach{
                    $G^{u,t}_{top} \in G^{U,t}_{top}$
                }{
                    \If{$G^{u,t}_{top} \neq \varnothing$}{
                                        $MQ^{U,t} \gets MQ^{U,t} \cup G^{u,t}_{top}\left[1\right]$\;
                    $G^{u,t}_{top} \gets G^{u,t}_{top}\left[2:\right]$
                    }

                }
        }
    \end{algorithm}
\end{figure}

\textbf{2) Task offloading and service image caching:} After the microservice task queue is constructed, our DLA-HDRL approach first transmits the state information of the head task to the offloading agent. Then the offloading agent in DLA-HDRL employs a layer-aware cross-attention mechanism to perceive layer sharing relationships between the targeted task image and cached images across various servers. Subsequently, it makes decisions on which edge server to offload the task to execute. Following this, the caching agent deployed on the edge server utilizes a dependency-aware multi-head cross-attention mechanism to comprehensively perceive dependencies among different task images in the service chain. Finally, it determines how to perform cached image replacement on this execution server.
\subsection{Layer-Aware Cross-Attention-Based Workflow Offloading Agent}
Because in the time-variant edge environment, it is hard to acquire the deterministic probability of the state transition , we use a 3-tuple $\left(\mathcal{S}_o, \mathcal{A}_o, \mathcal{R}_o\right)$ to represent the Markov Decision Process (MDP) of the workflow offloading agent, where $\mathcal{S}_o=\{s^{\tilde{t}}_o\}_{\tilde{t}\in\tilde{T}}$ is the state space, $\mathcal{A}_o=\{a^{\tilde{t}}_o\}_{\tilde{t}\in\tilde{T}}$ is the action space, $\mathcal{R}_o=\{r^{\tilde{t}}_o\}_{\tilde{t}\in\tilde{T}}$ is the reward set, and $\tilde{T}$ is the set of agent decision time slots. The detailed definitions are as follows:

\textbf{1) Offloading agent state $s^{\tilde{t}}_o$:} 
To fully perceive the layer sharing relationships among different microservice images in problem $\mathbf{P}$, we define the state of the offloading agent in time slot $\tilde{t}$ as follows:
\begin{equation}
    s^{\tilde{t}}_o=\left(\tilde{T}^{a,\tilde{t}}_{1:|E|}, \tilde{T}^{e,\tilde{t}}_{1:|E|}, \tilde{T}^{r,\tilde{t}}_{1:|E|},\tilde{T}^{p,\tilde{t}}_{1:|E|},\tilde{l}^{u,t}_{\tilde{t}},\tilde{L}^{\tilde{t}}_{1:|E|},\sigma^{u,t}_{\tilde{t}},\sigma^{\tilde{t}}_{1:|E|}\right)
\end{equation}
where $\tilde{T}^{a,\tilde{t}}_{1:|E|},\tilde{T}^{e,\tilde{t}}_{1:|E|}$, and $\tilde{T}^{p,\tilde{t}}_{1:|E|}$ denote the data transmission delay, computational delay, and microservice image pull delay for task $mt^{u,t}_{\tilde{t}}$ offloaded to each server, respectively, which can be calculated by Eqs.  \eqref{2}, \eqref{10} and \eqref{5}, respectively. $\tilde{T}^{r,\tilde{t}}_{1:|E|}=T^{r,\tilde{t}}_{1:|E|}-t$ is the available time of the computing resources of each server for task $mt^{u,t}_{\tilde{t}}$. $\tilde{l}^{u,t}_{\tilde{t}} \in \{1,\dots,|L|\}$ and $\tilde{L}^{\tilde{t}}_{1:|E|}=\left(\tilde{l}^{\tilde{t}}_{1, 1},\dots\tilde{l}^{\tilde{t}}_{1,\beta},\dots,\tilde{l}^{\tilde{t}}_{|E|,\beta}\right)$ are the type of image relied upon for task $mt^{u,t}_{\tilde{t}}$ and types of images cached by each server, respectively. $\sigma^{u,t}_{\tilde{t}}$ and $\sigma^{\tilde{t}}_{1:|E|}$ represent the memory requirement of $mt^{u,t}_{\tilde{t}}$ and the memory capacity of each server, respectively.

\textbf{2) Offloading agent action $a^{\tilde{t}}_o$:} 
The workflow offloading agent determines the optimal edge server for task offloading according to the state information of task $mt^{u,t}_{\tilde{t}}$. Therefore, we can formulate the offloading decision for the task $mt^{u,t}_{\tilde{t}}$ as the following discrete action:
\begin{equation}
    a^{\tilde{t}}_o\in \{1,2,\dots,|E|\}
\end{equation}

\textbf{3) Offloading agent reward $r^{\tilde{t}}_o$:} 
Considering that the objective of problem $\mathbf{P}$ is to minimize the system offloading completion time, while the goal of the DRL agent is to maximize the long-term reward, we formulate the immediate reward $r^{\tilde{t}}_o$ for performing the offloading action $a^{\tilde{t}}_o$ on task  $mt^{u,t}_{\tilde{t}}$ as follows:
\begin{equation}\label{reward}
    r^{\tilde{t}}_o = -\left(T^r_{mt^{u,t}_{\tilde{t}}}-t\right)
\end{equation}
where $T^r_{mt^{u,t}_{\tilde{t}}}$ represents the offloading completion time of task $mt^{u,t}_{\tilde{t}}$, which is calculated by Eq. \eqref{11}.

Considering that the simple fully-connected layers usually struggle to uncover the complex layer sharing relationships among different images, \textbf{we design a Layer-Aware Cross-Attention (LACA) mechanism to assist the agent in computing the attention of task images towards those cached on edge servers.} The detailed procedure is described in Algorithm \ref{LACA}.
The proposed LACA mechanism first implements an embedding layer $\mathcal{H}_o$ that translates each image into a vector $h_{\tilde{t}}\in\mathbb{R}^{1\times1\times\mathbb{H}}$ (line 2). To ensure consistency of the feature structure in the state representation, the LACA mechanism extracts the embedding of the image with index $b$ from each server to construct the matrix $H^b_{\tilde{t}} \in \mathbb{R}^{1\times|E|\times\mathbb{H}}$ (lines 3-7). Subsequently, the LACA mechanism translates the image embedding vector $h^{u,t}_{\tilde{t}}$ into the query matrix $q^b_o\in\mathbb{R}^{1\times1\times d_o}$ through a fully-connected layer $\mathcal{Q}_o$, while mapping the matrix $H^b_{\tilde{t}}$ to the key matrix $k^b_o\in\mathbb{R}^{1\times|E|\times d_o}$ and the value matrix $v^b_o\in\mathbb{R}^{1\times|E|\times d_o}$ through dedicated fully-connected layers $\mathcal{K}_o$ and $\mathcal{V}_o$. After that, we can calculate the attention vector $A^b_o\in \mathbb{R}^{1 \times 1 \times d_o}$ as follows:
\begin{equation}
\label{20}
    A^b_o = Softmax\left(\frac{q^b_o{k_o^b}^\top}{\sqrt{d_o}}\right)v^b_o
\end{equation}
Finally, the proposed LACA mechanism concatenates the attention vector $A^b_o$ with the state $\tilde{s}^{\tilde{t}}_o$, thus updating the state representation (lines 8-10).
\begin{figure}[h]
    \removelatexerror
    \begin{algorithm}[H]
        \SetAlgoLined %显示end
        \caption{LACA Mechanism}\label{LACA}
        \LinesNumbered
        \KwIn{State $s^{\tilde{t}}_o$}
        \KwOut{Updated state $\tilde{s}^{\tilde{t}}_o$ after attention computation}
        $\tilde{s}^{\tilde{t}}_o \gets s^{\tilde{t}}_o\left[1:4|E|\right]$\;
        $h^{u,t}_{\tilde{t}} \gets \mathcal{H}_o\left(\tilde{l}^{u,t}_{\tilde{t}}\right)$; $H^{\tilde{t}}_{1:|E|} \gets \mathcal{H}_o\left(\tilde{L}^{\tilde{t}}_{1:|E|}\right)$\;
        \For{$b=1:\beta$}{
            $H^{b}_{\tilde{t}} \gets \varnothing$\;
            \For{$e=1:|E|$}{
                $H^{b}_{\tilde{t}} \gets H^{b}_{\tilde{t}} \cup H^{\tilde{t}}_{e,b}$\;
            }
            $q^b_o \gets \mathcal{Q}_o\left(h^{u,t}_{\tilde{t}}\right)$; $k^b_o \gets \mathcal{K}_o\left(H^{b}_{\tilde{t}}\right)$; $v^b_o \gets \mathcal{V}_o\left(H^{b}_{\tilde{t}}\right)$\; 
            Calculate the attention vector $A^b_o$ by Eq. \eqref{20}\;
            $\tilde{s}^{\tilde{t}}_o \gets \tilde{s}^{\tilde{t}}_o \cup A^b_o$\;
        }
    \end{algorithm}
\end{figure}

Next, we need to determine an optimal offloading action in each decision time slot. Notice that the state space of the workflow offloading agent is continuous and the offloading action is discrete. We develop a new optimization approach. Our approach employs the Dueling Double Deep Q-Network (D3QN)\cite{D3QNNP_TCCN_2024} as the backbone, in which we successfully integrate the proposed LACA mechanis, further prompting the performance of D3QN. So, this LACA-promoted D3QN acts as the workflow offloading agent. Now, we can give the whole training process of DLA-HDRL in Algorithm \ref{OAgent}.

% NoisyNet\cite{NoisyNet_2018_ICLR} and Prioritized Experience Replay (PER)\cite{PER_ICLR_2016}. 

\begin{figure}[h]
    \removelatexerror
    \begin{algorithm}[H]
        \SetAlgoLined %显示end
        \caption{DLA-HDRL Training Process}\label{OAgent}
        \LinesNumbered
        \KwIn{Number of users $|U|$, number of episodes $\mathbb{E}$, time interval $\mathbb{T}$, training interval $\mathbb{S}$, batch size $\mathbb{B}$}
        \KwOut{Parameters of evaluation networks of offloading agent and caching agent $\theta^e_o$, $\theta^e_c$}
        Initialize D3QN models and PERs $M_o$, $M_c$\;

        \For{$episode=1:\mathbb{E}$}{
            $t\gets0; \tilde{t}\gets1; s_c \gets \varnothing; r_c \gets 0$\;
            Each of $|U|$ users independently generates a microservice workflow within the environment\;
            Construct $MQ^{U,t}$ based on TSP\;
            \While{$MQ^{U,t} \neq \varnothing$}{
                $mt^{u,t}_{\tilde{t}}\gets MQ^{U,t}\left[0\right]$, $MQ^{U,t}\gets MQ^{U,t}\left[1:\right]$\;
                Generate state $s^{\tilde{t}}_o$\;
                Workflow offloading agent select an action $a^{\tilde{t}}_o$ according to Eq. \eqref{25} and perform $a^{\tilde{t}}_o$\;
                Obtain $r^{\tilde{t}}_o$ and state $s^{\tilde{t}}_c$\;
                Microservice caching agent selects an action $a^{\tilde{t}}_c$ that maximizes the Q-value and perform $a^{\tilde{t}}_c$\; 
                Obtain $r^{\tilde{t}}_c, s^{\tilde{t}+1}_o$\;
                Store $\left(s^{\tilde{t}}_o,a^{\tilde{t}}_o,r^{\tilde{t}}_o,s^{\tilde{t}+1}_o\right) $ in $M_o$\;
                \If{$\tilde{t}>1$}{
                    Store $\left(s_c, a^{\tilde{t}}_c, r_c, s^{\tilde{t}}_c\right) $ in $M_c$\;
                }
                
                \If{$|M_o|,|M_c|\geqslant\mathbb{B}$ and $\tilde{t} \equiv 0 \pmod{\mathbb{S}}$}{
                    Sample $\mathbb{B}$ transitions from $M_o$ and $M_c$ respectively\;
                    Update priorities of transitions in both PER $M_o$ and $M_c$ according to Eq. \eqref{TD}\;
                    Update $\theta^{e}_o$ and $\theta^{e}_c$ according to Eq. \eqref{MSE}\;
                    Copy parameters from evaluation networks
                    to target networks every fourth update\;
                }
                
                $s_c \gets s^{\tilde{t}}_c; t \gets t+\mathbb{T}; s_c\gets s^{\tilde{t}}_c; r_c\gets r^{\tilde{t}}_c$\;
            }
        }
    \end{algorithm}
\end{figure}
The D3QN model of the offloading agent is first initialized (line 1), and its structure is depicted in Fig. \ref{fig:Approach}. To effectively learn both the state value and nuanced discrepancies between different actions, D3QN splits the network architecture into two branches to calculate the value function $V_{\theta^e_o}(s^{\tilde{t}}_o)$ and the advantage function $A_{\theta^e_o}(s^{\tilde{t}}_o, a^{\tilde{t}}_o)$, separately. Combining the value function and the advantage function enables the Q function to be computed as follows:
\begin{equation}\label{21}
    Q_{\theta^e_o}(s^{\tilde{t}}_o, a^{\tilde{t}}_o) = V_{\theta^e_o}(s^{\tilde{t}}_o) + A_{\theta^e_o}(s^{\tilde{t}}_o, a^{\tilde{t}}_o) + \frac{1}{|E|}\sum_{i=1}^{|E|}A_{\theta^e_o}(s^{\tilde{t}}_o, a_{o,i})
\end{equation}
where $a_{o,i}$ represents that the agent offloads the task to the server $e_i \in E$.

We use the proposed TSP algorithm to construct the microservice task queue $MQ^{U,t}$ from all users at the beginning of each episode (lines 4-5). The workflow offloading agent then obtains the state of the head task in $MQ^{U,t}$ from the environment (lines 7-8). 
Diverging from the $\epsilon-greedy$-based exploration commonly employed in traditional DRL, the workflow offloading agent utilizes NoisyNet \cite{NoisyNet_2018_ICLR} to transform its exploration mechanism into a learnable neural network layer. This approach facilitates efficient exploration within the solution space. Specifically, let $y=\omega x+b$ denote an advantage layer without noise, where $y\in\mathbb{R}^{|E|\times1},\ \omega\in\mathbb{R}^{|E|\times d_x},\ x\in\mathbb{R}^{{d_x}\times1}$, and $b\in\mathbb{R}^{|E|\times1}$ are the output, weight matrix, input, and bias, respectively. NoisyNet typically employs Gaussian noise as the noise parameter function. The output of the advantage layer after injecting Gaussian noise can be expressed as follows:
\begin{equation}
    y_\epsilon = \left(\mu^\omega+\sigma^\omega \odot\epsilon^\omega \right)x+\mu^b+\sigma^b\odot \epsilon^b
\end{equation}
where the parameters $\mu^\omega\in\mathbb{R}^{|E|\times d_x},\ \sigma^\omega\in\mathbb{R}^{|E|\times d_x},\ \mu^b\in\mathbb{R}^{|E|\times 1}$, and $\sigma^b\in\mathbb{R}^{|E|\times 1}$ are learnable while $\epsilon^\omega\in\mathbb{R}^{|E|\times d_x}$ and $\epsilon^b\in\mathbb{R}^{|E|\times 1}$ are non-learnable noise variables. Each element $\mu_{i,j}\in\mu$ is initialized by a sample of the uniform distribution in
the interval $\left[-1/\sqrt{d_x}, 1/\sqrt{d_x}\ \right]$, where $d_x$ represents the length of $x$. Each element $\sigma_{i,j}\in\sigma$ is initialized by a constant $0.5/\sqrt{|E|}$ \cite{NoisyNet_2018_ICLR}. Moreover, factorized Gaussian noises are generally applied to generate the noise variables. Therefore, $\epsilon^\omega$ and $\epsilon^b$ can be obtained as follows:
\begin{equation}
    \epsilon^\omega_{i,j}=f\left(\bar{\epsilon}_j\right)f\left(\bar{\epsilon}_i\right),\ \epsilon^b_{j}=f\left(\bar{\epsilon}_j\right)
\end{equation}
where $\bar{\epsilon}_i \sim \mathcal{N}\left(0,1\right)$ and $f\left(\bar{\epsilon}_i\right)=sign(\bar{\epsilon}_i)\sqrt{\bar{\epsilon}_i}$. Then, we employ NoisyNet to construct both the advantage layer and the value layer in the workflow offloading agent. The Eq. \eqref{21} can be rewritten as follows:
\begin{equation}
    Q^{\epsilon}_{\theta^e_o}(s^{\tilde{t}}_o, a^{\tilde{t}}_o) = V^{\epsilon}_{\theta^e_o}(s^{\tilde{t}}_o) + A^{\epsilon}_{\theta^e_o}(s^{\tilde{t}}_o, a^{\tilde{t}}_o) + \frac{1}{|E|}\sum_{i=1}^{|E|}A^{\epsilon}_{\theta^e_o}(s^{\tilde{t}}_o, a_{o,i})
\end{equation}

Subsequently, the workflow offloading agent selects an action that maximizes $Q^{\epsilon}_{\theta^e_o}(s^{\tilde{t}}_o, a^{\tilde{t}}_o)$ with state $s^{\tilde{t}}_o$. Considering that successful task execution requires offloading to edge server with sufficient memory resources, we constrain the action space in decision step $\tilde{t}$ to set $\bar{A}^{\tilde{t}}_o$, which comprises servers satisfying the memory requirement of task $mt^{u,t}_{\tilde{t}}$. Consequently, the action selection policy of the offloading agent is characterized as follows:
\begin{equation}\label{25}
    a^{\tilde{t}}_o = \mathop{\arg\max}\limits_{a\in\bar{A}^{\tilde{t}}_o}\ Q_{\theta^e_o}^\epsilon(s^{\tilde{t}}_o,a)
\end{equation}

Then, the environment executes the action $a^{\tilde{t}}_o$ and returns the reward $r^{\tilde{t}}_o$ and the next state $s^{\tilde{t}+1}_o$ (lines 9-12). After that, the agent stores the transition $(s^{\tilde{t}}_o,a^{\tilde{t}}_o,r^{\tilde{t}}_o,s^{\tilde{t}+1}_o) $ in the Prioritized Experience Replay (PER) $M_o$ with initial priority (line 13).
When the size of PER exceeds the batch size $\mathbb{B}$ and the number of decision steps exactly divides the training interval $\mathbb{S}$, the offloading agent samples a batch of transitions from $M_o$ weighted by priority for learning (lines 17-18 ). PER generates the
probability of the sampling transition $k$ as follows:
\begin{equation}
    P\left(k\right)=\frac{p^{\alpha}_k}{\sum_i p^{\alpha}_i}
\end{equation}
where $\alpha$ is a hyperparameter to introduce randomness into the selection of experiences and $p_k=|\delta_k|+\mathcal{B}$ represents the priority of transition $k$ in PER. $\mathcal{B}$ is a constant to prevent zero priority and $|\delta_k|$ is the temporal difference error which can be computed as follows:
\begin{equation}\label{TD}
    |\delta_k| = |Q^{\epsilon}_{\theta^{t}_o}(s^{\tilde{t}+1}_o,\mathop{\arg\max}\limits_{a\in\bar{A}^{\tilde{t}+1}_o}\ Q^{\epsilon}_{\theta^{e}_o}(s^{\tilde{t}+1}_o,a)-Q^{\epsilon}_{\theta^{e}_o}(s^{\tilde{t}}_o,a^{\tilde{t}}_o)|
\end{equation}
where $\theta^{t}_o$ is the target network parameter in the workflow offloading agent.

The value of target Q with NoisyNet and PER of the transition $(s^{\tilde{t}}_o,a^{\tilde{t}}_o,r^{\tilde{t}}_o,s^{\tilde{t}+1}_o) $
can be expressed as follows:
\begin{equation}
    \bar{Q}^{\tilde{t}}_o=r^{\tilde{t}}_o+\gamma \ Q^{\epsilon}_{\theta^{t}_o}(s^{\tilde{t}+1}_o,\mathop{\arg\max}\limits_{a\in\bar{A}^{\tilde{t}+1}_o}\ Q^{\epsilon}_{\theta^{e}_o}(s^{\tilde{t}+1}_o,a))
\end{equation}
Finally, all parameters of the evaluation network are updated through backpropagation to minimize the Mean Squared Error (MSE) loss function (line 20). We define the MSE loss function of the workflow offloading agent as follows:
\begin{equation}\label{MSE}
    L(\theta^{e}_o) = \frac{1}{\mathbb{B}} \sum_{k=1}^{\mathbb{B}}\left[\bar{Q}_o^{\tilde{t}_k}-Q^{\epsilon}_{\theta^{e}_o}(s^{\tilde{t}_k}_o,a^{\tilde{t}_k}_o)\right]^2
\end{equation}

\subsection{Denedency-aware Multi-Head Cross-Attention-Based Microservice Caching Agent}
The MDP of microservice caching agents deployed on edge servers can be described as a 3-tuple $(\mathcal{S}_c, \mathcal{A}_c, \mathcal{R}_c)$, where $\mathcal{S}_c=\{s^{\tilde{t}}_c\}_{\tilde{t}\in\tilde{T}}$ is the state space, $\mathcal{A}_c=\{a^{\tilde{t}}_c\}_{\tilde{t}\in\tilde{T}}$ is the action space, and $\mathcal{R}_c=\{r^{\tilde{t}}_c\}_{\tilde{t}\in\tilde{T}}$ is the set of rewards. The detailed definitions are as follows:

\textbf{1) Caching agent state $s^{\tilde{t}}_c$:} 
Given that the objective of service image caching is to reduce image pull delay for subsequent tasks in the task queue, we define the state $s^{\tilde{t}}_c$ for the agent performing service image caching on edge server $a^{\tilde{t}}_o$ in time slot $\tilde{t}$ as follows:
\begin{equation}
    s^{\tilde{t}}_c = \left(l^{u,t}_{\tilde{t}},\tilde{L}^{\tilde{t}}_{a^{\tilde{t}}_o},l^{t}_{\tilde{t}+1:\tilde{t}+\mathbb{L}}\right)
\end{equation}
where $l^{u,t}_{\tilde{t}}$ is the type of image of the task offloaded to the server $a^{\tilde{t}}_o$, $\tilde{L}^{\tilde{t}}_{a^{\tilde{t}}_o}=\left(\tilde{l}^{\tilde{t}}_{a^{\tilde{t}}_o,1},\dots,\tilde{l}^{\tilde{t}}_{a^{\tilde{t}}_o,\beta}\right)$ represents the types of images cached by $a^{\tilde{t}}_o$, and $l^{t}_{\tilde{t}+1:\tilde{t}+\mathbb{L}}$ is the types of images required by the next $\mathbb{L}$ subsequent tasks in the task queue.

\textbf{2) Caching agent action $a^{\tilde{t}}_c$:} 
The microservice caching agent determines which cached image on server $a^{\tilde{t}}_o$ should be replaced with the task image $l^{u,t}_{\tilde{t}}$. Therefore, we use $a^{\tilde{t}}_c \in \{1,\dots,\beta\}$ to represent the image replacement action.

\textbf{3) Caching agent reward $r^{\tilde{t}}_c$:} 
Considering that both image caching and task offloading ultimately aim to minimize the total task completion time, we also use Eq. \eqref{reward} to define the reward function of the microservice caching agent.

The core challenge of service image caching lies in effectively discovering and leveraging dependencies among task image, cached images on execution server, and images required for subsequent tasks. 
Given the typically long subsequent task queue, we integrate the advantages of cross-attention and multi-head attention, proposing a Dependency-Aware Multi-Head Cross-Attention (DAMH-CA) mechanism for mining long-range dependencies between different sequences. Its detailed procedure outlined in Algorithm \ref{DAMH-CA}. 
% Although the cross-attention mechanism can uncover dependencies between distinct sequences, it exhibits performance degradation when processing long sequences. To address this limitation, we integrate the strengths of multi-head attention and cross-attention, proposing a Dependency-Aware Multi-Head Cross-Attention (DAMH-CA) mechanism, and its detailed procedure outlined in Algorithm \ref{DAMH-CA}.

\begin{figure}[h]
    \removelatexerror
    \begin{algorithm}[H]
        \SetAlgoLined %显示end
        \caption{The Proposed DAMH-CA Mechanism}\label{DAMH-CA}
        \LinesNumbered
        \KwIn{State $s^{\tilde{t}}_c$, number of heads $\mathcal{N}$}
        \KwOut{Updated state $\tilde{s}^{\tilde{t}}_c$ after attention computation}
        $Mask\gets\varnothing$;$A_c\gets\varnothing$\;
        \For{$i=1:\mathbb{L}$}{
            \eIf{$l^{t}_{\tilde{t}+i} \notin \{1,\dots,|L|\}$}{
                $Mask\gets Mask\ \cup 1$\;
            }
            {
                $Mask\gets Mask\ \cup 0$\;
            }
        }
        $\bar{h}^{u,t}_{\tilde{t}} \gets \mathcal{H}_c\left(\tilde{l}^{u,t}_{\tilde{t}}\right)$; $\bar{H}^{\tilde{t}}_{a^{\tilde{t}}_o} \gets \mathcal{H}_c\left(\tilde{L}^{\tilde{t}}_{a^{\tilde{t}}_o}\right)$\;
        $\bar{L}^t_{\tilde{t}+1:\tilde{t}+\mathbb{L}}\gets \mathcal{H}_c\left(l^t_{\tilde{t}+1:\tilde{t}+\mathbb{L}}\right)$\;
        $q_c\gets \mathcal{Q}_c\left(\bar{h}^{u,t}_{\tilde{t}} \oplus \bar{H}^{\tilde{t}}_{a^{\tilde{t}}_o} \right)$\;
        $k_c\gets \mathcal{K}_c\left(\bar{L}^t_{\tilde{t}+1:\tilde{t}+\mathbb{L}}\right)$;$v_c\gets \mathcal{V}_c\left(\bar{L}^t_{\tilde{t}+1:\tilde{t}+\mathbb{L}}\right)$\;
        \For{$n=1:\mathcal{N}$}{
            Calculate $A^n_c$ according to Eq. \eqref{MHCA}\;
            $A_c \gets A_c \cup A^n_c$\;
        }
        $\tilde{s}^{\tilde{t}}_c \gets \mathcal{O}\left(A_c\right)$\;
    \end{algorithm}
\end{figure}
The proposed DAMH-CA mechanism first constructs the masking sequence $Mask$ for the padded regions within subsequent task images, where padding is inserted due to failure to meet the length requirement $\mathbb{L}$ (lines 1-7). Then it implements an embedding layer $\mathcal{H}_c$ that translates each image into a vector $\bar{h}_{\tilde{t}}\in\mathbb{R}^{1\times1\times\mathbb{H}}$ (lines 8-9). After that, the proposed DAMH-CA mechanism concatenates $\bar{h}^{u,t}_{\tilde{t}}$ and $\bar{H}^{\tilde{t}}_{a^{\tilde{t}}_o}$ as the input to the fully-connected layer $\mathcal{Q}_c$ and translates it into the query matrix $q_c \in \mathbb{R}^{1\times(1+\beta)\times \mathcal{N} \times d_c}$. It also maps the matrix $\bar{L}^t_{\tilde{t}+1:\tilde{t}+\mathbb{L}} \in \mathbb{R}^{1\times\mathbb{L}\times\mathbb{H}}$ to key matrix $k_c\in\mathbb{R}^{1\times \mathbb{L} \times \mathcal{N} \times d_c}$ and value matrix $v_c\in\mathbb{R}^{1\times \mathbb{L} \times \mathcal{N} \times d_c}$ (lines 11-12). The attention vector $A^n_c\in \mathbb{R}^{1\times (1+\beta)\times d_c}$ for the attention head $n$ is calculated by:
\begin{equation}\label{MHCA}
    A^n_c = Softmax(Mask(\frac{q_{c,n}k_{c,n}^\top}{\sqrt{d_c}}))v_{c,n}
\end{equation}
where function $Mask(x)$ represents the replacement of values in $x$ where the mask value is equal to 1 with an infinitesimal value. Finally, the proposed DAMH-CA mechanism concatenates the attention vectors $A^n_c$ from all attention heads and maps them to the final output through a fully-connected layer $\mathcal{O}$ (lines 13-17), which ends the algorithm.   

The microservice caching agent deployed on edge servers requires making image caching decisions based on continuous state information. Similarly to the workflow offloading agent, we also employ a D3QN algorithm that combines the proposed DAMH-CA mechanism to address this problem. The overall training procedure of the microservice caching agent in DLA-HDRL has been presented in Algorithm \ref{OAgent}.

During the training of the microservice caching agent, we integrate the workflow offloading agent into the DRL environment as a component. After the workflow offloading agent selects an action to offload $mt^{u,t}_{\tilde{t}}$ to edge server $a^{\tilde{t}}_o$, the state $s^{\tilde{t}}_c$ is generated within the environment, comprising the task image, the cached image on server $a^{\tilde{t}}_o$, and the task queue images (line 10). The microservice caching agent then selects an image caching action corresponding to the highest Q-value output from the evaluation network (line 11). As the workflow offloading agent constitutes an integral component of the environment, the next state becomes available only after the workflow offloading agent executes its action in the next time slot. Consequently, we utilize the tuple $(s_c,a_c,r_c,s^{\tilde{t}}_c)$ to record a transition and store it in the PER (lines 14-16). Finally, in alignment with the workflow offloading agent, we train the evaluation network of the microservice caching agent using the MSE loss function.

\subsection{Computational Complexity}
The computational complexity of a single inference using DLA-HDRL can be comprehensively analyzed by considering the decision-making process of the workflow offloading agent and the decision-making process of the microservice caching agent. 

1) The inference of the workflow offloading agent begins with computing the embedding vectors of the task image and images cached by servers, which has a complexity of $O(\mathbb{H}+|E| \cdot \beta\cdot \mathbb{H})=O(|E| \cdot \beta\cdot \mathbb{H})$. It then uses the proposed LACA mechanism to compute the attention vector, whose complexity depends on matrix multiplications and non-linear transformations. Specifically, the construction of the matrix $q^b_o$, $k^b_o$, and $v^b_o$ has a complexity of $O((2|E|+1)\cdot d_o\cdot\mathbb{H})=O(|E| \cdot d_o \cdot \mathbb{H})$, the complexity of the matrix multiplication between $q^b_o$ and $k^b_o$ is  $O(|E|\cdot d_o)$, the complexity of the normalized softmax operation is $O(|E|)$, and the final attention computation has a complexity of $O(|E|\cdot d_o)$. In conclusion, the total computational complexity of the proposed LACA can be represented as $O(|E| \cdot \beta\cdot \mathbb{H}+\beta \cdot(|E| \cdot d_o \cdot \mathbb{H}+|E|\cdot d_o+|E|+|E|\cdot d_o))=O(\beta \cdot |E|\cdot d_o\cdot\mathbb{H})$. The proposed neural network architecture for the workflow offloading agent employs progressively narrower layers, causing the matrix multiplication in the initial fully-connected layer (with the highest input-output dimensionality) to dominate forward propagation complexity. This computational complexity is quantified as $O((4|E|+\beta \cdot d_o)\cdot d_f)=O(\beta \cdot d_o\cdot d_f)$, where $d_f$ is the output dimensionality of the fully-connected layer. 

2) The inference of the microservice caching agent begins with using the proposed DAMH-CA mechanism to explore the dependencies between distinct long sequences. It first constructs a masking sequence with a computational complexity of $O(|L|)$. Then, it computes the embedding vectors with a computational complexity of $O(\mathbb{H}+\beta \cdot \mathbb{H}+\mathbb{L}\cdot\mathbb{H})=O(\beta\cdot\mathbb{H})$. It subsequently constructs the matrix $q_c$, $k_c$, and $v_c$ with a computational complexity of $O((1+\beta+\mathbb{L})\cdot\mathcal{N}\cdot d_c\cdot\mathbb{H})=O(\beta\cdot\mathcal{N}\cdot d_c\cdot\mathbb{H})$ and computes attention vectors for all heads with a computational complexity of $O(\mathcal{N}\cdot(1+\beta)\cdot\mathbb{L}\cdot d_c)$. Therefore, the total computational complexity of the proposed DAMH-CA mechanism can be calculated as $O(|L|+\mathbb{L}\cdot\mathbb{H}+\mathbb{L}\cdot\mathbb{H}\cdot \mathcal{N} \cdot d_c + \mathcal{N}\cdot(1+\beta)\cdot\mathbb{L}\cdot d_c)=O(\mathbb{L}\cdot\mathbb{H}\cdot \mathcal{N} \cdot d_c)$. Finally, the microservice caching agent selects an image replacement action, which has a computational complexity of $O((1+\beta)\cdot d_c\cdot d_f)$, where $d_f$ is the output dimensionality of the fully-connected layer. 

In summary, the total computational complexity of a single inference using DLA-HDRL can be derived as $O(\beta \cdot |E|\cdot d_o\cdot\mathbb{H}+\beta\cdot \mathcal{N}\cdot d_c\cdot\mathbb{H})$. Since $d_o$, $\mathbb{H}$, $\mathcal{N}$, and $d_c$ are fixed hyperparameters, the time complexity of the proposed DLA-HDRL for a single inference is linear with respect to $O(\beta \cdot |E|)$.

\section{Experiment}
\label{Sec:Experiment}

To thoroughly validate the performance of our proposed DLA-HDRL approach, we established our experimental platform based on real-world microservice data and edge environment data. In this section, we first introduce our experiment settings, and then we show the
experimental results and provide the analyses. 
All experiments were implemented using PyTorch 2.0.0 with CUDA 11.8 on an Ubuntu 20.04 system equipped with 32GB RAM, an Intel Core i7 CPU, and an NVIDIA RTX4090 GPU.

\subsection{Data Collection and Datasets}
Given the absence of an open-source public dataset that exhibits the layer sharing relationships among different microservice images, we compiled a dataset containing the download frequencies and layer structures of microservice images from a cloud platform\footnote{https://cloud.tencent.com} and released it on GitHub. Specifically, we first aggregated the weekly download frequencies of 50 commonly used microservice images from DockerHub. Subsequently, we employed the Dive image analysis tool\footnote{https://github.com/wagoodman/dive} to analyze the file hash values and sizes of all image layers. If two distinct microservice images contained image layers with identical file hash values, this demonstrated that they share this particular layer. Using the microservice image download frequency data from this dataset, we performed the generation of corresponding microservice images for user-generated services through the roulette wheel selection method\cite{RouletteWheel_TCCN_2025}. To ensure realistic experimental conditions, we base our edge network construction on the EUA dataset \cite{EUA_TPDS_2020}, which provides the geographical distribution of real-world edge servers in Australia. Besides, to make the experiment more practical, we employ the
Alibaba cluster trace dataset\cite{Alibabadataset_SOCC_2021} to generate the topological structures of microservice workflows. It contains substantial microservice call graphs within
more than 10 clusters. We randomly selected 4500 call graphs as the training set and 500 call graphs as the test set to evaluate the performance of our DLA-HDRL approach. 
\begin{figure*}[t]
\centering
\begin{subfigure}{0.325\textwidth}
    \includegraphics[width=\linewidth]{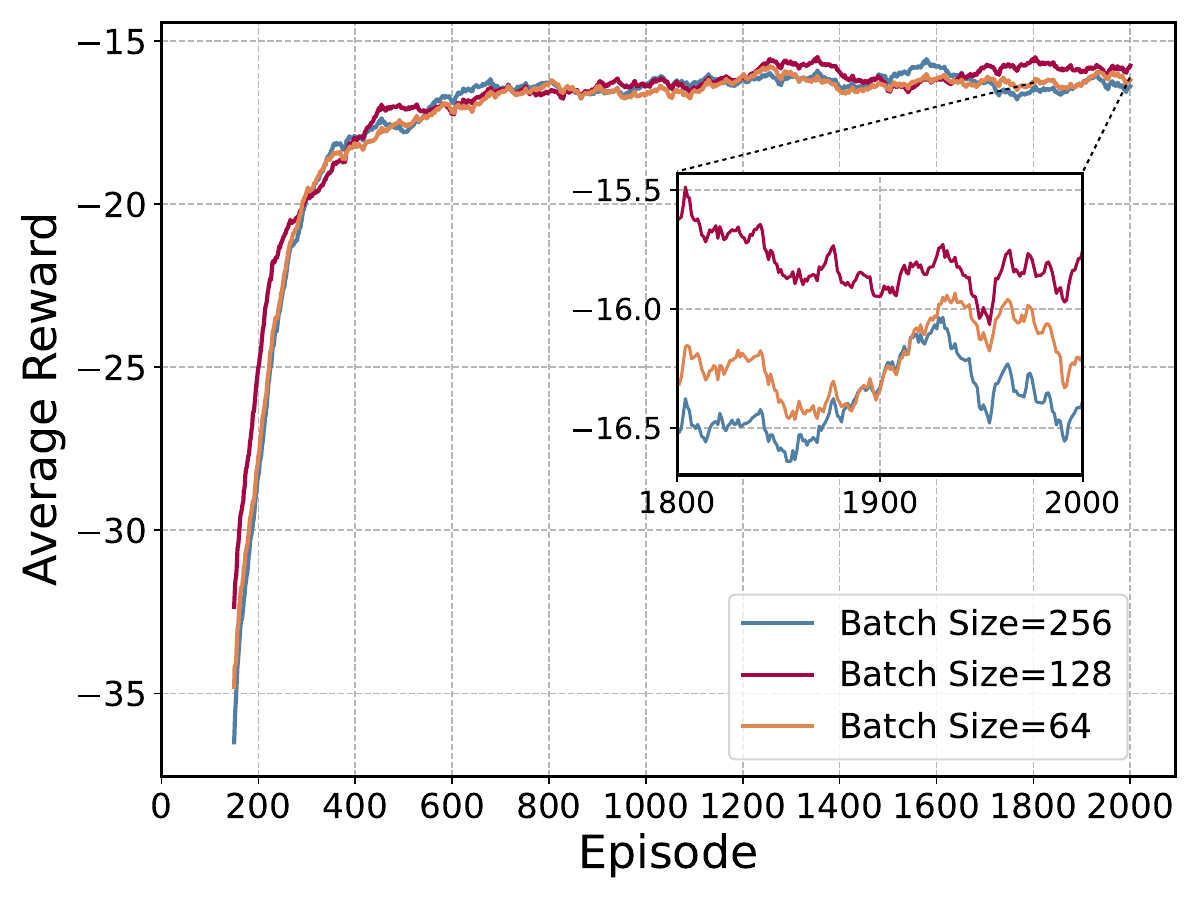}
    \caption{Reward with different batch sizes}
    \label{Convergence:sub1}
\end{subfigure}
\hfill
\begin{subfigure}{0.325\textwidth}
    \includegraphics[width=\linewidth]{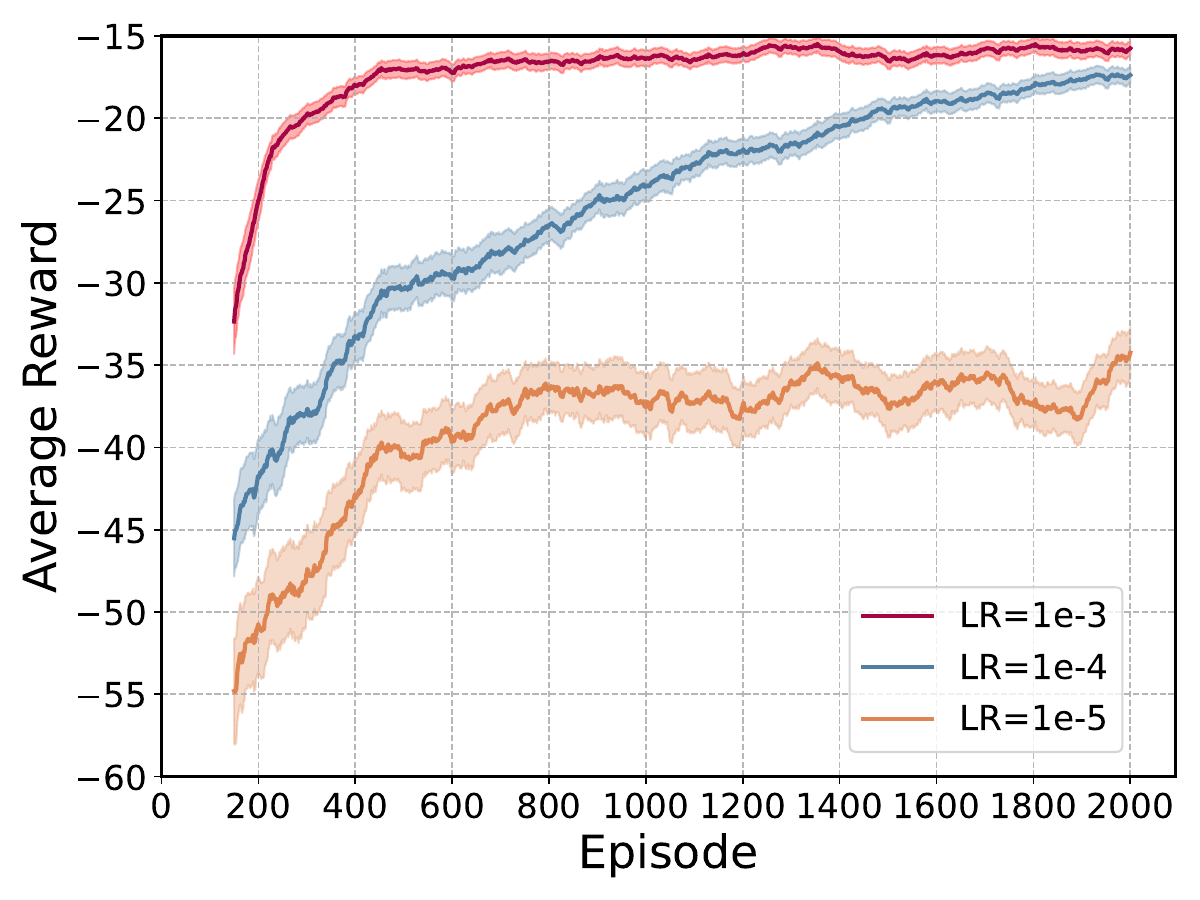}
    \caption{Reward with different learning rates}
    \label{Convergence:sub2}
\end{subfigure}
\hfill
\begin{subfigure}{0.325\textwidth}
    \includegraphics[width=\linewidth]{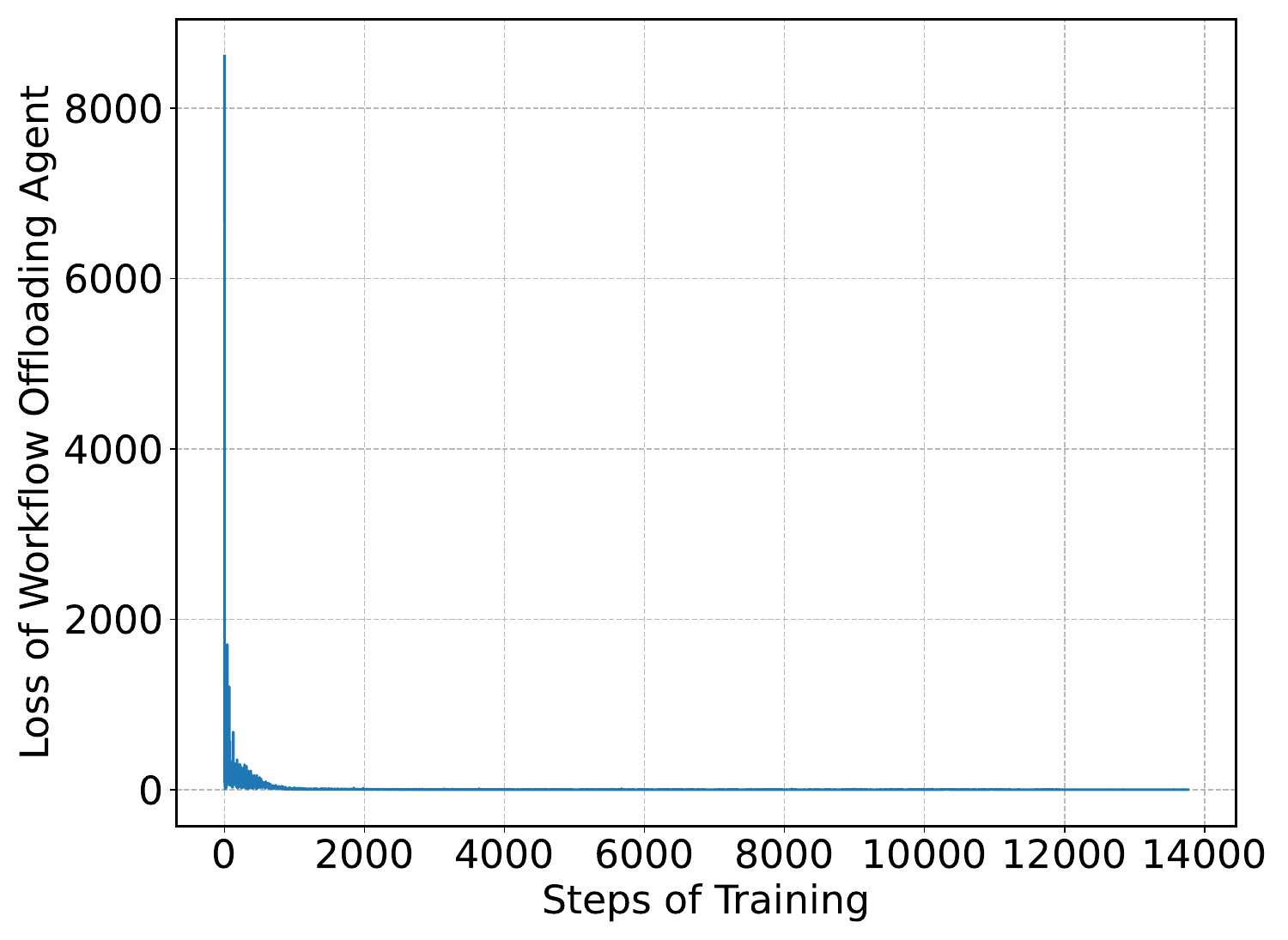}
    \caption{Loss of workflow offloading agent}
    \label{Convergence:sub3}
\end{subfigure}
% 第二行子图（增加垂直间距）
\vspace{0.5cm} % 行间距调整

\begin{subfigure}{0.325\textwidth}
    \includegraphics[width=\linewidth]{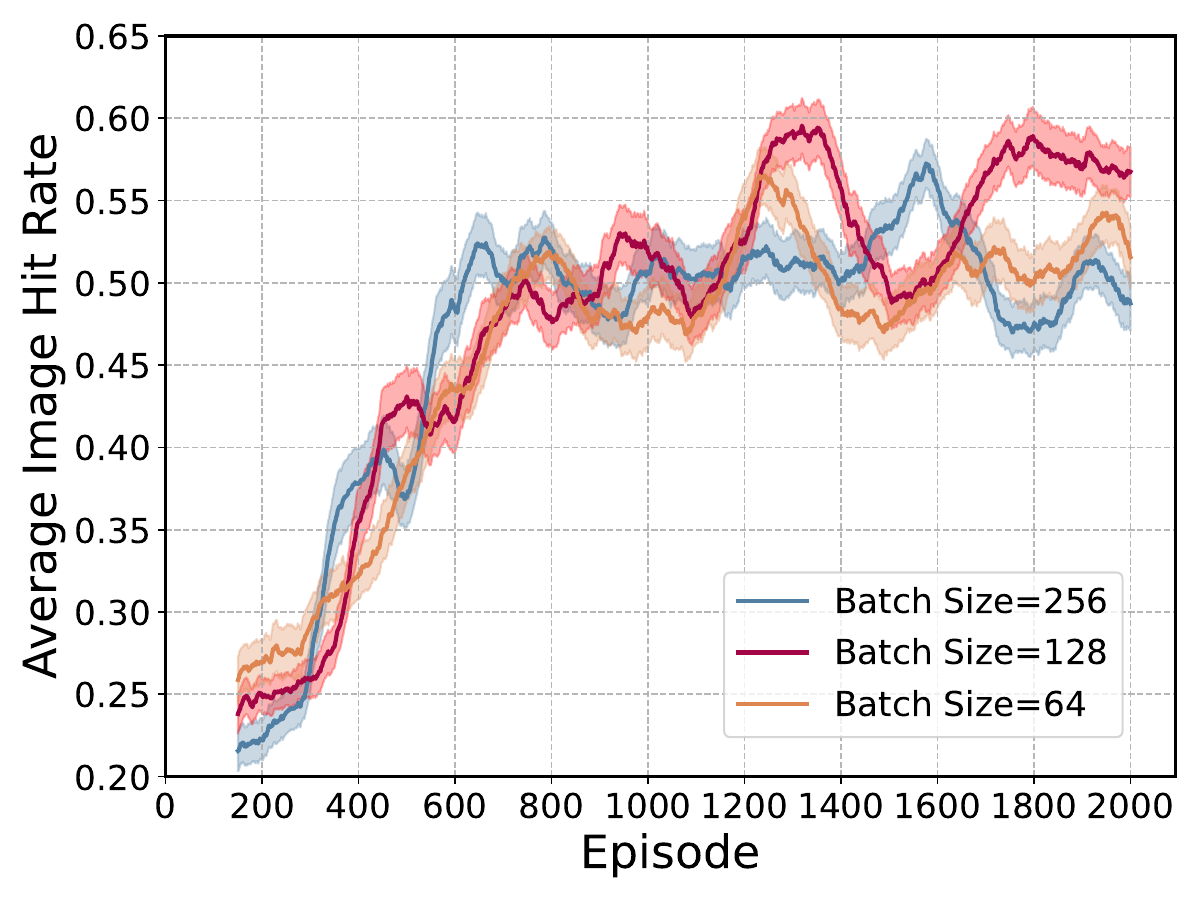}
    \caption{Hit rate with different batch sizes}
    \label{Convergence:sub4}
\end{subfigure}
\hfill
\begin{subfigure}{0.325\textwidth}
    \includegraphics[width=\linewidth]{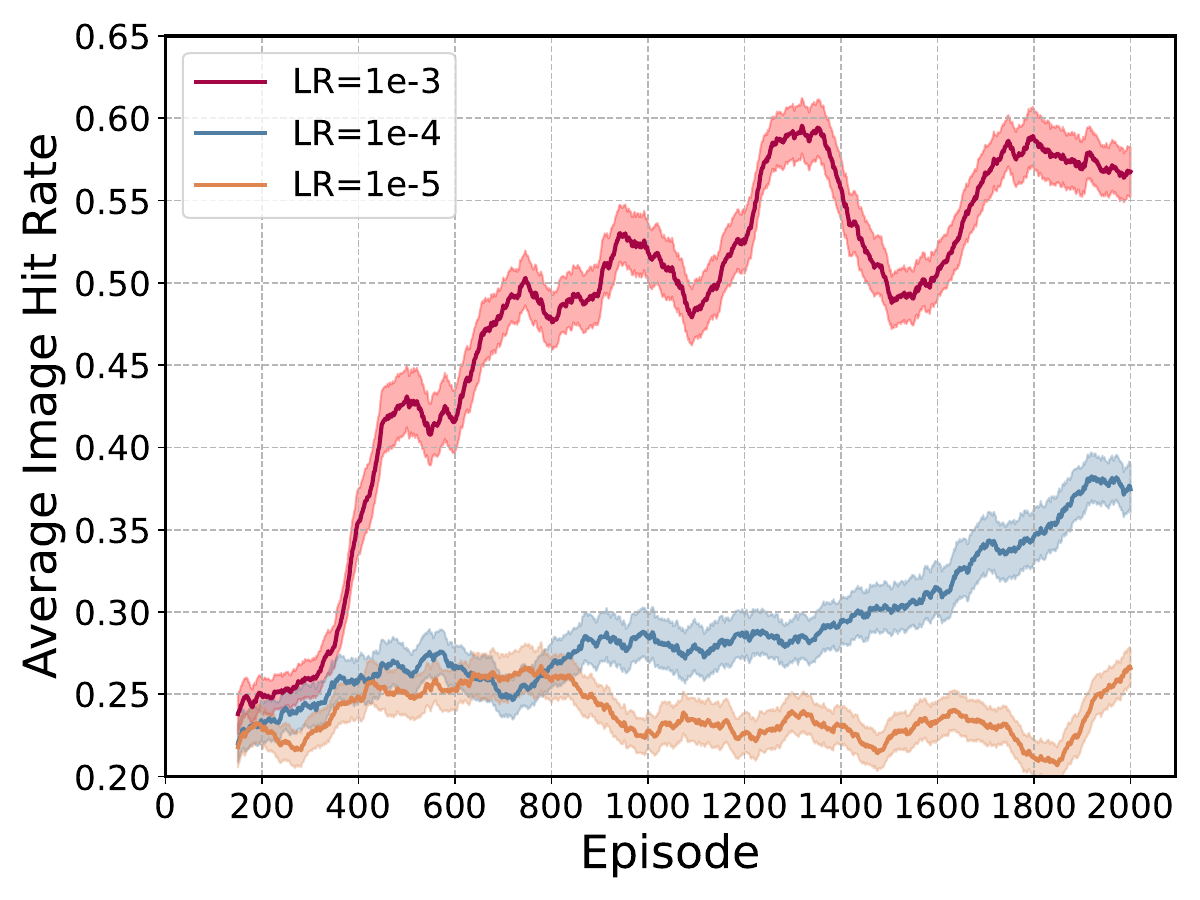}
    \caption{Hit rate with different learning rates}
    \label{Convergence:sub5}
\end{subfigure}
\hfill
\begin{subfigure}{0.325\textwidth}
    \includegraphics[width=\linewidth]{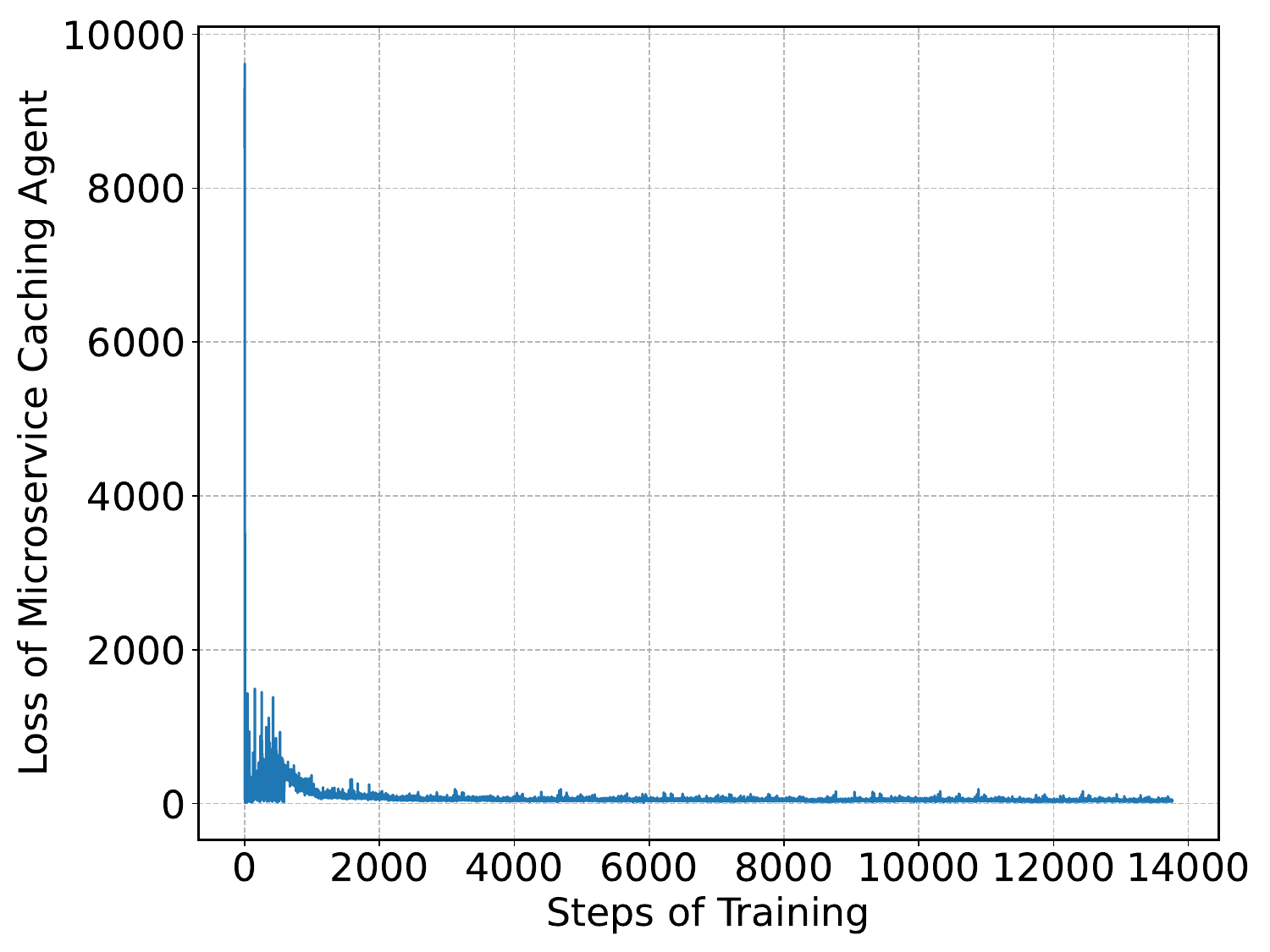}
    \caption{Loss of microservice caching agent}
    \label{Convergence:sub6}
\end{subfigure}
\caption{Performance of average system reward and average microservice image hit rate with different batch sizes and
learning rates}
\label{Convergence}
\end{figure*}

\subsection{Environmental and Parameter Settings}
Our experimental platform consists of $|E|$ adjacent edge servers located in Sydney within the EUA dataset. Each server can cache $\beta$ microservice images. During the training phase, each of the $|U|$ users generates a microservice workflow every $t$ seconds, with each user located at a different position. The workflow offloading agent then determines the task offloading strategy, and the microservice caching agent determines the image caching strategy. The parameter settings of the experimental platform are detailed in Table \ref{table:PARAMETER}. An episode concludes upon completion of offloading all tasks from the task queue generated during time slot $t$.

\begin{table}[htbp]
  \centering
  \caption{PARAMETER SETTINGS}
  \label{table:PARAMETER}
  \resizebox{\linewidth}{!}{
    \begin{tabularx}{\linewidth}{c>{\centering\arraybackslash}X>{\centering\arraybackslash}X}
      \toprule
      \multicolumn{1}{c}{} & \multicolumn{1}{c}{Parameter} & \multicolumn{1}{c}{Value} \\
      \midrule
      \multirow{17}{*}{\rotatebox{90}{Configuration of Edge Environment}} 
      & $|L|$ & 50 \\
      & $|U|$ & $\left[10, 15, 20\right]$ \\
      & $|E|$ & $\left[4, 5, 6\right]$ \\
      & $\beta$ & $\left[2, 4, 6\right]$ \\
      & $\alpha_{e_i}$ & $450\sim750$m\cite{EUA_TPDS_2020} \\
      & $d^{u,t}_n$ & $12.5 \sim 50$MB\cite{Uplinksize_TSC_2025} \\
      & $\tilde{d}^{u,t}_n$ & $0.25\sim1$MB \\
      & $\kappa^{u,t}_n$ & $100\sim300$CPU cycles/bit\cite{ComputationalIntensity_TMC_2024} \\
      & $\xi_{e_i}$ & $30\sim50$G CPU cycles/s \cite{ComputationalIntensity_TMC_2024} \\
      & $\sigma^{u,t}_n$ & $0.1 \sim 0.8$ (Normalized memory size)\cite{Alibabadataset_SOCC_2021} \\
      & $\sigma_{e_i}$ & $0.7\sim 1$ (Normalized memory size) \\
      & $B$ & 20MHz\cite{TransRate_IOTJ_2022} \\
      & $\rho$ & $0.5$W\cite{backgroundnoise_JSAC_2018} \\
      & $\eta$ & -4 \cite{pathloss_JSAC_2024} \\
      & $\lambda^2$ & $2\times10^{-13}$W \cite{backgroundnoise_JSAC_2018} \\
            & $t$ & $300$s\cite{DeployMStoExecutionTasks_IOTJ_2024} \\
       & $R^\nu_{e_i}, R^{e}_{e_i,e_j}$ & $512\sim1000$Mbps\cite{TransRate_IOTJ_2022} \\
      \midrule
      \multirow{8}{*}{\rotatebox{90}{DLA-HDRL}} 
      & $\mathbb{E}$ & 2000 \\
      & $d_f$ & 32 \\
      & Optimizer & Adam \\
      & $\mathbb{S}$ & 10 \\
      & $\alpha$ & 0.6 \\
      & $\mathcal{B}$ & 1e-5 \\
      & $\mathbb{H}$ & 16 \\
      & $\mathbb{L}$ & 10 \\

      \bottomrule
    \end{tabularx}%
  }
\end{table}

\subsection{Baselines}
For performance comparison, we compare the proposed DLA-HDRL
approach with the following baselines:

\textbf{1) Dueling Double Deep Q-Network with NoisyNet and PER (D3QNNP)\cite{D3QNNP_TCCN_2024}}: D3QNNP is one of the most efficient value-based DRL algorithms. The action space of the D3QNNP is jointly defined by task offloading and microservice caching. NoisyNet is employed to enhance the exploration capabilities of the agent, while PER is utilized to improve the convergence efficiency during training.

\textbf{2) Soft Actor-Critic (SAC)\cite{DependencyAwareMicroserviceDelpyment_TMC_2024}}: SAC is one of the most advanced DRL algorithms that simultaneously optimizes policy entropy maximization and cumulative reward objectives through a decoupled actor-critic architecture.

\textbf{3) Proximal Policy Optimization (PPO)\cite{PPO_SAC}}: PPO is a classical policy-based DRL algorithm that optimizes policy parameters through a clipped surrogate objective function while leveraging importance sampling to enhance training stability.

\textbf{4) Heuristic Offloading and Caching (HOC)}: We constructed the HOC method to compare the performance of the proposed DLA-HDRL approach with traditional heuristic methods. The HOC method employs a greedy strategy to offload tasks to the edge server with the earliest available computing resources and subsequently utilizes the LRU (Least Recently Used) algorithm for microservice image caching.

\subsection{Convergence Performance}
We first analyze the convergence performance of our approach, where we take the average reward and the average image hit rate as the evaluation criteria, with different batch sizes and learning rates, as shown in Fig. \ref{Convergence}. We have the following observations and findings: (1) Figs. \ref{Convergence:sub1} and \ref{Convergence:sub4} present the average reward results and the average hit rate results during training at batch sizes of 64, 128, and 256, respectively. When the batch size is set to 128, the final converged average reward and average hit rate apparently outperform those achieved with the other two configurations. This phenomenon occurs because a larger batch size tends to stabilize gradient updates but may reduce generalization capability, while a smaller batch size introduces higher stochastic noise, potentially hindering convergence efficiency. (2) Figs. \ref{Convergence:sub2} and \ref{Convergence:sub5} present the average reward results and the average hit rate results during training at learning rates of 1e-3, 1e-4, and 1e-5, respectively. When the learning rate is set to 1e-3, both the average reward and average hit rate achieve superior convergence compared to the other two values. When the batch size is set to 128 and the learning rate is set to 1e-3, we notice that the average reward converges after 1200 episodes, while the average hit rate exhibits minor oscillations of approximately 0.1. This is because in different microservice workflows, the sizes of different microservice images may have extreme variations. For instance, the size of the ZooKeeper image is 313 MB, while the size of the Python image is only 44.9 MB. Significant variations in image sizes necessitate that the agent account for additional factors (e.g., computing delay and data transmission delay)  during task offloading with a small image to ensure efficient offloading decisions. Consequently, our DLA-HDRL approach can fully account for all components within the state space, ensuring the convergence stability of rewards. (3) Additionally, Figs. \ref{Convergence:sub3} and \ref{Convergence:sub6} depict the loss function results of the workflow offloading agent and the microservice caching agent, respectively. We observe that the loss values for both agents converge to a minimum value at the end of training, demonstrating the full convergence of our DLA-HDRL approach.

\begin{figure*}[!t]
\centering
\begin{subfigure}{0.325\textwidth}
    \includegraphics[width=\linewidth]{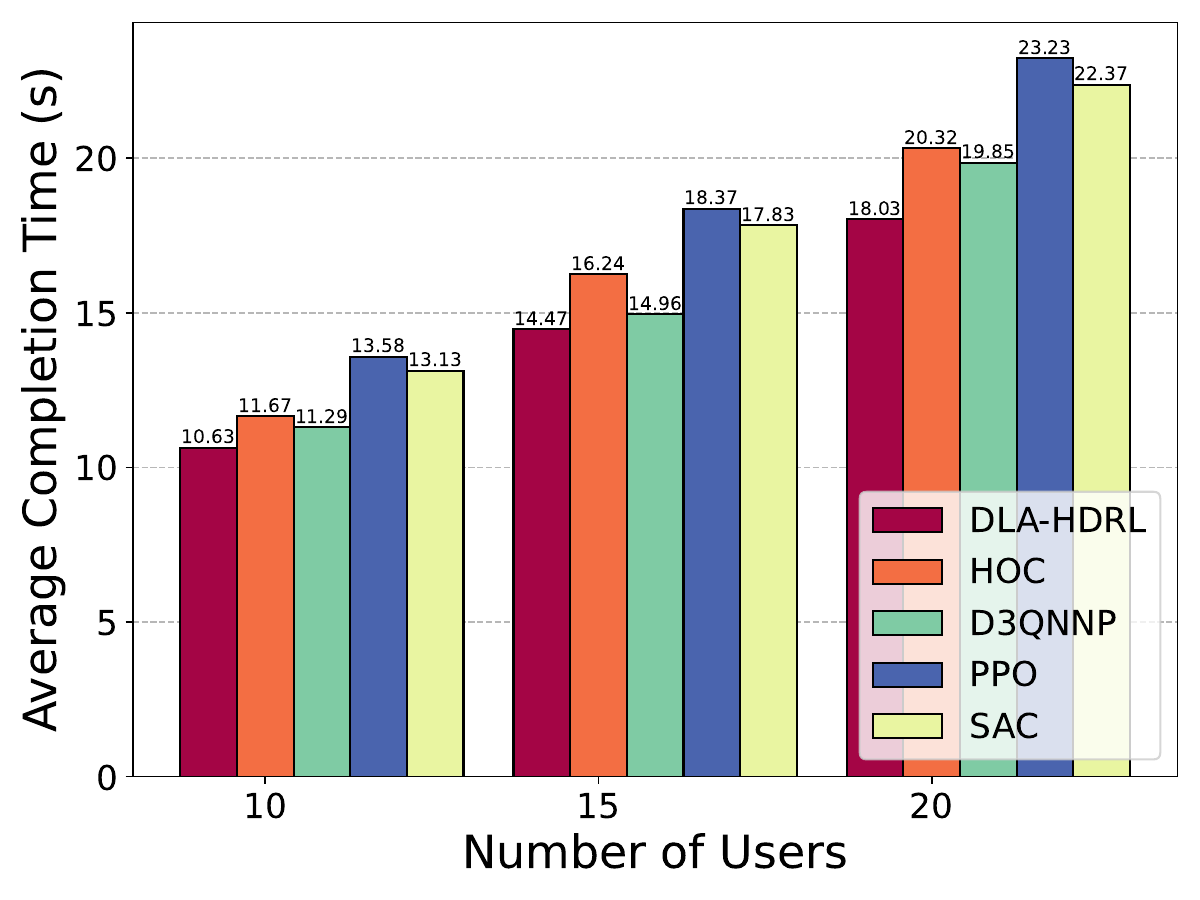}
    \caption{Average completion time with different user numbers}
    \label{TestSet:sub1}
\end{subfigure}
\hfill
\begin{subfigure}{0.325\textwidth}
    \includegraphics[width=\linewidth]{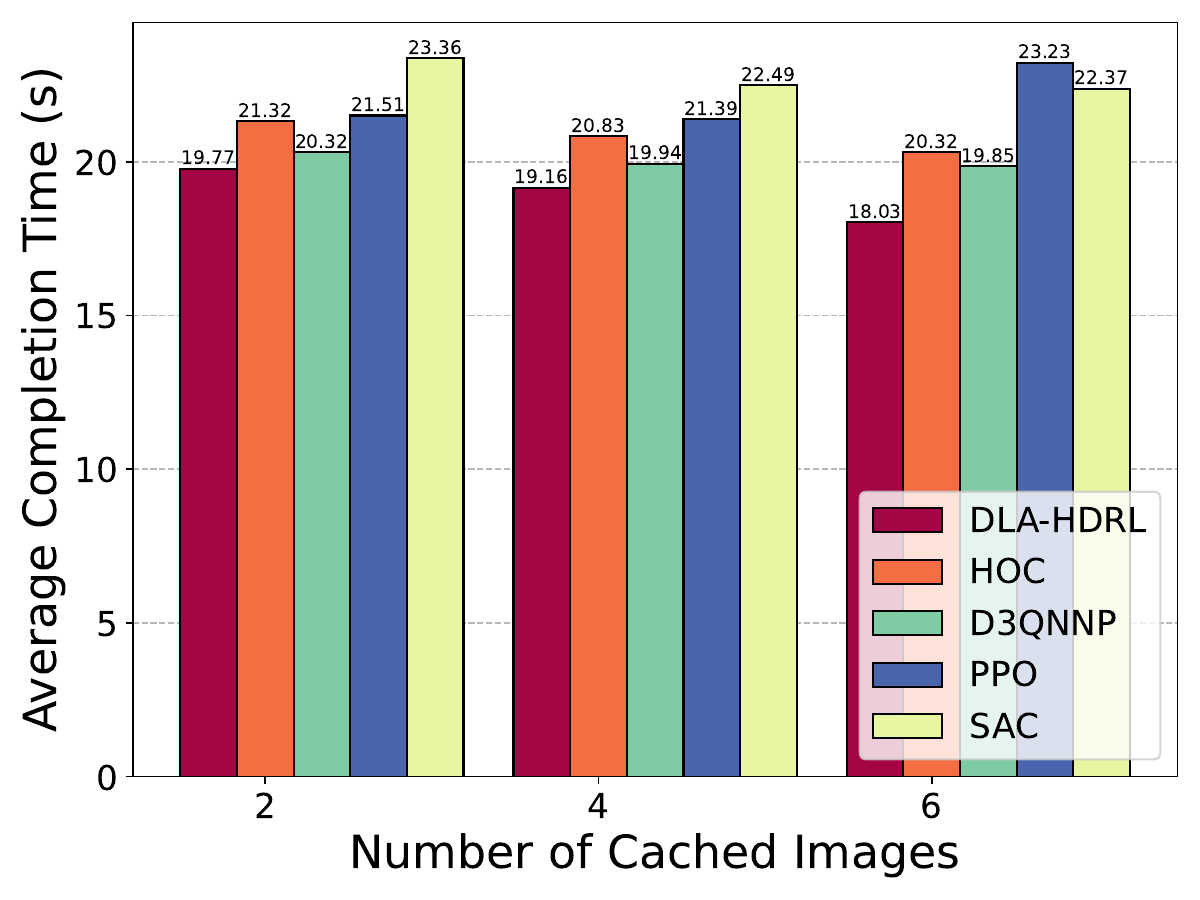}
    \caption{Average completion time with different numbers of cached images}
    \label{TestSet:sub2}
\end{subfigure}
\hfill
\begin{subfigure}{0.325\textwidth}
    \includegraphics[width=\linewidth]{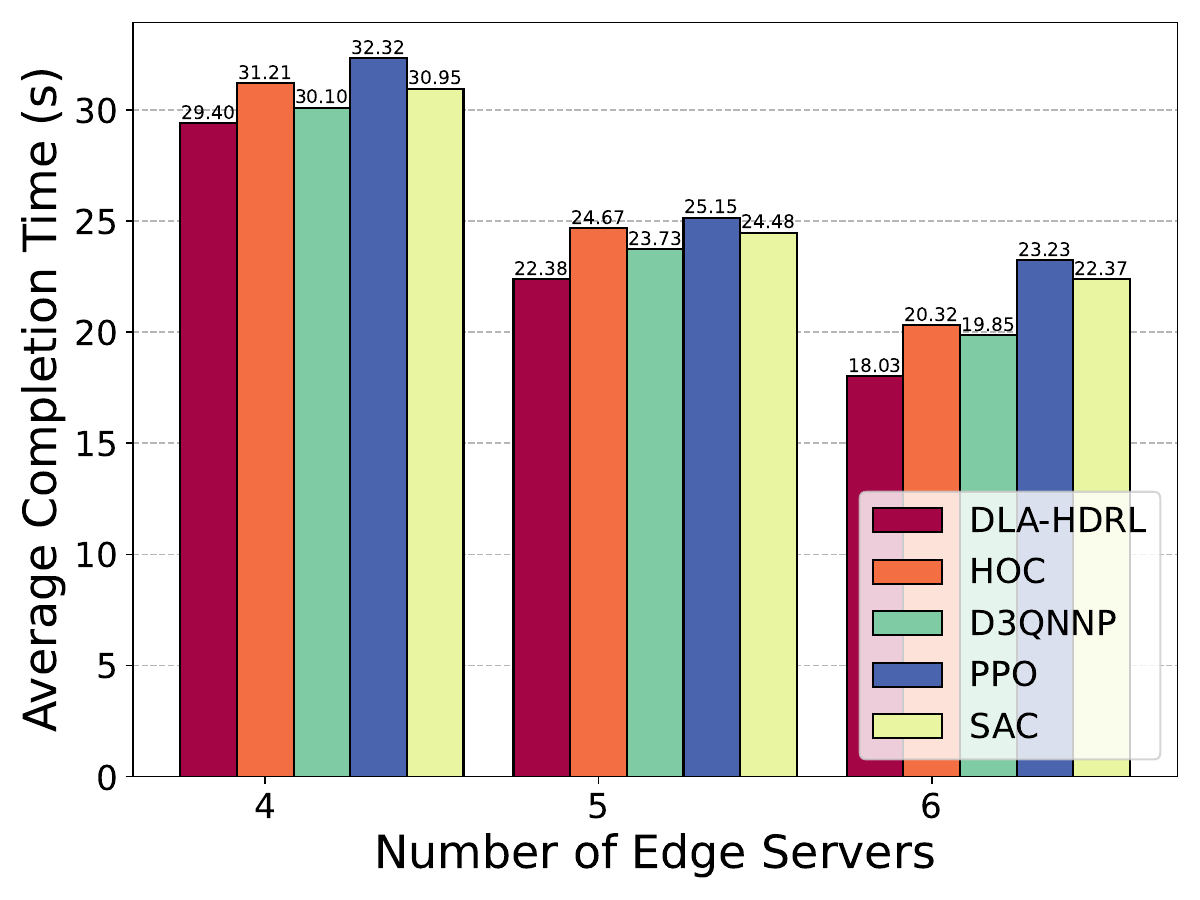}
    \caption{Average completion time with different edge server numbers}
    \label{TestSet:sub3}
\end{subfigure}
% 第二行子图（增加垂直间距）
\vspace{0.5cm} % 行间距调整

\begin{subfigure}{0.325\textwidth}
    \includegraphics[width=\linewidth]{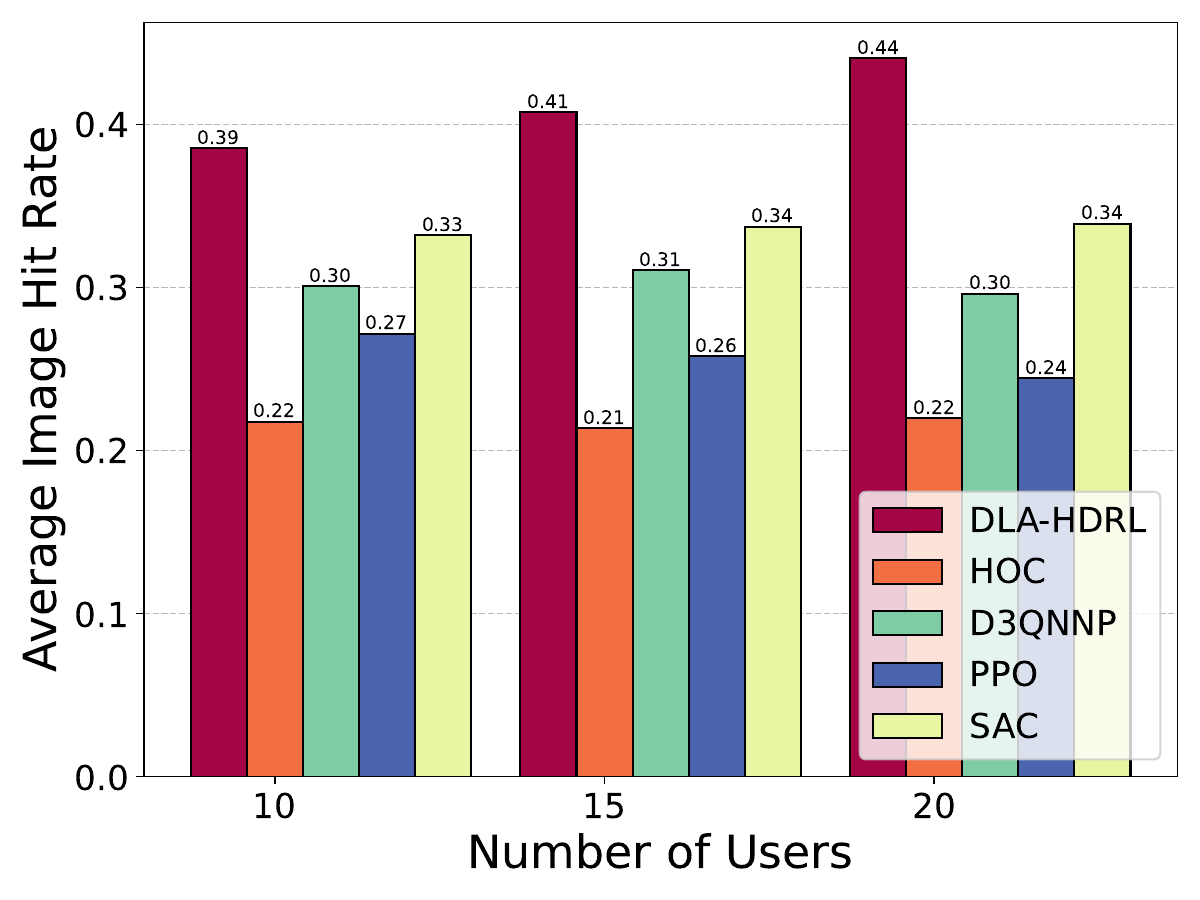}
    \caption{Hit rate with different user numbers}
    \label{TestSet:sub4}
\end{subfigure}
\hfill
\begin{subfigure}{0.325\textwidth}
    \includegraphics[width=\linewidth]{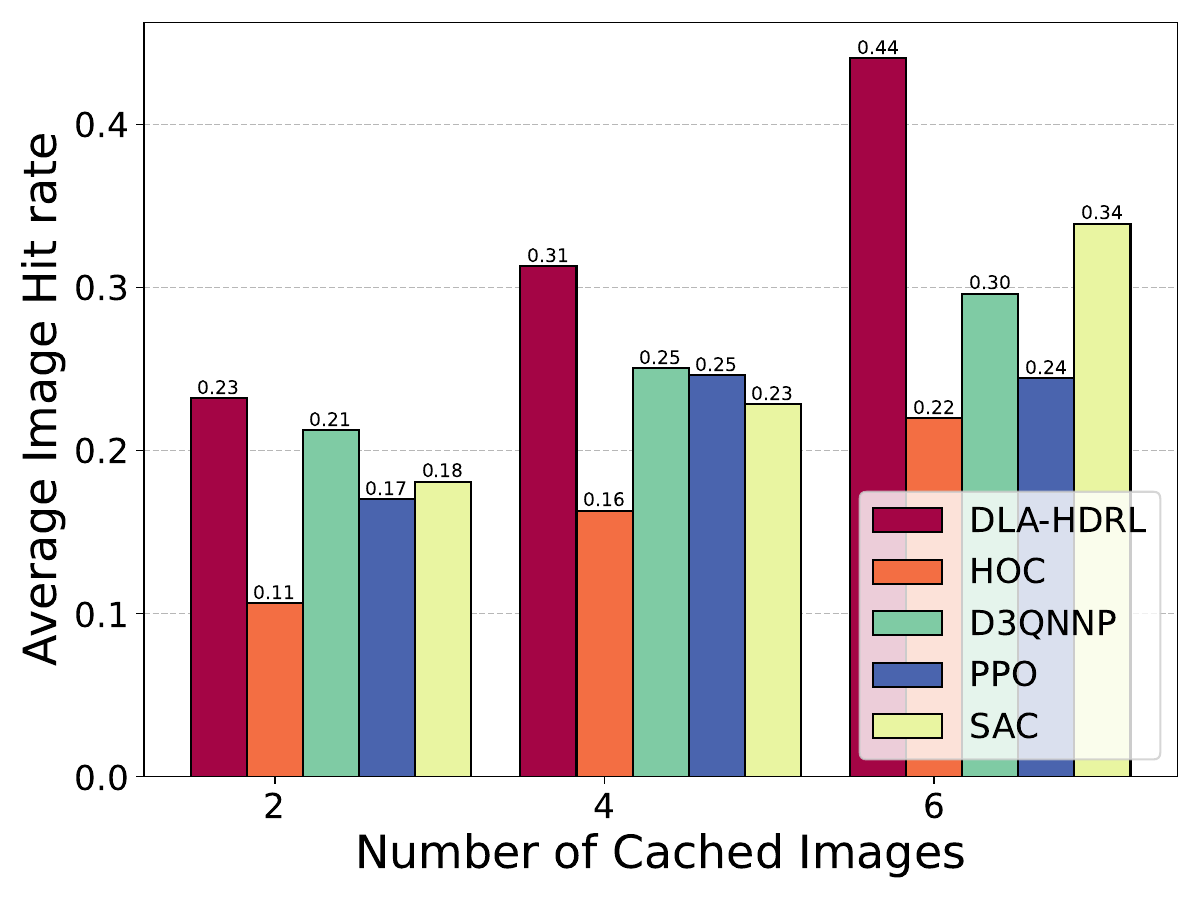}
    \caption{Hit rate with different numbers of cached images}
    \label{TestSet:sub5}
\end{subfigure}
\hfill
\begin{subfigure}{0.325\textwidth}
    \includegraphics[width=\linewidth]{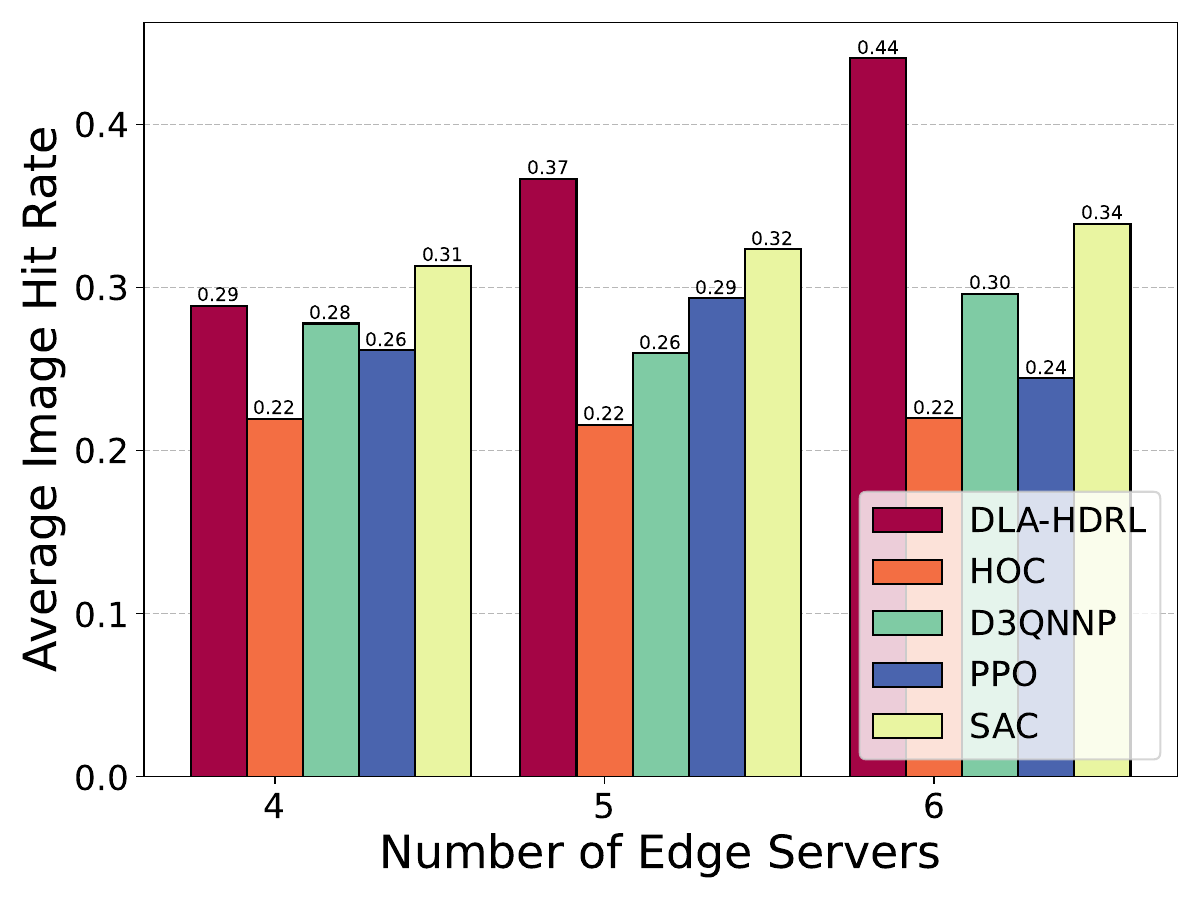}
    \caption{Hit rate with different edge server numbers}
    \label{TestSet:sub6}
\end{subfigure}
\caption{Performance of average system reward and average microservice image hit rate with different user numbers, numbers of images cached by each edge server, and edge server numbers.}
\label{TestSet}
\end{figure*}

\subsection{Performance Comparison}
We compare the performance of the proposed DLA-HDRL approach with baselines.  The results are displayed in Fig. \ref{TestSet}.

1) From Figs. \ref{TestSet:sub1} and \ref{TestSet:sub4}, we observe that when the number of users varies between 10 and 20, the proposed DLA-HDRL approach reduces the task completion time (Eq. \ref{11}) by 8.87\% to 11.27\%, 3.26\% to 9.19\%, 21.22\% to 22.38\%, and 18.82\% to 19.41\%, compared to the HOC, D3QNNP, PPO, and SAC algorithms, respectively.
Simultaneously, our approach achieved improvements in the average microservice image hit rate ranging from 77.11\% to 100.5\%, 28.09\% to 48.68\%, 41.92\% to 80.26\%, and 16.11\% to 29.87\%, respectively, compared to the same baselines.
Furthermore, we found that as the number of users increased, the performance advantage of our DLA-HDRL approach over baselines became increasingly pronounced. The reason for this phenomenon is that DLA-HDRL employs the LACA mechanism to calculate the similarity between the task image and images cached by edge servers. This enables DLA-HDRL to take full advantage of the layer sharing characteristic of microservice images to significantly reduce task completion time by increasing the image hit rate. Besides, our experimental results reveal that while the image hit rate of baselines remains stable with an increasing number of users, our DLA-HDRL approach demonstrates a progressive improvement in image hit rate. This advantage derives from our approach using the proposed DAMH-CA mechanism to make caching decisions. This mechanism characterizes the long-range dependencies among task images, images cached by servers, and images in task queue. Consequently, this enhanced capability of the DAMH-CA mechanism translates directly into an increased image hit rate as the length of the task queue increases. 

2) Figs. \ref{TestSet:sub2} and \ref{TestSet:sub5} illustrate that as the number of microservice images cached by each edge server $\beta$ increases from 2 to 6, our DLA-HDRL approach reduces the average task completion time from 7.29\% to 11.27\%, 2.69\% to 9.19\%, 8.08\% to 22.38\%, and 14.79\% to 19.41\% compared to the HOC, D3QNNP, PPO, and SAC algorithms, respectively. Concurrently, it improves the average image hit rate from 91.98\% to 117.79\%, 9.11\% to 48.68\%, 27.24\% to 80.26\%, and 28.34\% to 37.18\%, respectively, over the same baseline algorithms. We also observe that as $\beta$ increases, the performance improvements of DLA-HDRL in terms of both average task completion time and average image hit rate become increasingly pronounced. In contrast, the PPO algorithm even exhibits performance degradation when $\beta$ is set to 6. This phenomenon occurs because an increase in the value of $\beta$ expands both the state space and the action space for DRL. This expansion makes it difficult for traditional DRL algorithms to converge to satisfactory performance levels. Our DLA-HDRL approach addresses this challenge by employing an HDRL architecture, which decouples state space and action space. Furthermore, it leverages the LACA and DAMH-CA mechanisms to enhance feature extraction for the agent.

3) We can also observe from Figs. \ref{TestSet:sub3} and \ref{TestSet:sub6} that when the number of edge servers ranges from 4 to 6, DLA-HDRL can reduce the average task completion time from 5.79\% to 11.27\%, 2.32\% to 9.19\%, 9.02\% to 22.38\%, and 4.99\% to 19.41\%, compared to the traditional HOC, D3QNNP, PPO, and SAC algorithms, respectively. Furthermore, our approach improves the average image hit rate by 31. 63\% to 100.5\%, 3. 94\% to 48.68\% and 10. 37\% to 80.26\% compared to the same baselines. When the number of edge servers is 4, the proposed DLA-HDRL approach achieves a slightly lower hit rate than the SAC algorithm, but note that, our approach still attains optimal task completion time among all baselines. This occurs because while improving the hit rate reduces the delay of pulling images for microservices, the task completion time is also influenced by other factors (e.g., computing delay and data transmission delay). While ensuring a high hit rate through our proposed LACA and DAMH-CA mechanisms, DLA-HDRL additionally employs HDRL architecture to decouple the state spaces of task offloading and microservice caching. This architecture reduces the interference of image information to delay information when making offloading decisions, thereby enabling the agent to make decisions that can minimize the task completion time. Furthermore, the synergy between the proposed attention mechanisms and the HDRL architecture further allows DLA-HDRL to achieve enhanced performance with an increasing number of edge servers.

\subsection{Sensitivity Analysis to System Scale}
\begin{figure*}[!t]
\centering
\begin{subfigure}{0.325\textwidth}
    \includegraphics[width=\linewidth]{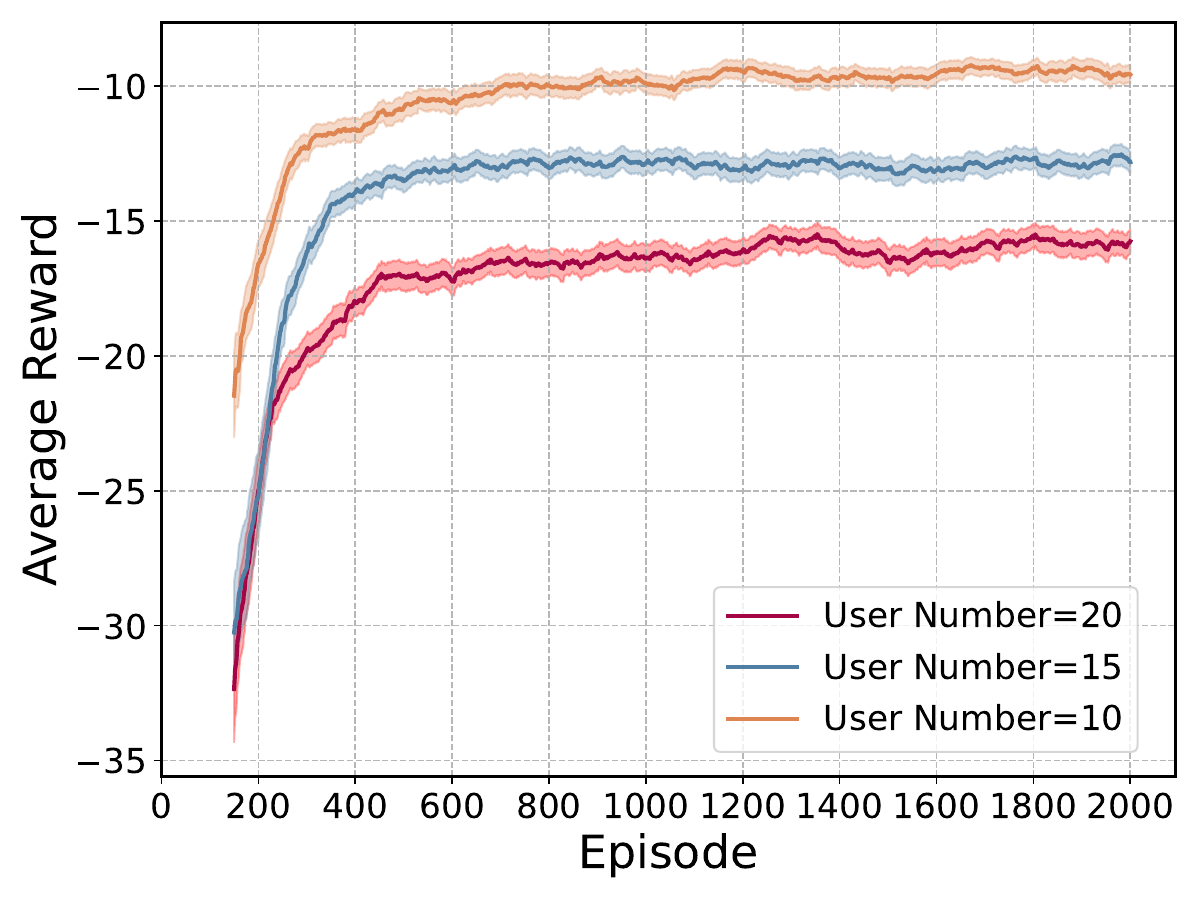}
    \caption{Reward with different user numbers}
    \label{Sensitivity:sub1}
\end{subfigure}
\hfill
\begin{subfigure}{0.325\textwidth}
    \includegraphics[width=\linewidth]{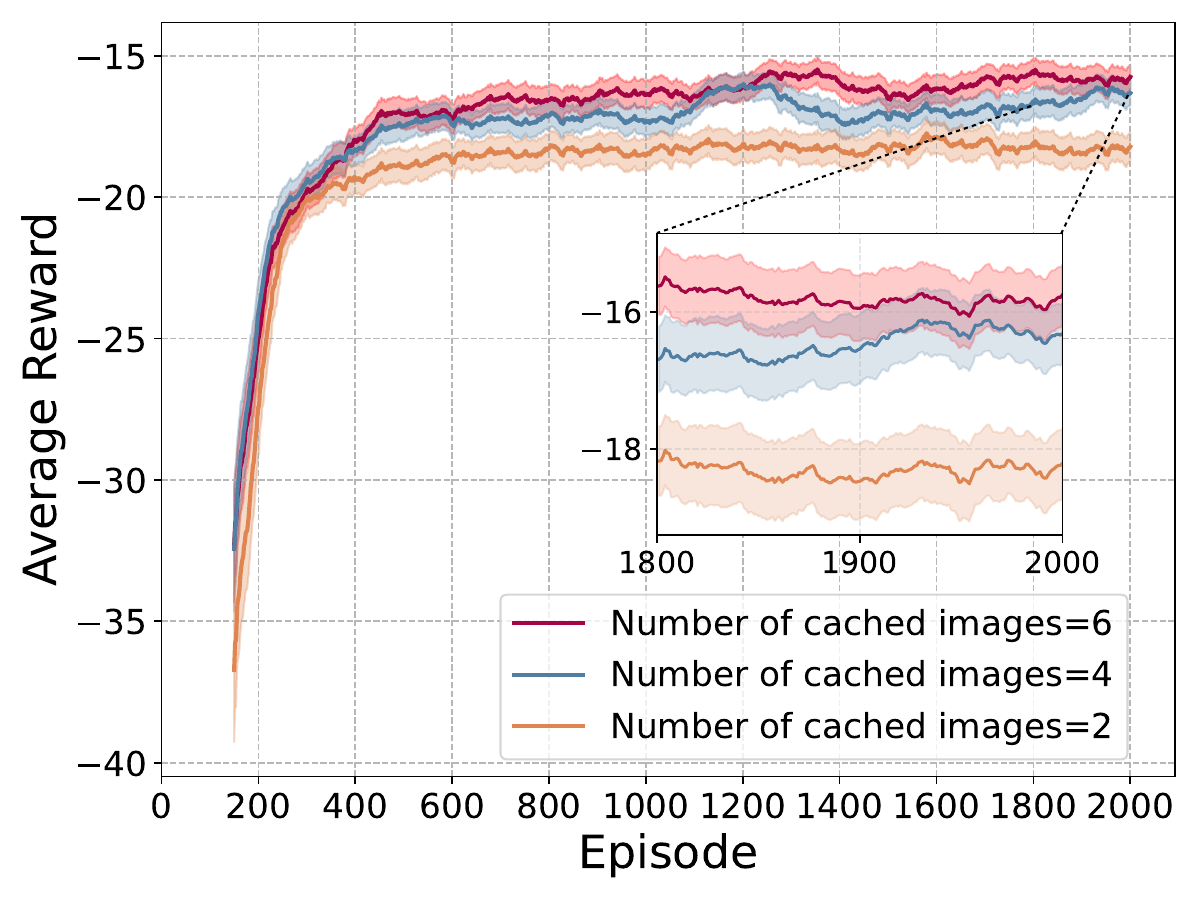}
    \caption{Reward with different numbers of cached images}
    \label{Sensitivity:sub2}
\end{subfigure}
\hfill
\begin{subfigure}{0.325\textwidth}
    \includegraphics[width=\linewidth]{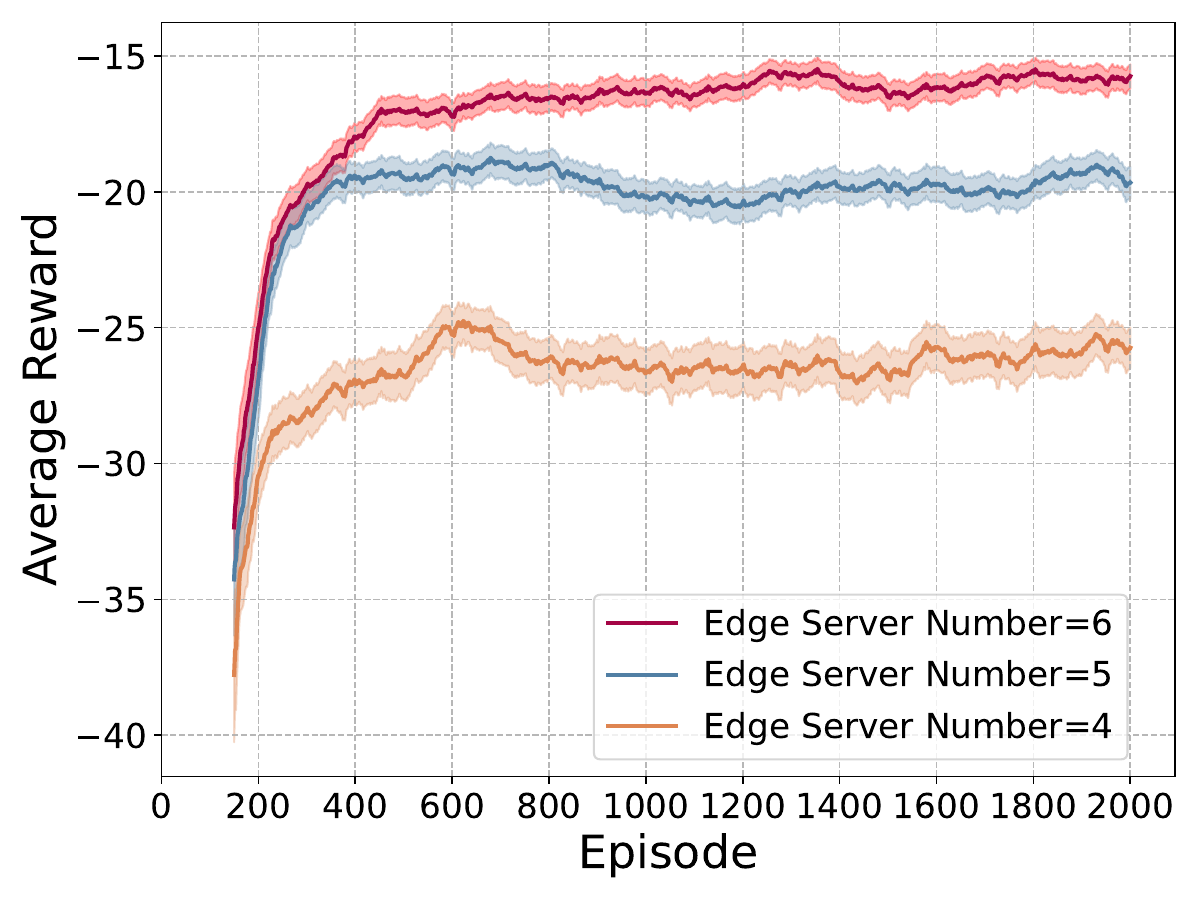}
    \caption{Reward with different edge server numbers}
    \label{Sensitivity:sub3}
\end{subfigure}
% 第二行子图（增加垂直间距）
\vspace{0.5cm} % 行间距调整

\begin{subfigure}{0.325\textwidth}
    \includegraphics[width=\linewidth]{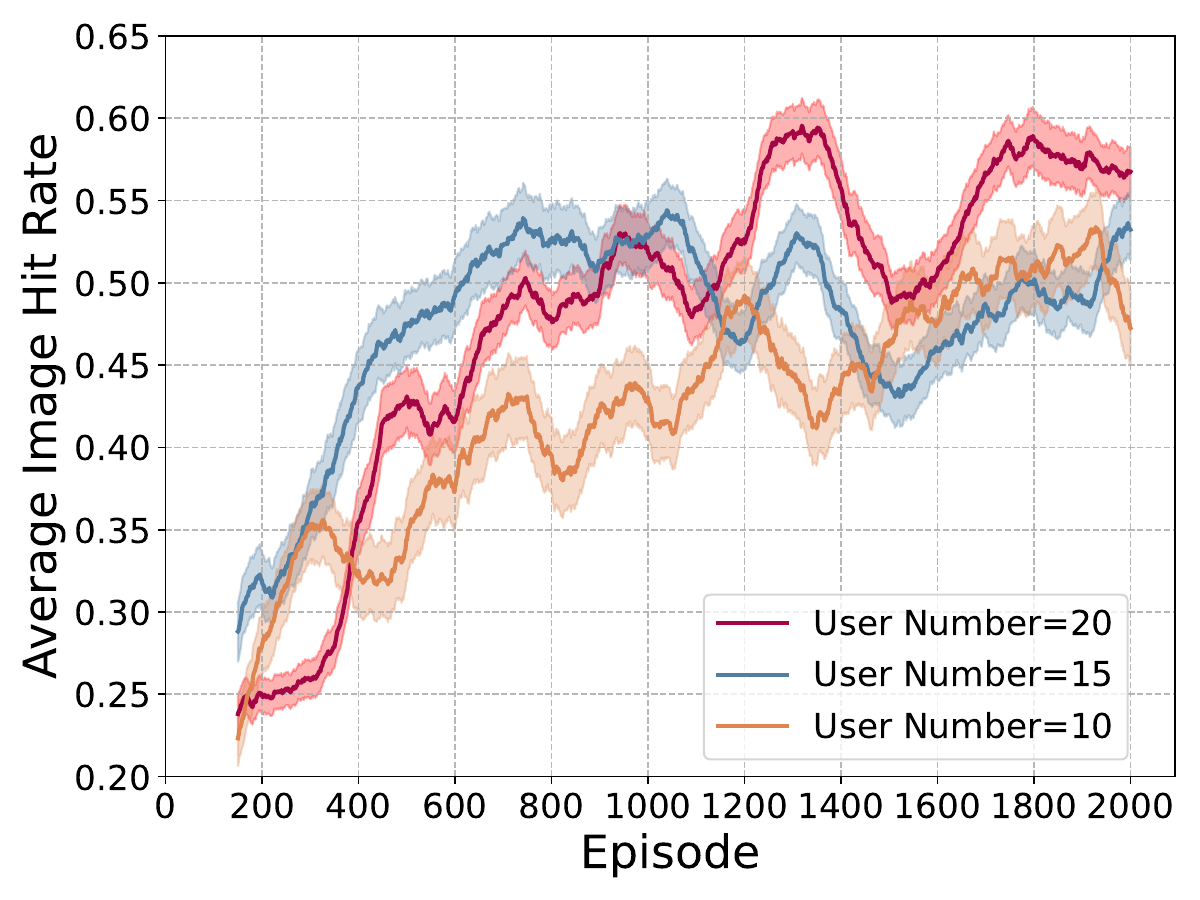}
    \caption{Hit rate with different user numbers}
    \label{Sensitivity:sub4}
\end{subfigure}
\hfill
\begin{subfigure}{0.325\textwidth}
    \includegraphics[width=\linewidth]{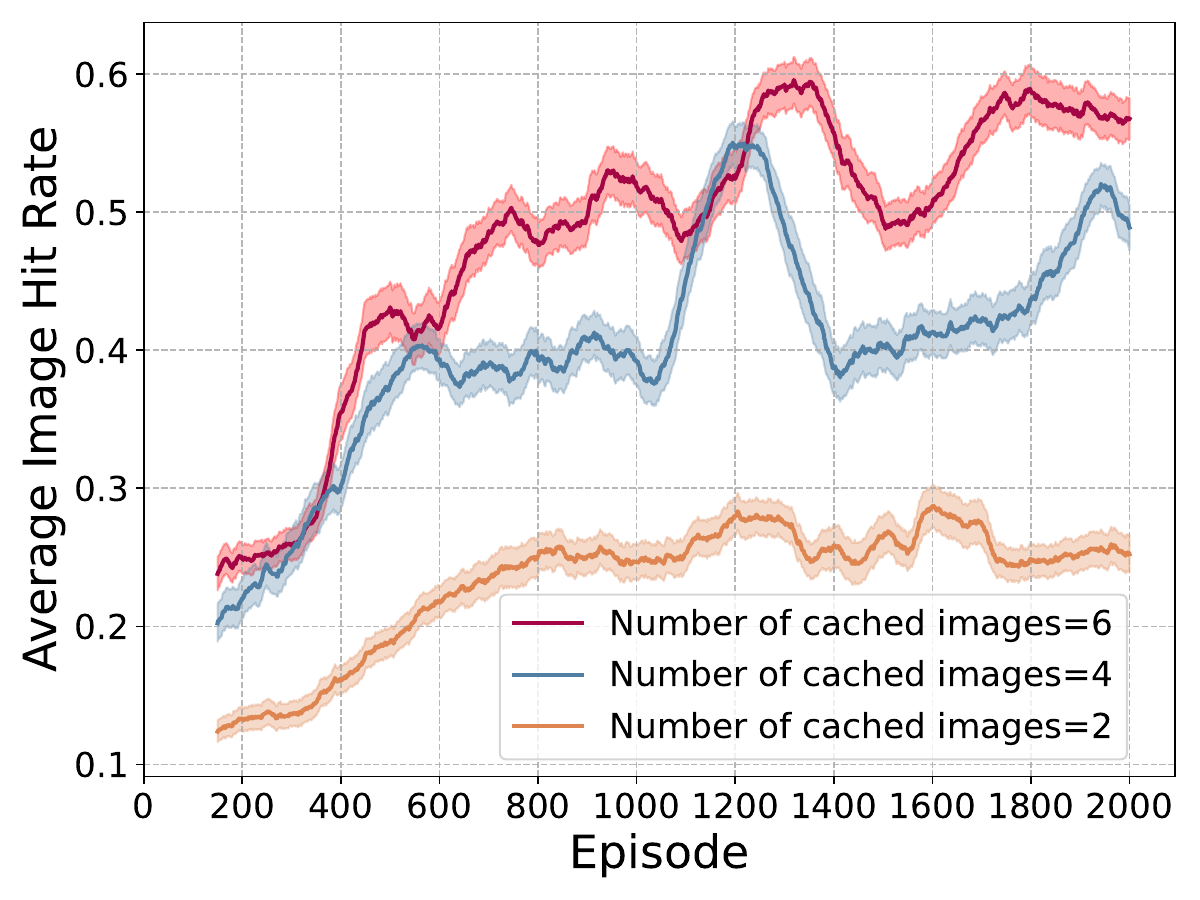}
    \caption{Hit rate with different numbers of cached images}
    \label{Sensitivity:sub5}
\end{subfigure}
\hfill
\begin{subfigure}{0.325\textwidth}
    \includegraphics[width=\linewidth]{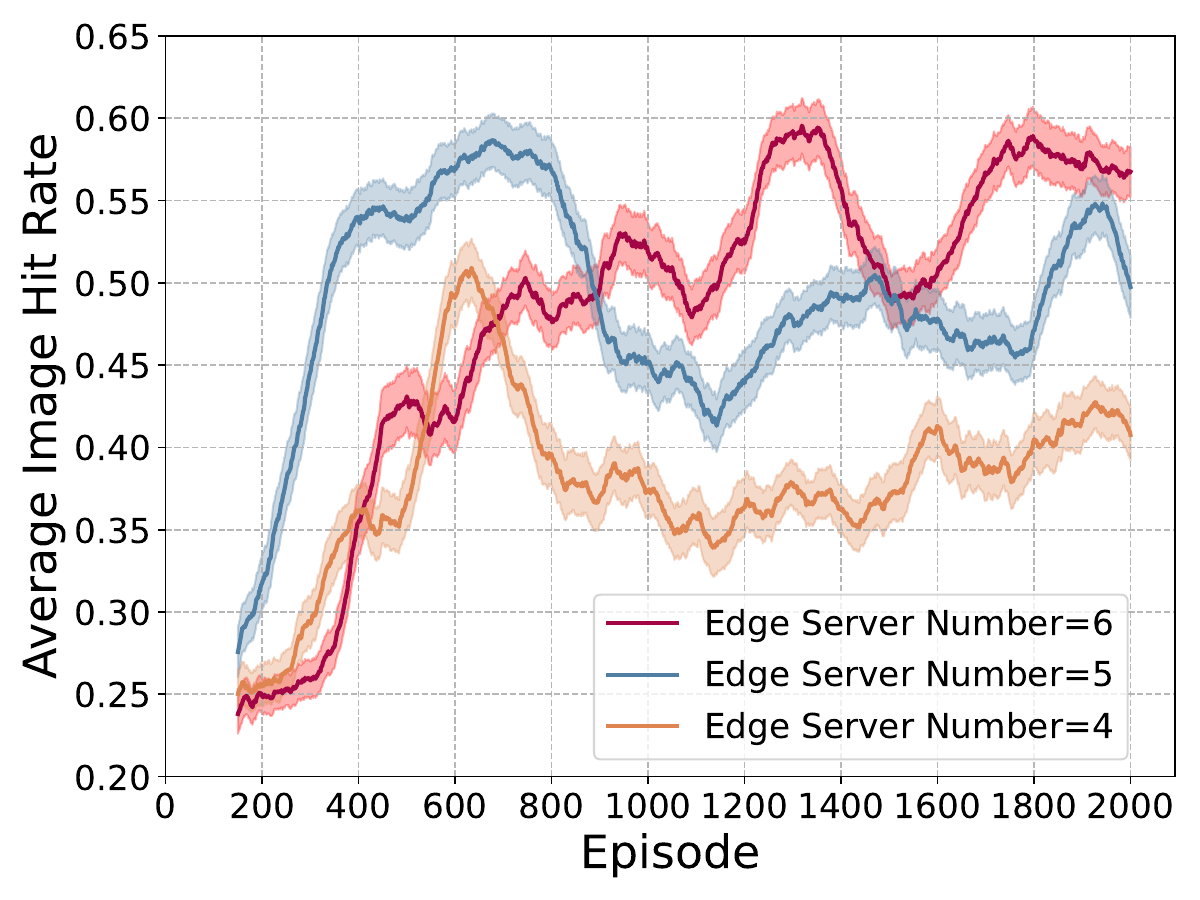}
    \caption{Hit rate with different edge server numbers}
    \label{Sensitivity:sub6}
\end{subfigure}
\caption{Performance of average system reward and average microservice image hit rate with different user numbers, numbers of images cached by each edge server, and edge server numbers.}
\label{Sensitivity}
\end{figure*}

Given the highly dynamic and time-varying nature of the edge environment, analyzing algorithm sensitivity to system changes is essential for its performance evaluation. Therefore, we conducted sensitivity analysis of our DLA-HDRL approach for several key variables of system scales, including the number of users $|U|$, the number of images cached by each edge server $\beta$, and the number of edge servers $|E|$, as shown in Fig. \ref{Sensitivity}. We have the following observations and inferences. 

1) As observed in Figs. \ref{Sensitivity:sub1} and \ref{Sensitivity:sub4}, the proposed DLA-HDRL approach consistently demonstrates full convergence after a certain number of episodes, regardless of changes in the number of users in edge. Furthermore, we observe that as the number of users increases, the final converged average reward value decreases, while the average image hit rate increases. This occurs because a higher number of users leads to a longer task queue per time slot, which consequently results in longer average task offloading completion times per episode. The increase in the average image hit rate with more users is attributed to our proposed DAMH-CA mechanism, which can effectively characterize long-range dependencies among microservice images for different tasks within an extended task queue. This capability enables efficient image caching decisions even as the task queue becomes longer. 

2) Figs. \ref{Sensitivity:sub2} and \ref{Sensitivity:sub5} illustrate the average reward results and average hit rate results of our DLA-HDRL approach as the number of microservice images cached by each server $\beta$ is from 2 to 6. For different values of $\beta$, the average reward of DLA-HDRL initially experiences a sharp increase and subsequently stabilizes over episodes. As the value of $\beta$ increases, both the converged average reward and the average hit rate increase. This improvement occurs because an edge server caching a larger number of images is enabled to store a greater variety of image layers locally. Consequently, it can better exploit the layered structure of microservice images to reduce the image pull delay. 

3) Figs. \ref{Sensitivity:sub3} and \ref{Sensitivity:sub6} depict the sensitivity of our DLA-HDRL approach to the number of edge servers from 4 to 6. Our experiments indicate that regardless of the number of edge servers, the average reward curve of our proposed method consistently achieves convergence by the end of training. Furthermore, as the number of edge servers decreases, both the converged average reward and the average hit rate of DLA-HDRL exhibit a significant decline. This performance occurs because a reduction in the number of edge servers diminishes the capacity of the system for caching microservice images and reduces its overall computational capability for task execution. Consequently, this leads to a substantial increase in the average task completion time while the average hit rate is reduced.

In summary, our experimental results demonstrate that DLA-HDRL has strong robustness in dynamic edge environments.

\subsection{Ablation Study for DLA-HDRL}

\begin{figure*}[!t]
\centering
\begin{subfigure}{0.245\textwidth}
    \includegraphics[width=\linewidth]{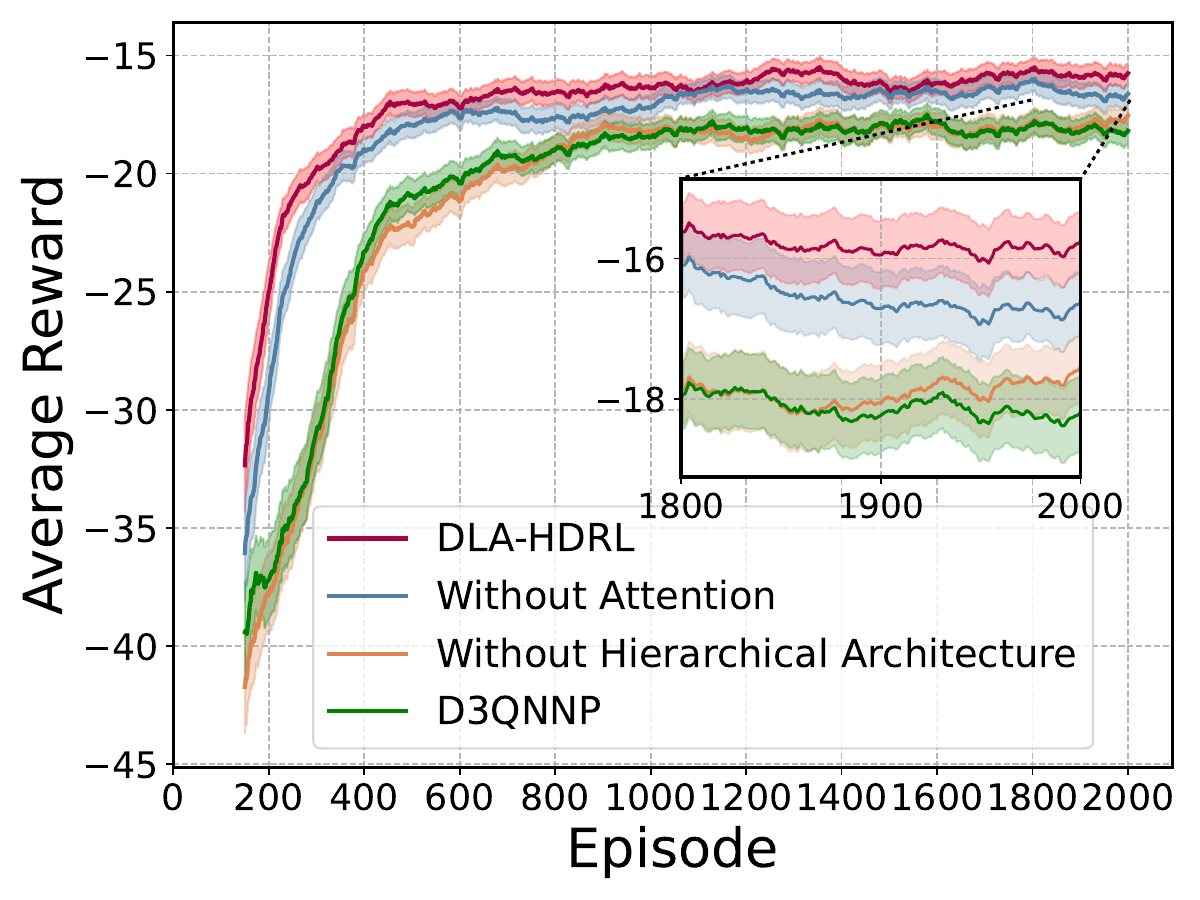}
    \caption{Average reward in training set}
    \label{Ablation:sub1}
\end{subfigure}
% \hspace{0.01\textwidth} % 替换 \hfill，手动控制间距
\begin{subfigure}{0.245\textwidth}
    \includegraphics[width=\linewidth]{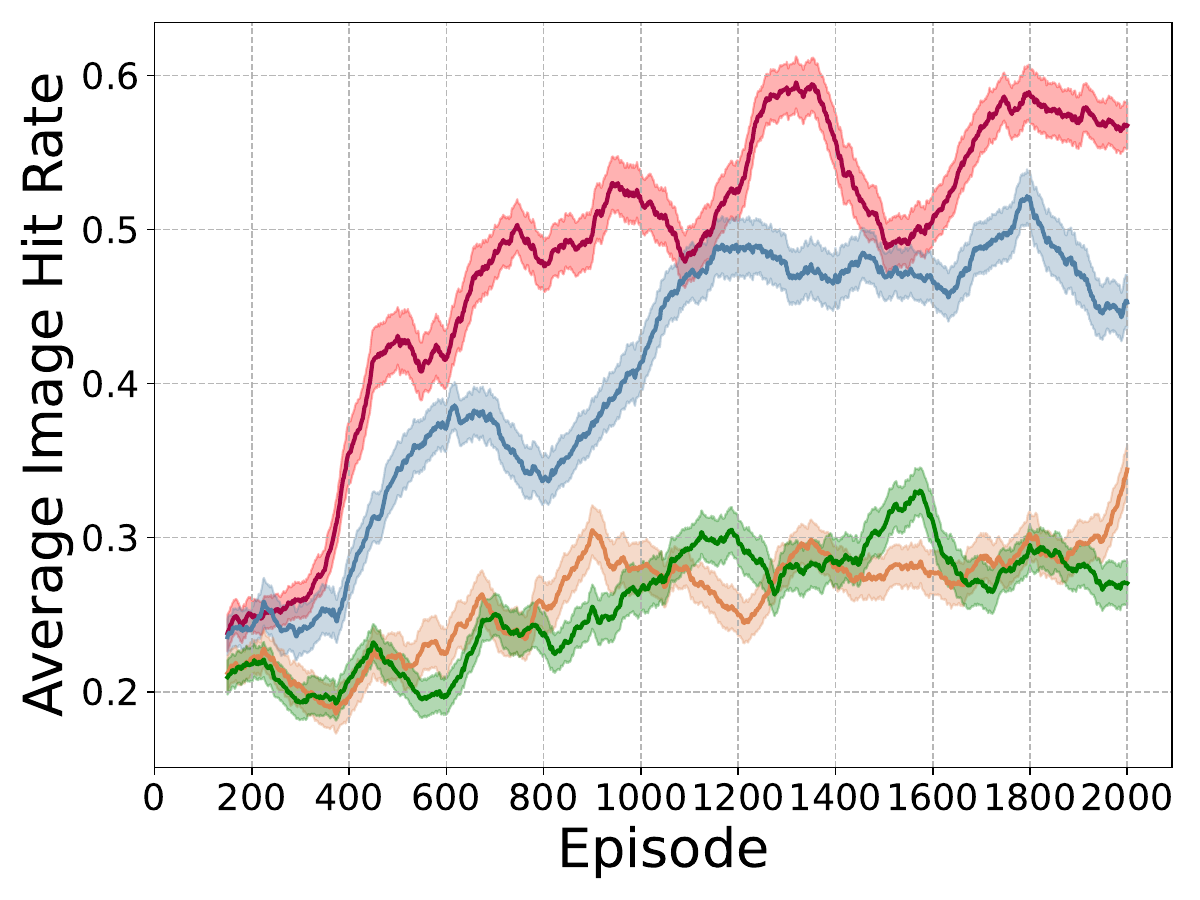}
    \caption{Average hit rate in training set}
    \label{Ablation:sub2}
\end{subfigure}
\begin{subfigure}{0.245\textwidth}
    \includegraphics[width=\linewidth]{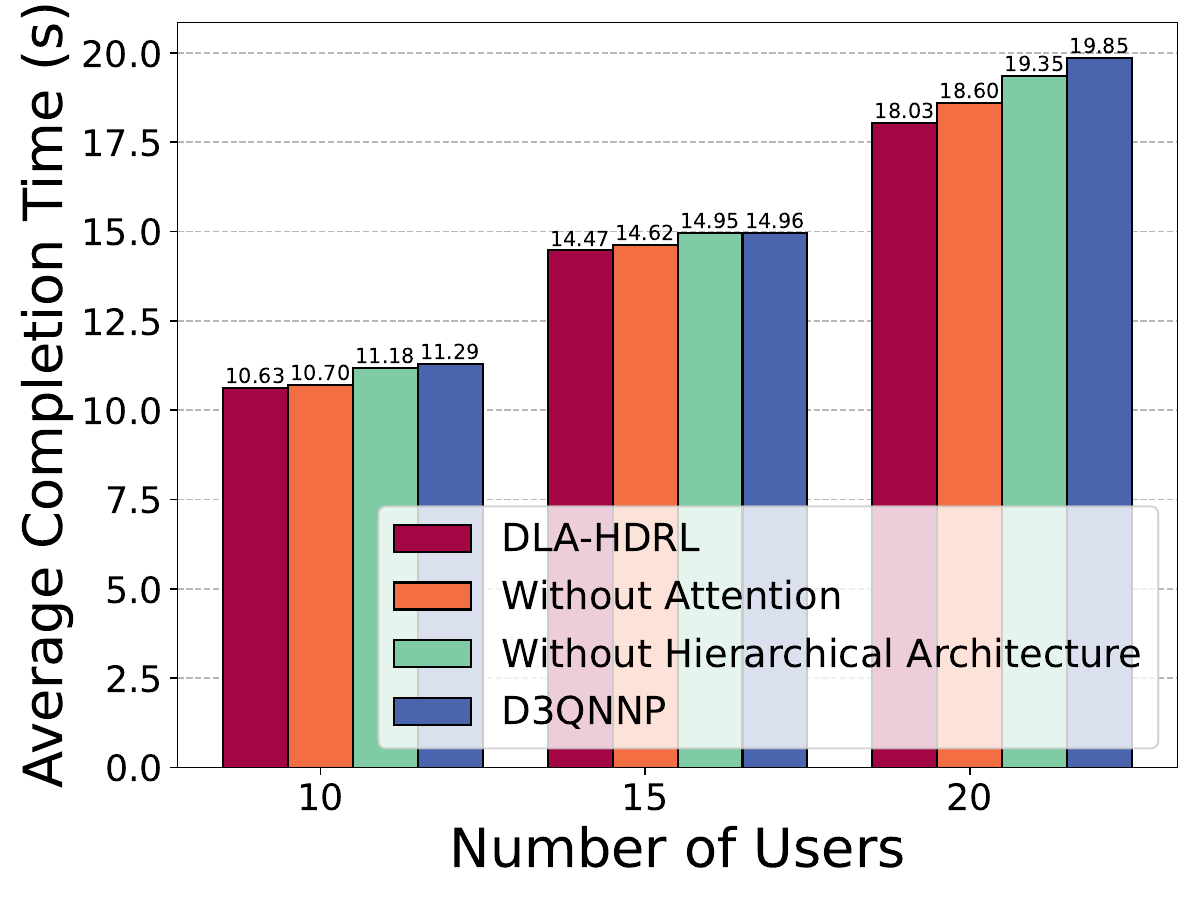}
    \caption{Average completion time in test set}
    \label{Ablation:sub3}
\end{subfigure}
% \hspace{0.01\textwidth} % 替换 \hfill，手动控制间距
\begin{subfigure}{0.245\textwidth}
    \includegraphics[width=\linewidth]{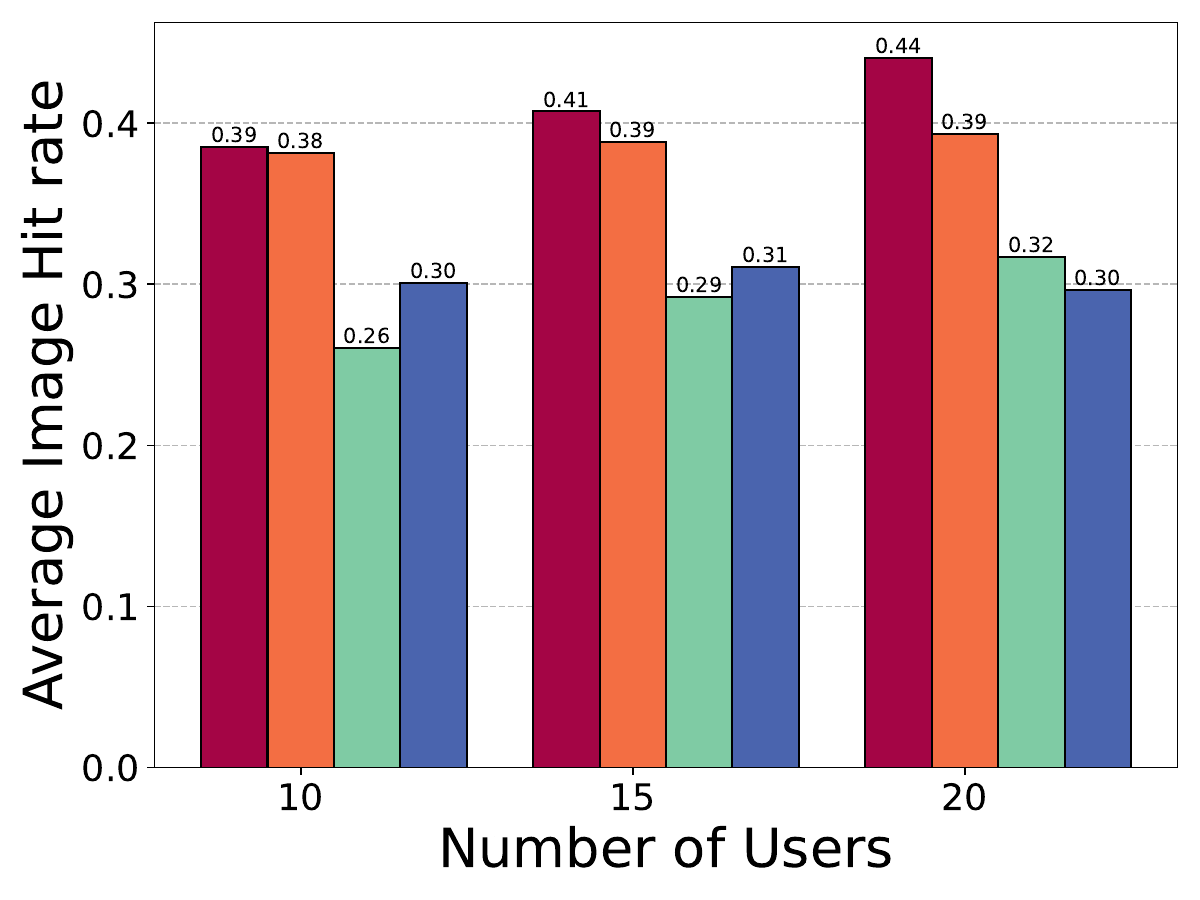}
    \caption{Average image hit rate in test set}
    \label{Ablation:sub4}
\end{subfigure}
\caption{Ablation study on attention mechanisms and HDRL architecture in our DLA-HDRL approach.}
\label{Ablation}
\end{figure*}

We performed an ablation study to validate the effectiveness of the attention mechanisms and the HDRL architecture in our DLA-HDRL approach. This study evaluated the DLA-HDRL approach with different configurations: 1) without attention mechanisms, 2) without hierarchical architecture, and 3) without attention mechanisms and hierarchical architecture (equal to D3QNNP), as shown in Fig. \ref{Ablation}.

Figs. \ref{Ablation:sub1} and \ref{Ablation:sub2} present the average reward and the average image hit rate, achieved by different algorithms after convergence of training. It is evident that across the four algorithm configurations, the convergence values for both average reward and average image hit rate follow the order (from highest to lowest): DLA-HDRL, DLA-HDRL without attention, DLA-HDRL without hierarchical architecture, and DLA-HDRL without attention and hierarchical architecture. Furthermore, Figs. \ref{Ablation:sub3} and \ref{Ablation:sub4} demonstrate that when the number of users changes, the DLA-HDRL approach reduces the average task completion time by up to 3.05\%, 6.8\%, and 9.19\% compared to the other three configurations, respectively. Simultaneously, it increases the average image hit rate by up to 11.92\%, 38.96\%, and 48.68\%. These results robustly validate that both the attention mechanism and the hierarchical architecture within the DLA-HDRL approach enhance the performance of our approach in task offloading and service caching. Additionally, we observe that the average image hit rate exhibits no improvement in DLA-HDRL without attention and the D3QNNP method as user numbers increase. In contrast, DLA-HDRL and DLA-HDRL without hierarchical architecture show significant improvements in the average image hit rate. This observation further verifies that the proposed attention mechanism effectively characterizes long-range dependencies as the task queue length increases, thereby improving the performance of the two approaches.

\section{Conclusion and Future Work}
\label{Sec:Conclusion}
In this paper, we investigated the joint optimization of microservice workflow offloading and service image caching for edge environments, aiming to provide users with low-delay computational services. Faced with this complex joint optimization problem, we proposed the DLA-HDRL approach based on the HDRL architecture, which decouples the large-scale state and action spaces of the problem to facilitate efficient agent convergence. Furthermore, considering the layer sharing characteristics of microservice images, we proposed a Layer-Aware Cross-Attention (LACA) mechanism. This mechanism reduces the delay of pulling images by computing the attention of task images to images cached by servers. Subsequently, for microservice workflow data represented as DAGs, we proposed a Dependency-Aware Multi-Head Cross-Attention (DAMH-CA) mechanism that effectively assists the agent in extracting long-range dependency features among tasks. Experimental results conducted on a platform comprising both our constructed dataset and real-world data demonstrate that our DLA-HDRL approach achieved superior performance compared to baseline methods across diverse edge computing environments.

With the increasing maturity of LLM technologies, in the future, we plan to explore the mechanism and application of LLMs in this problem and also intend to employ LLMs in other related problems for microservices, such as service deployment, service orchestration, and resource scheduling.
% 参考文献部分
\bibliographystyle{IEEEtran} % 使用IEEEtran参考文献样式
\bibliography{ref} % 引用ref.bib文件（不需要.bib扩展名）

\section{Biography Section}

\begin{IEEEbiography}[{\includegraphics[width=1in,height=1.25in,clip,keepaspectratio]{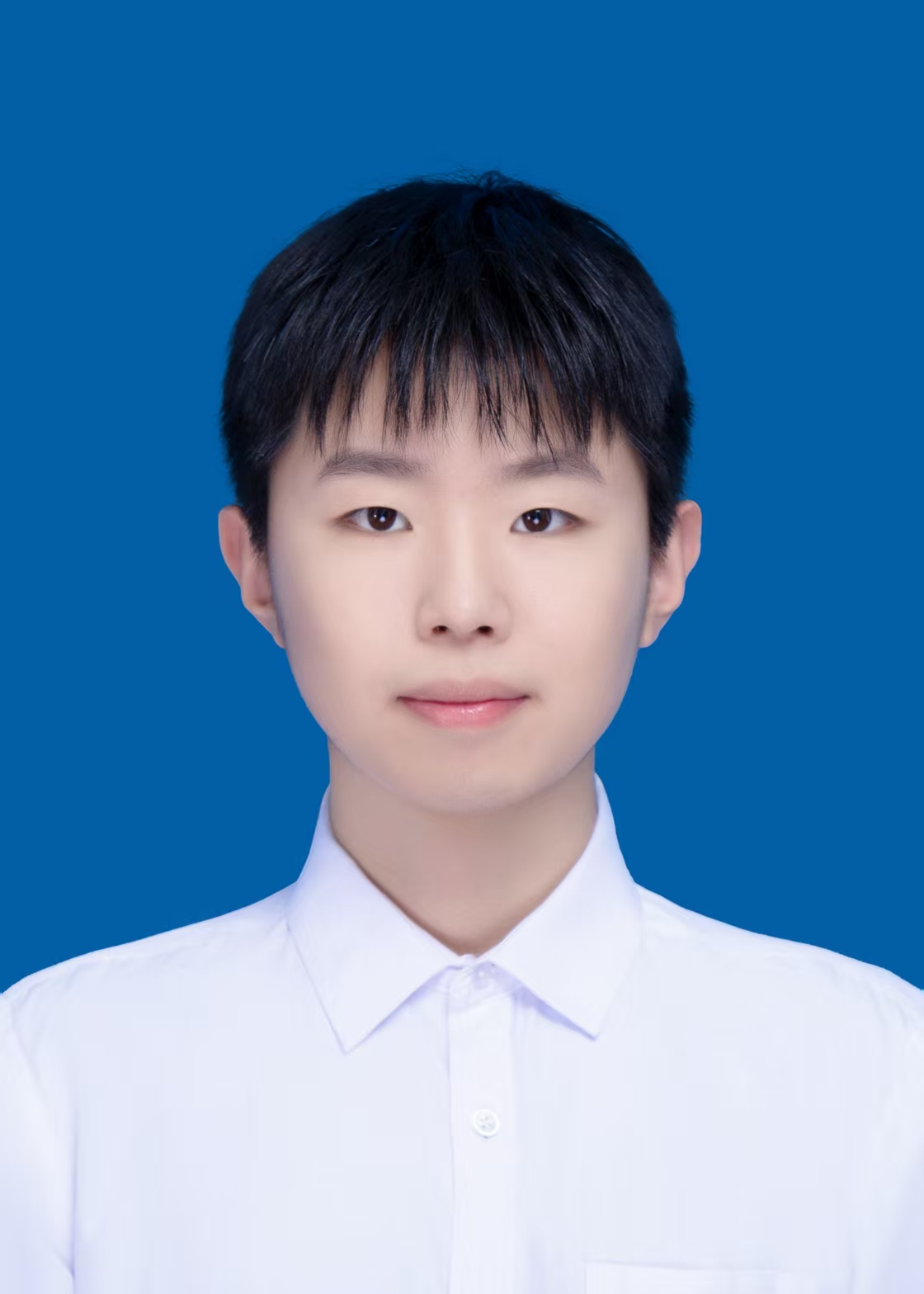}}]{Zhongxiao Wang}
 is a postgraduate student at School of Computer Science and Technology, Xidian University. He is expected to obtain his master degree in 2028. His research interests include edge computing and distributed computing.
\end{IEEEbiography}

% \vspace{3pt}
\begin{IEEEbiography}[{\includegraphics[width=1in,height=1.25in,clip,keepaspectratio]{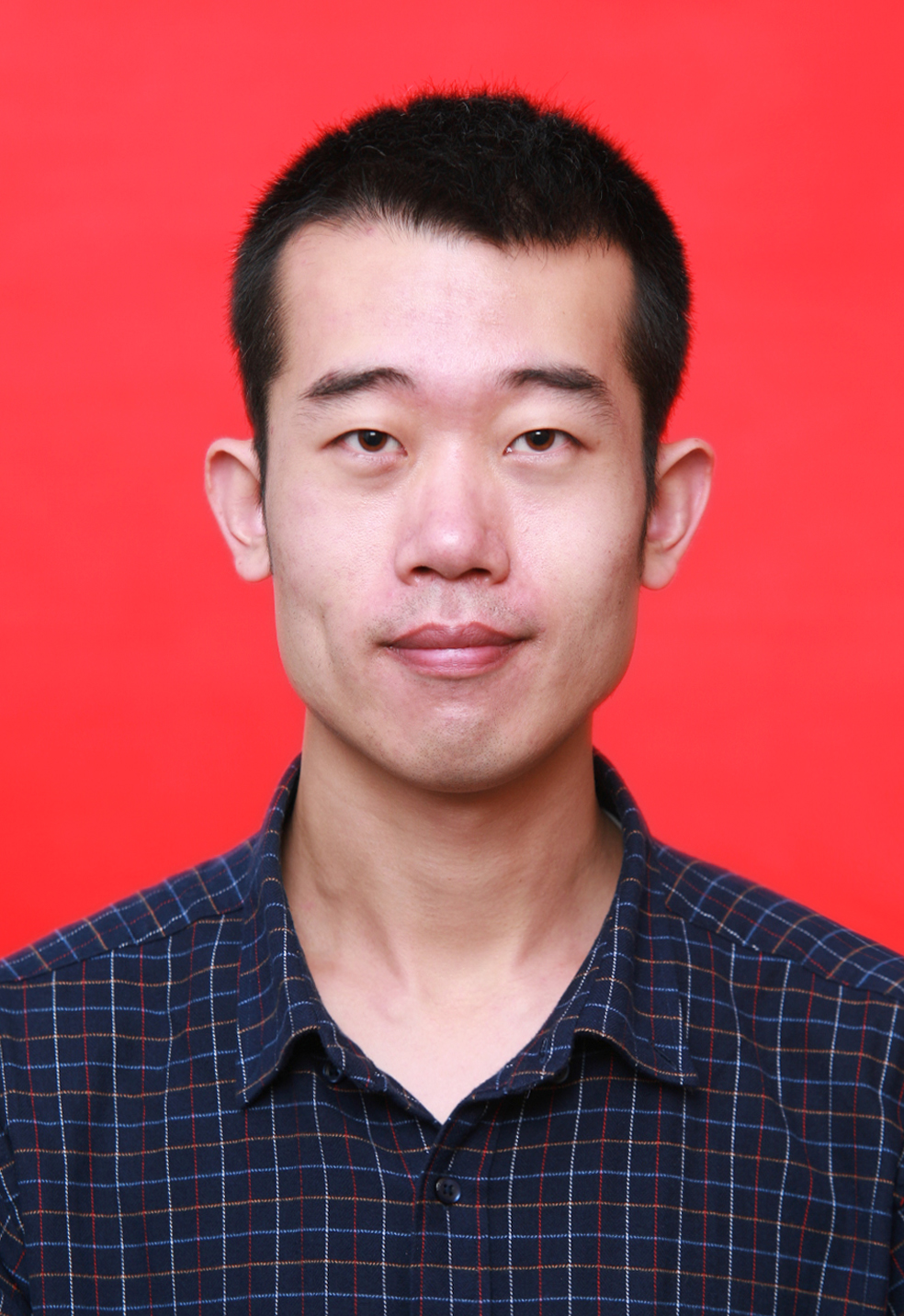}}]{Yueshen Xu} (Member, IEEE)
is an associate professor in School of Computer Science and Technology, Xidian University. He received his Ph.D. degree from Zhejiang University, and was a visiting scholar in University of Illinois at Chicago. His research focuses on edge computing, cyber-physical-social systems, and distributed computing. He has published more than 80 papers in prestigious conference and journals such as IEEE TSC, IEEE IoT-J, IEEE TITS, IEEE TII, IEEE TVT, IEEE TNSE, IEEE TCCN, IEEE TGCN, IEEE TETCI, IEEE TCSS, Info. Sci., FGCS, WWW, ICWS, and ICSOC. He has several ESI highly-cited papers, more than 3000 citations, and 28 as his h-index. He was selected as one of the  ``World's Top 2\% Scientists in 2023'' by the research team of Stanford University.
\end{IEEEbiography}

% \vspace{3pt}
\begin{IEEEbiography}[{\includegraphics[width=1in,height=1.25in,clip,keepaspectratio]{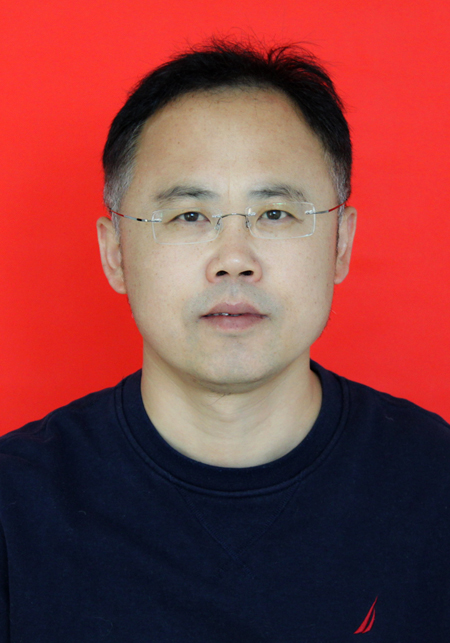}}]{Qingshan Li}
is a Professor of the School of Computer Science and Technology, Xidian University. His research spans multiple cutting-edge areas including autonomous systems, software engineering, open-source software, cloud-edge-fog computing, and large language models. He has authored or co-authored over 130 papers published in top-tier conferences (e.g., ICSE, ESEC/FSE, ASE) and premier journals (e.g., ACM/IEEE Transactions), along with three monographs. He also holds more than 40 granted or pending patents. He has chaired program or organization committees for several prestigious conferences such as ACM SIGIR 2020, CCF China Software Conference 2021, CCF China Open Source Conference 2022, and CCF China Software Conference 2024.
\end{IEEEbiography}

% \vspace{3pt}
\begin{IEEEbiography}[{\includegraphics[width=1in,height=1.25in,clip,keepaspectratio]{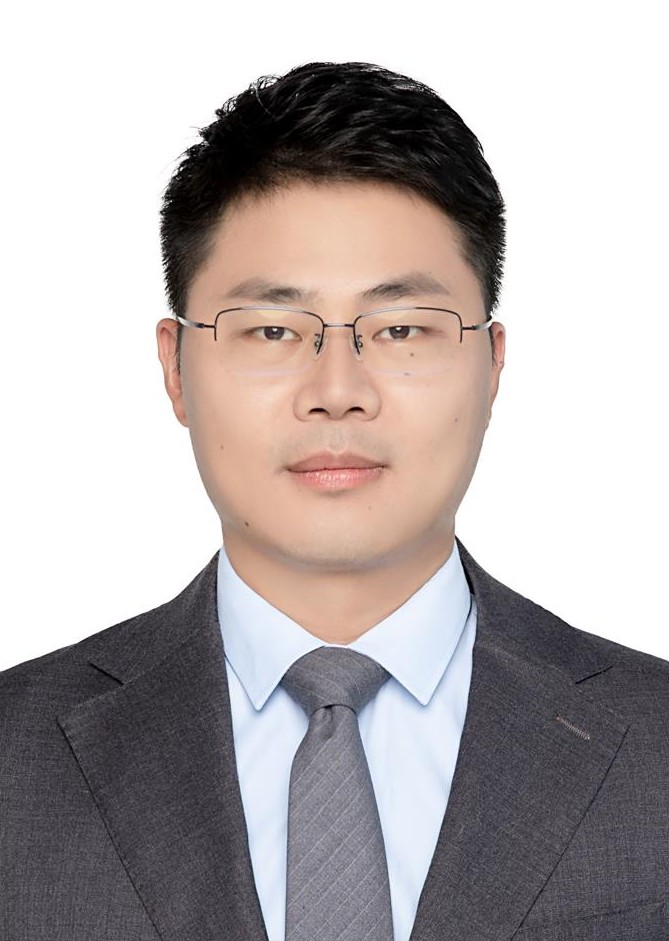}}]{Xinkui Zhao}
is a ZJU 100-Young professor at Zhejiang University. He received his PhD degree from the College of Computer Science and Technology at Zhejiang University, and then worked as the director of container services at Huawei Cloud. His research interests include cloud-native architecture and intelligent operating systems. He has led several enterprise cloud-native platforms and authored or coauthored more than 50 papers published on leading conferences and journals.
\end{IEEEbiography}

% \vspace{3pt}
\begin{IEEEbiography}[{\includegraphics[width=1in,height=1.25in,clip,keepaspectratio]{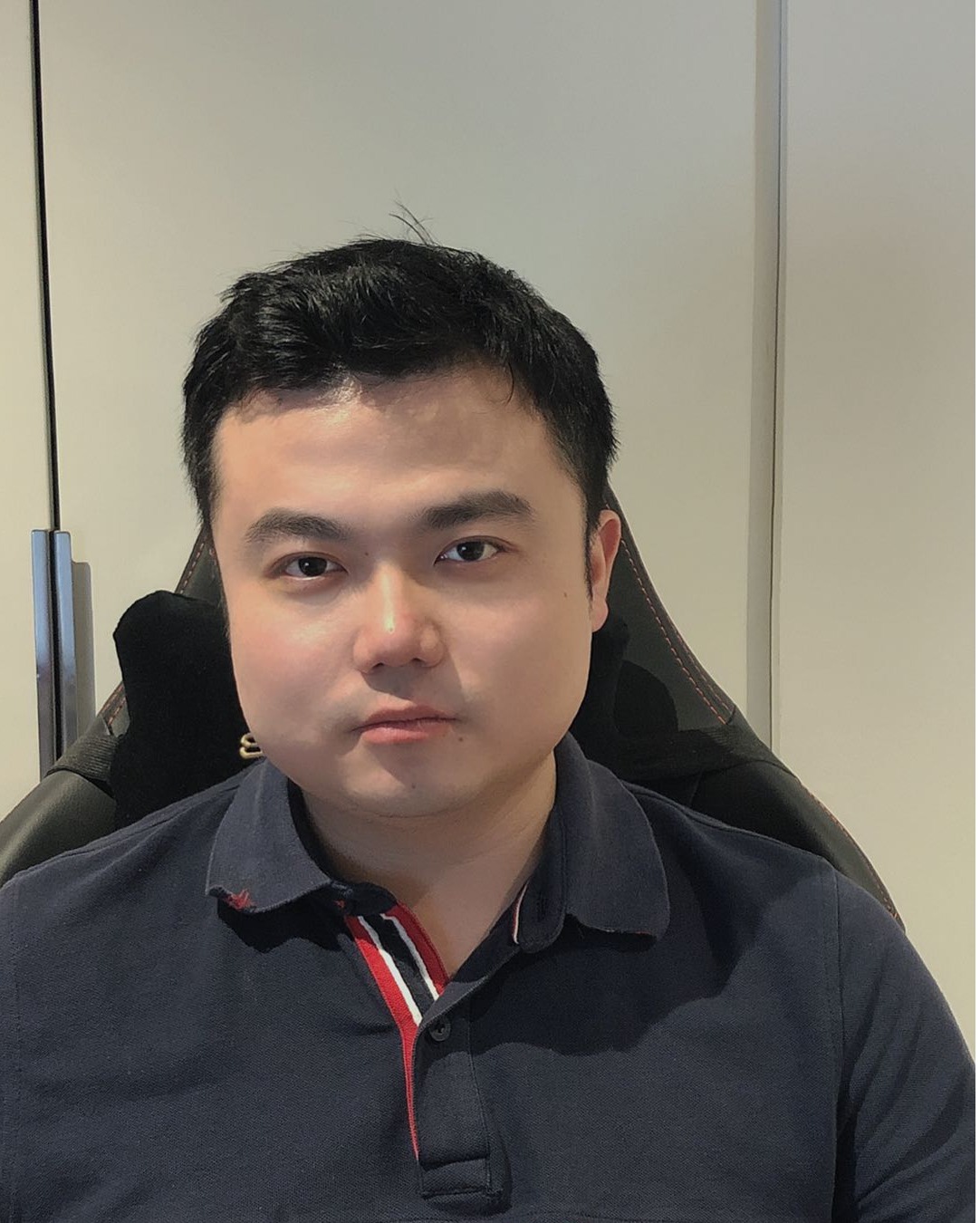}}]{Wei Shao}  (Member, IEEE) is an IWY (Impossible Without You) Research Scientist at CSIRO Data61, an Adjunct Lecturer at the University of New South Wales (UNSW), a Visiting Researcher at the University of California, Davis (UC Davis), and an Adjunct Fellow at RMIT University. Previously, he served as a Postdoctoral Researcher at UC Davis and Arizona State University. He completed his Ph.D. degree in Computer Science in 2018 from the RMIT University, Australia. His research interests encompass Cybersecurity, Graph Neural Networks, Spatio-temporal Data Mining, Reinforcement Learning Applications, and the Internet of Things (IoT). Wei has authored over 60 research papers published in leading conferences and journals. His contributions to academia have been recognized with accolades such as the Distinguished Paper Award at UbiComp and the Best Reviewer Award at KDD. Additionally, he received CSIRO’s Early Career in Science Award for his outstanding scientific contributions to Australian research.
\end{IEEEbiography}

% \vspace{3pt}
\begin{IEEEbiography}[{\includegraphics[width=1in,height=1.25in,clip,keepaspectratio]{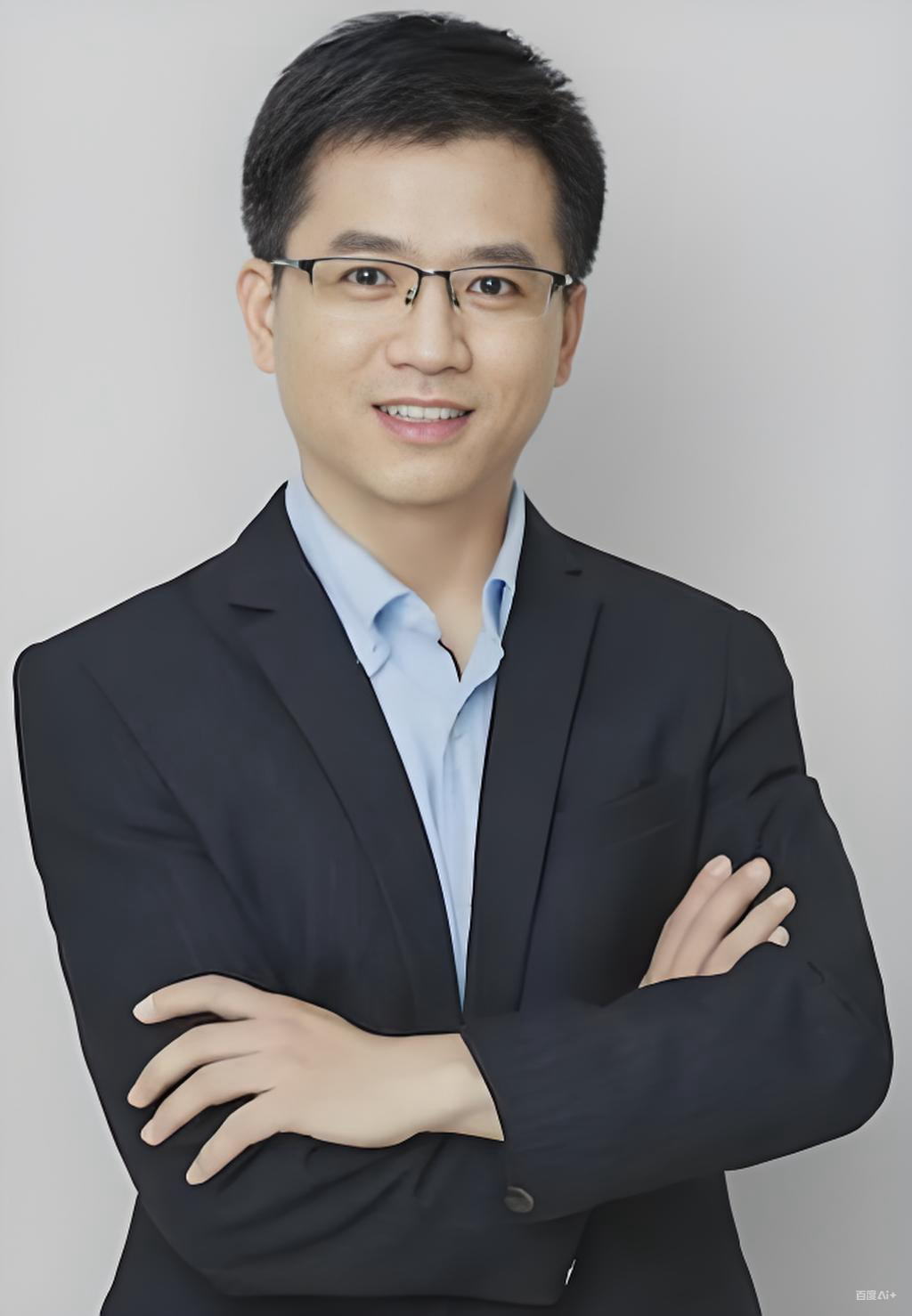}}]{Shuiguang Deng} (Senior Member, IEEE)
 is a full professor at the College of Computer Science and Technology in Zhejiang University. He received the BS and PhD both in Computer Science from Zhejiang University in 2002 and 2007, respectively. His research interests include Service Computing, Mobile Computing, and Edge Computing. Up to now he has published more than 100 papers in journals such as IEEE TOC, TPDS, TSC, TCYB, and TNNLS, and refereed conferences. He is the Associate Editor of the journal IEEE Trans. on Services Computing and IET Cyber-Physical Systems Theory \& Applications. He is a senior member of IEEE.
\end{IEEEbiography}

% \vspace{3pt}
\begin{IEEEbiography}[{\includegraphics[width=1in,height=1.25in,clip,keepaspectratio]{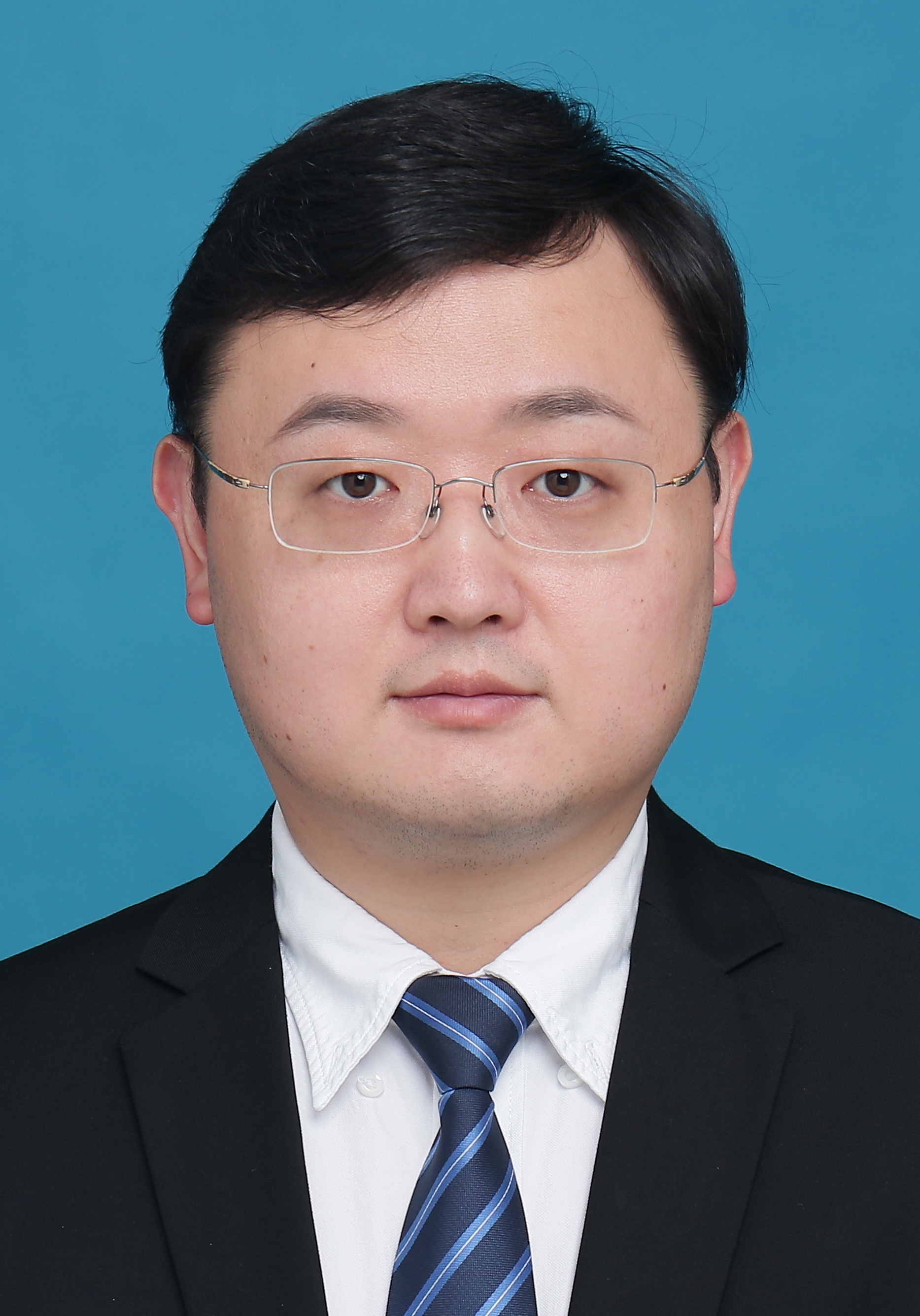}}]{Rui Li} (Member, IEEE) obtained the bachelor degree from Xidian University, at the Department of Applied Mathematics. He obtained his Ph.D. degree from Xi'an Jiaotong University, at the Department of Computer Science and Technology in 2014. He is currently a professor in School of Computer Science and Technology, Xidian University. His interested topics include social computing and mobile computing. He is a member of academic organizations, including IEEE, CCF, and ACM. He has near forty papers that have been published in related journals and conferences.
\end{IEEEbiography}

\vfill
\end{document}